\documentclass[a4paper,fleqn,usenatbib]{mnras}

\usepackage[T1]{fontenc}
\usepackage{ae,aecompl}

\usepackage{graphicx}	
\usepackage{amsmath}	
\usepackage{amssymb}	

\title[MUSE view of NGC~5102]{Dominant dark matter and a counter rotating disc: MUSE view of the low luminosity S0 galaxy NGC~5102\thanks{Based on observations collected at the European Organisation for Astronomical Research in the Southern Hemisphere under ESO programme 60.A-9308(A).} \thanks{Based on observations made with the NASA/ESA Hubble Space Telescope, and obtained from the Hubble Legacy Archive, which is a collaboration between the Space Telescope Science Institute (STScI/NASA), the Space Telescope European Coordinating Facility (ST-ECF/ESA) and the Canadian Astronomy Data Centre (CADC/NRC/CSA).}}

\author[M. Mitzkus, M. Cappellari and C.J. Walcher]{
Martin Mitzkus$^{1}$\thanks{E-mail: mmitzkus@aip.de},
Michele Cappellari$^{2}$ and
C. Jakob Walcher$^{1}$
\\
\\
$^{1}$Leibniz-Institut f\"ur Astrophysik Potsdam (AIP), An der Sternwarte 16, 14482 Potsdam, Germany\\
$^{2}$Sub-department of Astrophysics, Department of Physics, University of Oxford, Denys Wilkinson Building, Keble Road,\\
 Oxford OX1 3RH, UK
}

\date{14 October 2016}

\pubyear{2016}

\begin{document}
\label{firstpage}
\pagerange{\pageref{firstpage}--\pageref{lastpage}}
\maketitle

\begin{abstract}

The kinematics and stellar populations of the low-mass nearby S0 galaxy NGC~5102 are studied from integral field spectra taken with the Multi-Unit Spectroscopic Explorer (MUSE). The kinematic maps reveal for the first time that NGC~5102 has the  characteristic $2\sigma$ peaks indicative of galaxies with counter-rotating discs. This interpretation is quantitatively confirmed by fitting two kinematic components to the observed spectra. Through stellar population analysis we confirm the known young stellar population in the centre and find steep age and metallicity gradients. We construct axisymmetric Jeans anisotropic models of the stellar dynamics to investigate the initial mass function (IMF) and the dark matter halo of the galaxy. 
The models show that this galaxy is quite different from all galaxies previously studied with a similar approach: even within the half-light radius, it cannot be approximated with the self-consistent mass-follows-light assumption. Including an NFW dark matter halo, we need a heavy IMF and a dark matter fraction of $0.37\pm0.04$ within a sphere of one $R_{\rm e}$ radius to describe the stellar kinematics. The more general model with a free slope of the dark matter halo shows that slope and IMF are degenerate, but indicates that a light weight IMF (Chabrier-like) and a higher dark matter fraction, with a steeper (contracted) halo, fit the data better. Regardless of the assumptions about the halo profile, we measure the slope of the {\em total} mass density to be $-1.75\pm 0.04$. This is shallower than the slope of $-2$ of an isothermal halo and shallower than published slopes for more massive early type galaxies.

\end{abstract}




\section{Introduction}

NGC~5102 is a low luminosity S0 galaxy in the Centaurus group \citep{2002A&A...385...21K} at a distance of $4.0\pm0.2$\,Mpc, as determined via the surface brightness fluctuation method by \citet{2001ApJ...546..681T}. There is some uncertainty about the distance to NGC\,5102: values based on the Planetary Nebula luminosity function tend to find shorter distances of 3.1\,Mpc \citep{1994AJ....108.1610M}. The distances obtained from tip of the red-giant branch measurements scatter: \citet{2002A&A...385...21K} find a distance of 3.4\,Mpc, \citet{2008AJ....135.1636D} measures 3.2\,Mpc and \citet{2015ApJ...802L..25T} find 3.74\,Mpc. The choice of the distance \citep[throughout this paper we will use $D = 4.0\pm0.2$\,Mpc by][]{2001ApJ...546..681T} does not influence our conclusions but sets the scale of our models in physical units. Specifically, lengths and masses scale as $D$, while $M/L_{\ast}$ scales as $D^{-1}$.

The photometry of NGC~5102 shows the typical features of an S0 galaxy: a bulge that follows a $r^{1/4}$ profile and a disc with a flatter profile at larger radii \citep{1979ApJ...231..354P, 2015ApJ...799...97D}. The central deviation from the $r^{1/4}$ profile builds the nucleus. Other observations show that NGC~5102 is an unusual S0 galaxy: The nucleus has an unusually blue colour \citep[first mentioned by Freeman in ][]{1971QJRAS..12..305E} for S0 galaxies and is interpreted as a star cluster with an age of $\sim10^8$\,yr \citep[e.g. ][]{1975ApJ...202....7G, 1976AJ.....81..795V, 1979ApJ...231..354P}. 

The second peculiarity about NGC~5102 is the relatively large \ion{H}{I} content of $3.3\times10^8$\,M$_{\sun}$ \citep{1975ApJ...202....7G}. \citet{1993A&A...269...15V} map the spatial and spectral properties of the \ion{H}{I} and find a ring like structure with a prominent central depression with a radius of 2.2\,kpc. The rotation curve is aligned with the stellar disc and flattens at lager radii at 95\,km\,s$^{-1}$ \citep{1993A&A...269...15V, 2015MNRAS.452.3139K}. The \ion{H}{I} profile with at 50\% intensity is $W_{50}=200$\,km\,s$^{-1}$ \citep{1993A&A...269...15V, 2015MNRAS.452.3139K}, given that the flat rotation curve regime is reached this gives a measure of the maximum radial rotational velocity $V_m = W_{50}/2$.

The last notable unusual finding is the significant gas emission structure, first detected in H$\alpha$ by \citet{1976AJ.....81..795V}. Later \citet{1994AJ....108.1610M} imaged the gas emission in H$\alpha$ and [\ion{O}{III}] finding a ring like structure as well. Based on the line ratios \citet{1994AJ....108.1610M} conclude that the gas is ionized by a shock, moving with $\sim60$\,km\,s$^{-1}$. This structure has a diameter of 1.3\,kpc and an estimated age of $\sim10^7$\,yr. Based on this age estimate the authors conclude that this super shell is not associated with the blue nuclear star cluster. From the mass and the energy of the shell \citet{1994AJ....108.1610M} conclude that a few hundred supernovae are enough to form this super shell.

Observing NGC~5102 in X-ray \citet{2005ApJ...625..785K} find that the number of X-ray point sources is smaller than expected from galaxies of similar morphological type. The significance of these results is unclear due to contamination with background AGN and low number statistics. One of the X-ray point sources coincides with the optical centre of NGC~5102, hinting at weak AGN activity. This is in contrast to the findings of other authors that do not see AGN activity in NGC~5102 \citep[e.g. ][]{2004AJ....127.3338B}. The diffuse X-ray component is centrally peaked and centred on the optical galaxy centre. The authors suggest that this is the super shell detected by \citet{1994AJ....108.1610M} and their mass and energy estimates roughly agree with the values of \citet{1994AJ....108.1610M}.

The star formation history (SFH) of NGC~5102 has been investigated many times.
Using HST's 'Faint Object Camera' \citet{1997A&A...326..528D} detected a population of resolved blue stars find in the centre of NGC~5102. The authors interpret these as young stars from a star formation burst that occurred $1.5\times10^7$\,yr ago, but they cannot exclude that these stars are much older post AGB stars. In line with this \citet{2010AJ....139..984B} find a stable star formation rate (SFR) over the last 0.2\,Gyr with a recent (20\,Myr) star formation burst in the centre. A star formation burst a few $10^7$\,yr ago is in remarkable agreement with the estimated time-scale of the super bubble detected by \citet{1994AJ....108.1610M}.
Using long slit spectra, with the slit oriented along the minor axis of the galaxy, \citet{2015ApJ...799...97D} determine a luminosity weighted SSP age for the nucleus and bulge of $1^{+0.2}_{-0.1}$ and $2^{+0.5}_{-0.2}$\,Gyr respectively.
From Lick index measurements \citet{2015ApJ...799...97D} find a (spatially) roughly constant metallicity of $Z=0.004$ for NGC~5102. 

For the disc \citet{2010AJ....139..984B} find a declining SFR over the last 0.2\,Gyr and a recently quiescent SFR. 
\citet{2008AJ....135.1636D} determines the SFR of the disc in the last 10\,Myr to be 0.02\,M$_{\sun}$\,yr$^{-1}$. However during intermediate epochs the SFR was higher and 20\% of the disc mass formed within the last 1\,Gyr.

\citet{2010AJ....139..680D} finds hints for a merger in NGC~5102: at a projected distance of 18\,kpc from the galaxy centre along the minor axis he finds a concentration of AGB stars and suggests this might be the remnant of a companion galaxy. On the one hand this interpretation is supported by the warped \ion{H}{I} in the outer disc \citep{1993A&A...269...15V}. \citet{2010AJ....139..680D} points out that this warp is also traced by the OAGB stars. On the other hand the author mentions that this interpretation is challenged by the current isolation of NGC~5102.  \citet{2007AJ....133..504K} determine the nearest neighbour being ESO383-G087 at a projected distance of $0.3$ to $0.4$\,Mpc. 
\citet{2015ApJ...799...97D} determines an SSP age of $10\pm2$\,Gyr for the disc and finds the stellar population to be metal poor at $Z=0.004$, like the one in the nucleus and bulge. 

The mass of NGC~5102 was estimated by different authors: \citet{1993A&A...269...15V} find the total mass inside a radius of 6.1\,kpc from the \ion{H}{I} rotation curve to be $1.2\times10^{10}$\,M$_{\sun}$. \citet{2008AJ....135.1636D} subtract a mean dark matter density and derive a stellar mass of $7.0\times10^{9}$\,M$_{\sun}$ within the same region. 
\citet{2010AJ....139..984B} estimate the stellar mass by combining the photometry of NGC\,5102 with the $M_{\ast}/L_B$ models by \citet{2001ApJ...550..212B} and find it to be $(5.6\pm0.8)\times10^9$\,M$_{\sun}$. 

In this paper we analyse new high-quality integral-field observations of the NGC~5102 to study the stellar population and kinematics of this low-mass galaxy in more detail. We construct dynamical models to study the dark matter content.

\section{Observations and data reduction}

NGC~5102 was observed with the new second generation VLT instrument MUSE \citep{2010SPIE.7735E..08B} during the Science Verification run (programme 60.A-9308(A), PI: Mitzkus)  in the night from the 23rd to the 24th of June 2014.
The data were taken in the wide filed mode with a $1\arcmin\times 1$\arcmin\ field of view and a plate scale of 0\farcs2.
The normal mode covers the spectral range from 4750--9340\,\AA\ with a fixed sampling of 1.25\,\AA, the spectral FWHM is 2.5\,\AA\ over the whole wavelength range. 

The data set consists of 4 dithered exposures, each rotated by $90\degr$ with 960\,s exposure time, centred on NGC~5102. In  total we have a time on target of 3840\,s. The complicated image slicer system of MUSE leaves residual structures in the data. We applied the dither and rotation pattern to smoothly distribute these residuals. 
The observations where split in two observing blocks, scheduling between the two object exposures of each block a 300\,s offset sky exposure.
The calibration frames where taken according to the standard ESO calibration plan for MUSE.

The MUSE NGC~5102 data were reduced using the dedicated MUSE Data Reduction System \citep[DRS,][]{2012SPIE.8451E..0BW}  version 1.0.1 through the EsoRex\footnote{EsoRex stands for ESO Recipe Execution Tool, a command line based tool to execute data reduction software for ESO-VLT data.} program. 
For most of the reduction steps we followed the standard procedure as outlined in the MUSE Data Reduction Cookbook shipped with the DRS.
The sky subtraction for this data set is complicated because NGC~5102 extends over the full field of view, thus the sky cannot be estimated from the object exposure. 
The sky continuum and the sky lines where measured from the afore mentioned sky exposures  and then subtracted from the object exposures that were taken directly before and after the sky exposure.

The response curve is usually measured from the spectrophotometric standard star observed during the night the data were obtained. This would be the spectrophotometric standard star CD-32, but the reference calibration spectrum of this star is labelled as 'bad resolution, should not be used' in the MUSE data reduction cookbook.
Therefore we used the standard star GD153, observed in the following night. 
All other steps to measure and apply the response curve follow the standard procedure described in the MUSE Data Reduction Cookbook.

\section{NGC~5102 kinematics}

\subsection{Voronoi binning}

To make sure that all spectra have sufficient signal to noise ratio (S/N) for an unbiased kinematic extraction, especially at large radii, we used the Voronoi binning method\footnote{Available from http://purl.org/cappellari/software} \citep{2003MNRAS.342..345C} to co-add spatially adjacent spectra. 
We use a white light image in the wavelength range from 4800--6000\,\AA\ as signal input  and the square root of the mean pipeline propagated variances in the same wavelength range as noise input to the Voronoi binning routine. Because the pipeline uncertainties are not always reliable, due to the difficulty of propagating covariances, we verified the S/N after the binning. We derived the S/N as the ratio between the mean signal and the rms of the residual from a \texttt{pPXF} fit (see Sec.~\ref{subsec:One_component_kinematic_fit}). For the spectral range from 4760 to 7400\,\AA\ the median is $\text{S/N}=113$ per spectral pixel of 50.6 km/s (see table~\ref{tab:kinematic_fit_parameter} for details).
Finally all spectra falling on to the same Voronoi bin are added up, resulting in nearly 1300 Voronoi binned spectra.

\subsection{One component kinematic fit}
\label{subsec:One_component_kinematic_fit}

\begin{figure*}
 \includegraphics[width=\textwidth]{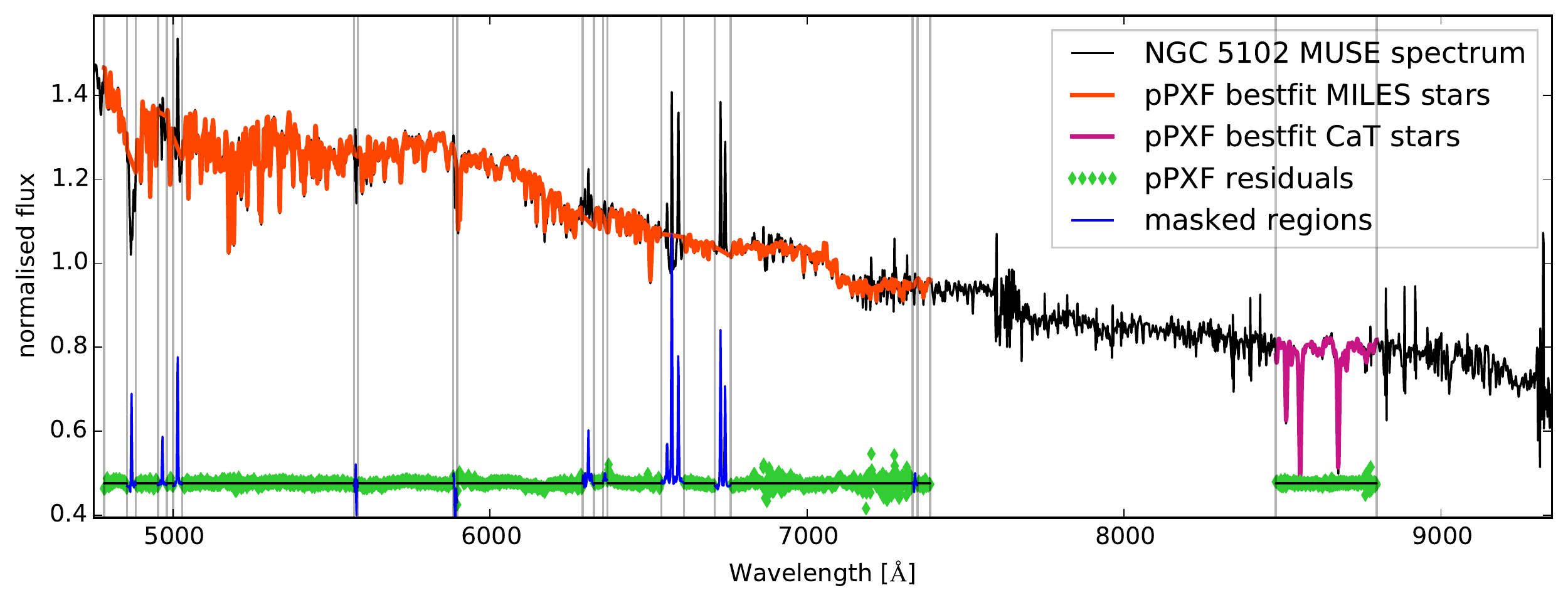}
 \caption{The collapsed NGC~5102 MUSE spectrum is shown in black. The best-fitting \texttt{pPXF} model, a linear combination of the MILES (red) or CaT (magenta) stars is shown in the respective wavelength ranges. The residuals between observed and model spectrum are shown in green, in masked regions these are blue. The grey vertical lines mark the masked intervals.}
 \label{fig:bestfit_4760-6800_bestfit}
\end{figure*}

We determine the line-of-sight velocity distribution (LOSVD) by using the penalized pixel fitting code$^2$
 \citep[\texttt{pPXF}, ][]{2004PASP..116..138C}. In \texttt{pPXF} the LOSVD is parametrized as a Gauss-Hermite function, allowing to measure departures from a pure Gaussian LOSVD. The input spectrum is fitted with a linear combination of spectra from a template library and additive or multiplicative polynomials might be used to correct for continuum mismatch. As template library we use the $\sim1000$ stellar spectra of the MILES library \citep{2006MNRAS.371..703S, 2011A&A...532A..95F} for fits in the 4760--7400\,\AA\ wavelength range, where the blue cut is set by the spectral coverage of MUSE and the red cut by the MILES library.
For the wavelength region 8440--8810\,\AA\ around the \ion{Ca}{ii} triplet (referred to as \ion{Ca}{ii} triplet region in the following) the $\sim700$ stars from the CaT library \citep{2001MNRAS.326..959C} are used. 
The MUSE spectral FWHM is 2.5\,\AA, which matches the instrumental resolution of the MILES stars \citep{2011A&A...531A.109B, 2011A&A...532A..95F}. Therefore this library is not convolved with an instrumental FWHM. Since the spectral resolution of the CaT library is higher (1.5\,\AA), we degrade the resolution of these stars to match the MUSE spectral resolution of 2.5\,\AA.

We logarithmically rebin the galaxy (for the MILES range we use a velocity step of 50.6\,km\,s$^{-1}$, for the \ion{Ca}{II} region the velocity step is 42.5\,km\,s$^{-1}$) and the template spectra before the fit and mask regions of gas emission (including H$\alpha$ and H$\beta$) and telluric emission lines. We use additive polynomials to account for the response function and template mismatch, as well as to account for spatial variation in the stellar line strength. For the  4760--7400\,\AA\ region we use a 7th order polynomial, the much shorter \ion{Ca}{ii} triplet region is fitted with a first order polynomial. For the fitting we use a two step approach: The collapsed spectrum (adding all MUSE spectra together) is fitted with all stars in the corresponding library. From this fit a template is computed by computing the weighted sum of the stellar spectra, using the weights \texttt{pPXF} determined. The collapsed NGC~5102 spectrum and the two template spectra are shwon in Fig.~\ref{fig:bestfit_4760-6800_bestfit} This spectrum is then used to fit the LOSVD of each Voronoi bin. 
Given the significant population gradient we observe in NGC~5102 (see Sec.~\ref{sec:Population_fitting}) we checked that the kinematics are not biased by the fixed template and computed the kinematics also from a fit where the full set of MILES stars is used as input template to \texttt{pPXF}. These extractions are qualitatively similar to the kinematics obtained with a fixed best fit template, but are noisier and less symmetric. Therefore we base the further discussion on the kinematics obtained with the fixed best fit template.

\begin{figure*}
 \includegraphics[width=0.49\textwidth]{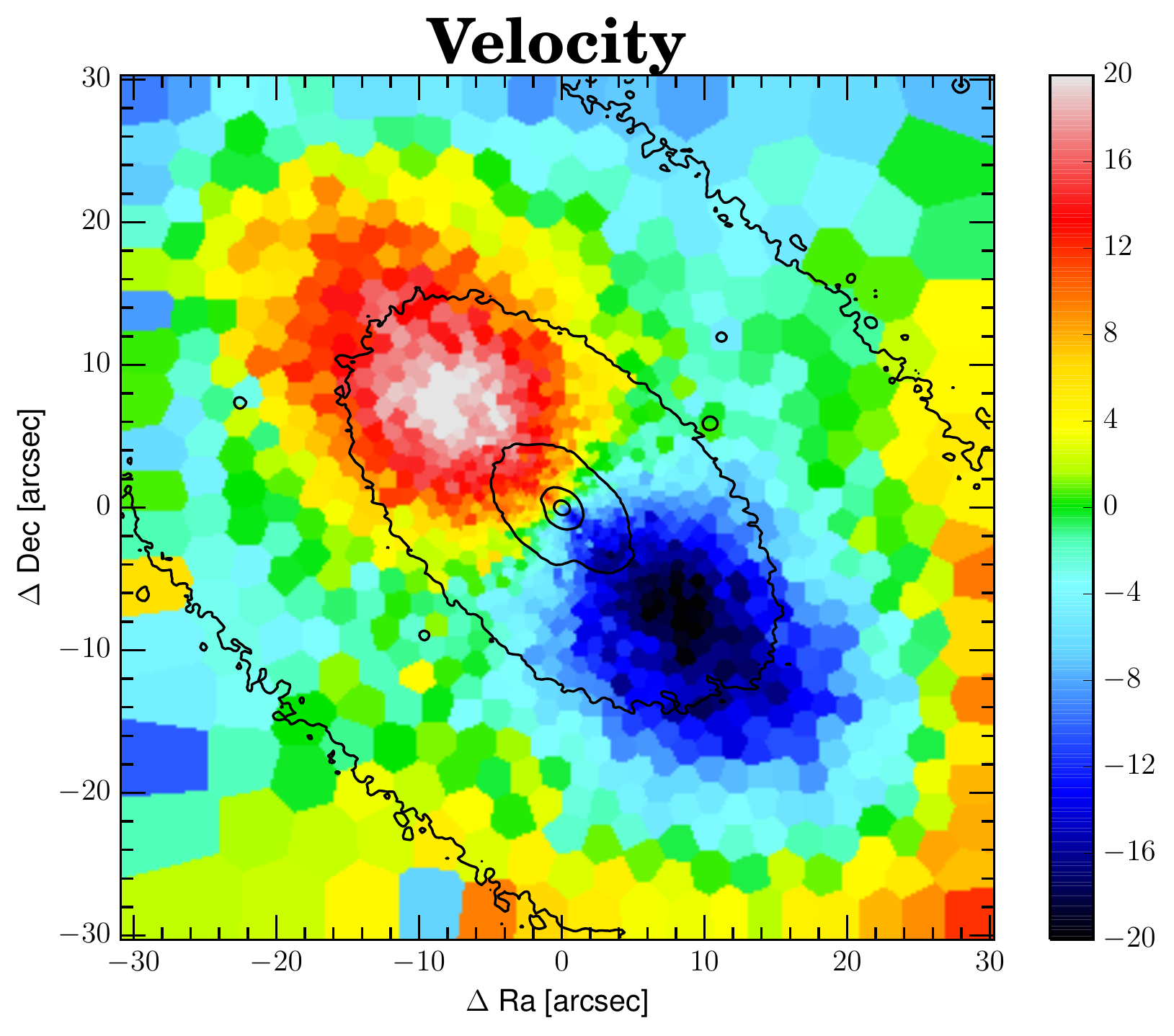}
 \includegraphics[width=0.49\textwidth]{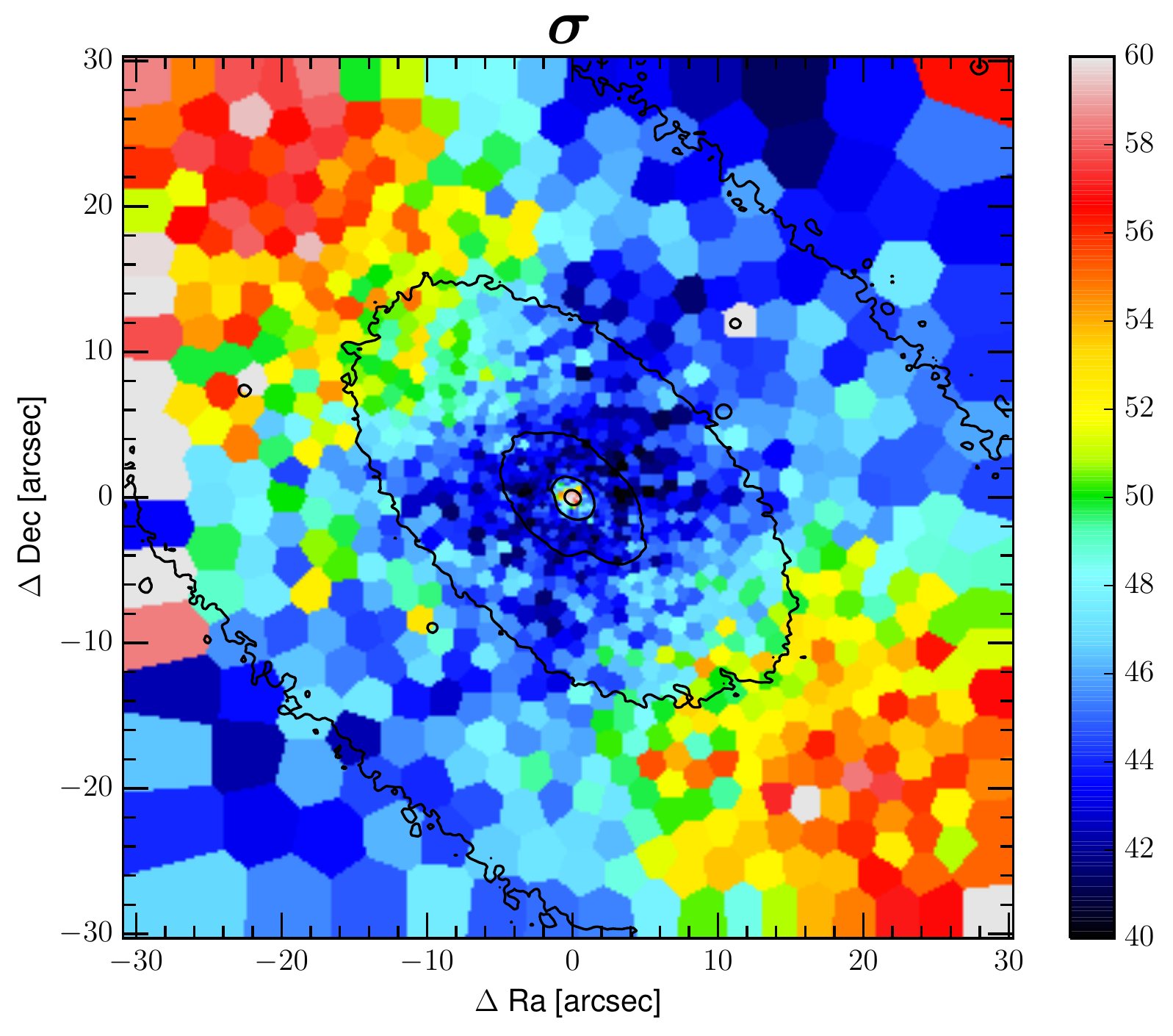}
 \includegraphics[width=0.49\textwidth]{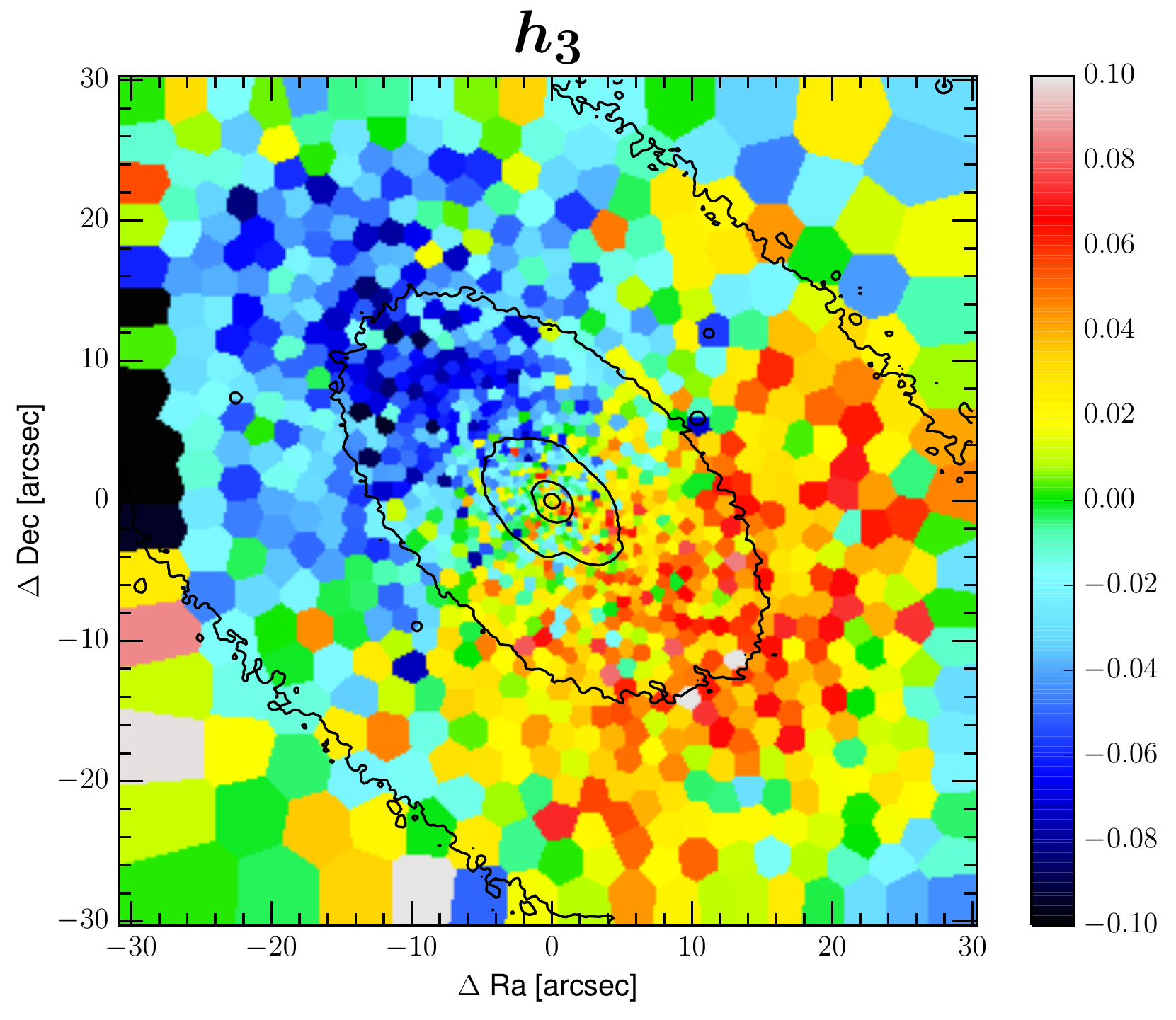}
 \includegraphics[width=0.49\textwidth]{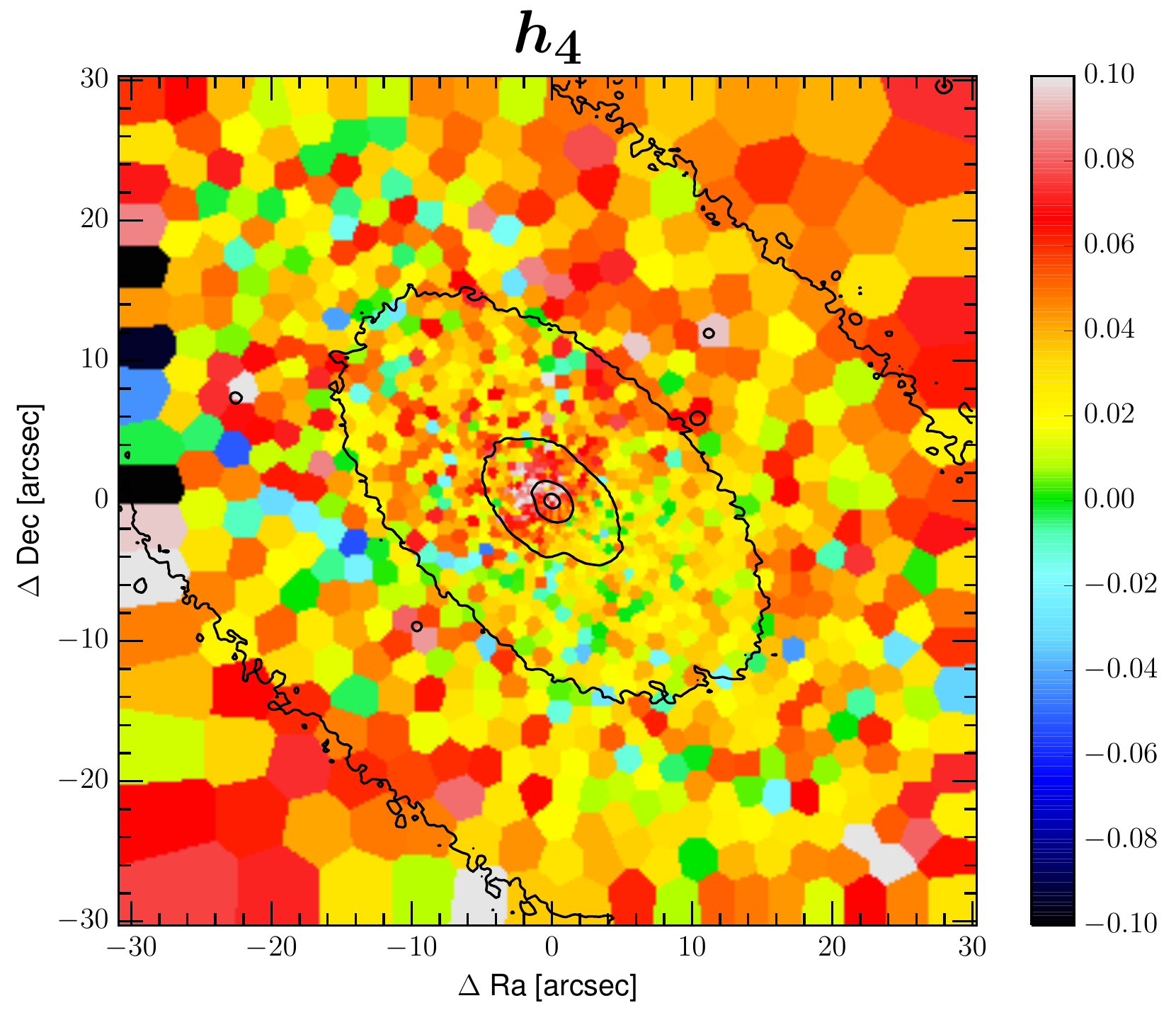}
 \caption{From top left to bottom right: Velocity, velocity dispersion $\sigma$, $h_3$ and $h_4$ fields of the 4 moment kinematic fit to the 4760--7400\,\AA\ wavelength region. The median value of $-0.019$ has been subtracted from the $h_3$ map. The asymmetry of the $h_3$ field is indicative for residual template mismatch. To exclude that this is caused by the single template spectrum used to extract this kinematics we checked the $h_3$ map of the full MILES spectrum fit for structure in the $h_3$ map. This map has an even stronger structure and is less symmetric. Black lines are the isophotes of the galaxy's integrated light.}
 \label{fig:blue_kinematik_4_mom}
\end{figure*}

We fit a two moment ($V$ and $\sigma$) and a four moment ($V$, $\sigma$ and the Gauss-Hermite $h_3$ and $h_4$ parameters) LOSVD to the data. 
The uncertainties on the spectrum are assumed to be constant with wavelength. 
The reason for this is that the pipeline propagated errors contain noise themselves and this extra noise can potentially increase the uncertainties in the kinematics.
The assumption of constant errors can be cross checked against the residuals (difference between input and best fit spectrum) of the fit: Fig.~\ref{fig:bestfit_4760-6800_bestfit} shows that these residuals are constant with wavelength, thus justifying the assumption of constant errors.

The resulting two dimensional maps of the LOSVD in the 4 moment extraction is shown in Fig.~\ref{fig:blue_kinematik_4_mom}.
The velocity field is plotted relative to the systemic velocity of $V_{\rm sys} = 474.5$\,km\,s$^{-1}$.
We compute the systemic velocity with the \texttt{fit\_kinematic\_pa} routine$^2$,
 a \texttt{Python} implementation of the method \citet{2006MNRAS.366..787K} describe in Appendix C to measure the global kinematic position angle.
The structure of the velocity field (Fig.~\ref{fig:blue_kinematik_4_mom}) is complicated: a clear rotation pattern is visible in the centre and there is an indication that the rotation reverses at larger radii. 
Looking at the velocity dispersion map we see a central peak and a rise of the dispersion towards the outer parts of the field of view along major axis.
This peculiar rise of the stellar velocity dispersion at large radii along the galaxy projected major axis, was first observed and interpreted by \citet{1992ApJ...400L...5R} in the S0 galaxy NGC~4550 with counter-rotating discs discovered by \citet{1992ApJ...394L...9R}.
Detailed dynamical models of this galaxy, as well as of the similar one NGC~4473, based on integral-field kinematics, were presented in \citet{2007MNRAS.379..418C}. They confirmed the original interpretation of this class of galaxies, having two sigma peaks along the projected major axis, as due to counter-rotating discs of comparable light contribution. 
The ATLAS$^{\rm 3D}$ survey \citep{2011MNRAS.413..813C} observed a volume-limited sample of 260 early type galaxies (ETGs) with stellar mass $M_\ast\ga1\times10^{10}$\,M$_{\sun}$ and found that 11 of them (4\%) belong to this class of counter-rotating disc galaxies, which they aptly named $2\sigma$ galaxies \citep{2011MNRAS.414.2923K}. Dynamical models of six of them, for a range of counter-rotating mass fraction, are shown in Fig. 12 of \citet{2016ARA&A..54..597C}  NGC~4473 is an example where the 2$\sigma$ peaks were predicted using dynamical models \citep{2007MNRAS.379..418C}, but only later actually observed with more spatially extended observations \citep[][ Fig.~1]{2013MNRAS.435.3587F}. To our knowledge this is the first time this rotation pattern is observed in NGC~5102. The previous dynamical modelling result strongly suggest this galaxy also contains two counter-rotating discs, and this interpretation is actually confirmed in Sec.~\ref{Mass-decomposition-of-counter-rotating-discs}.

The fit of the \ion{Ca}{ii} region is in broad agreement with the results obtained from the 4760--7400\,\AA\ wavelength range and the MILES library. In table \ref{tab:kinematic_fit_parameter} we compare key parameters of the four velocity extractions: the systemic velocity $V_{\rm sys}$, the amplitude of the inner rotation pattern $V_{\rm ampl}$, the position angle (PA), the additive offset of the dispersion maps $\Delta\sigma$ and the scaling factor $\delta_{\sigma}$ that minimizes the residuals between two dispersion maps (we compare to the blue 2 moment kinematics). The amplitude of the rotation is larger in the four moment kinematic extractions, indicating that the LOSVD is non-Gaussian. This is to be expected when two counter rotating components are present in the system. 
The velocity dispersion in this galaxy is expected to vary appreciably as a function of wavelength. This is because the width of the absorption lines will depend on the relative contribution the two counter-rotating components add to each wavelength range. This gets amplified by the quite different populations (see Sec.~\ref{sec:Population_fitting}).

\begin{table*}
\centering
 \caption{Parameters from the kinematic fits. }
 \label{tab:kinematic_fit_parameter}
 \begin{tabular}{c c c c c c c l}\hline
   & $V_{\rm sys}^a$ & $V_{\rm ampl}^b$ & $\Delta V_{\rm sampling}^c$ & PA$^d$ & $\Delta\sigma^e$ & $\delta_{\sigma}^f$ & S/N$^g$  \\
   & $[$km\,s$^{-1}]$ & $[$km\,s$^{-1}]$ & $[$km\,s$^{-1}]$ & $[$\degr$]$ & $[$km\,s$^{-1}]$ & -- & -- \\\hline
  MILES 2 moment & 473.3 & 17.5 & 50.6 & 45.5 & -- & -- & 113\\
  MILES 4 moment & 474.5 & 20.6 & 50.6 & 47.0 & 1.2 & 1.04 & 113 \\
  \ion{Ca}{II} triplet 2 moment & 472.7 & 16.9 & 42.5 & 44.5 & 7.3 & 1.05 & 109 \\
  \ion{Ca}{II} triplet 4 moment & 472.5 & 21.4 & 42.5 & 44.5 & 6.8 & 1.04 & 110 \\\hline
  \multicolumn{8}{p{13cm}}{$^a$ $V_{\rm sys}$ is the systemic velocity of NGC 5102, determined with the \texttt{fit\_kinematic\_pa} routine, a \texttt{Python} implementation of the method \citet{2006MNRAS.366..787K} describe in Appendix C to measure the global kinematic position angle. } \\
  \multicolumn{8}{p{13cm}}{$^b$ $V_{\rm ampl}$ is the amplitude of the rotation velocity, determined as $1/2$ of the sum of absolute values of the minimum and maximum values of the by symmetrized velocity field (using the \texttt{Python} routine \texttt{cap\_symmetrize\_velfield}).}\\
  \multicolumn{8}{p{13cm}}{$^c$ The logarithmically rebinned spectra have a constant sampling step in the velocity space. $\Delta V_{\rm sampling}$ is this sampling size.} \\
  \multicolumn{8}{p{13cm}}{$^d$ PA is the kinematic position angle, also determined with the aforementioned \texttt{fit\_kinematic\_pa} routine. } \\
  \multicolumn{8}{p{13cm}}{$^e$ $\Delta\sigma$ is the mean of the difference MILES 2 moment extraction dispersion and the dispersion from the kinematic extraction under consideration.} \\
  \multicolumn{8}{p{13cm}}{$^f$ $\delta_{\sigma}$ is the factor the velocity dispersion field under consideration needs to be scaled to minimize the absolute difference to the MILES 2 moment dispersion. }\\
  \multicolumn{8}{p{13cm}}{$^g$ S/N is as the ratio between the mean signal and the rms of the residual from a \texttt{pPXF} fit.}
 \end{tabular}
\end{table*}

\subsection{Fitting for two components}

As mentioned in the previous section the stellar kinematics indicate two counter rotating populations in NGC~5102.
The median {\em observed} velocity difference of the two stellar components (derived from the results presented in Fig.~\ref{fig:two-comp_vel}) is $\sim75$\,km\,s$^{-1}$. This suggests that the LOSVD of the two stellar components will not be resolvable by the MUSE spectral resolution of $\sigma_{\rm instr} \approx 65$\,km\,s$^{-1}$ around the strong H$\beta$ feature. This makes the actual extraction of two kinematic components particularly challenging from these data, even though the presence of two components is quite clear from the sigma maps. 
However in the red part of the spectrum the chances of separating the two components are actually quite good, given the $\sigma_{\rm instr} \approx 37$\,km\,s$^{-1}$. We therefore use the \ion{Ca}{II} triplet and the MIUSCAT library to separate the two components.
\texttt{pPXF} has the option to fit an arbitrary number of different kinematic components \citep[e.g.][]{2011MNRAS.412L.113C, 2013MNRAS.428.1296J}, so we use for each kinematic component an identical set of templates. Our template consists of those 8 MIUSCAT spectra that contribute most the integrated NGC~5102 spectrum.

Finding the suspected two kinematic solutions using \texttt{pPXF} requires precise starting values, because \texttt{pPXF} uses a local minimisation algorithm.
When two kinematic components are present, the single-component solution is necessarily, by symmetry, a saddle point in the $\Delta\chi^2$ landscape. Moreover one should expect a cross-like degeneracy centred on the single-component fit. Due to these degeneracies imprecise starting velocities can lead \texttt{pPXF} to converge prematurely to the single-component solution.
To make sure we do not miss the global minimum of the two-component fit, we need to employ a global rather than local optimization approach. The approach we adopt is straightforward but brute force. For each Voronoi bin spectrum we scan a region of starting velocities for both components: All combinations of starting velocities in the range from $V_{\rm sys}-100$ to $V_{\rm sys}+140$\,km\,s$^{-1}$ are tested in steps of $V_{\rm step} = 15$\,km\,s$^{-1}$ for both components. This results in 289 different starting velocity combinations that we fit. In each fit we constrain in \texttt{pPXF} the velocity range to $V_{i, \rm start} - V_{\rm step}/2 \leq V_i \leq V_{i, \rm start} + V_{\rm step}/2 $, where $i = {1,2}$ is the index for the two component. This means we force \texttt{pPXF} find the best solution in the interval around the starting value. 

\begin{figure}
 \includegraphics[width=0.455\columnwidth]{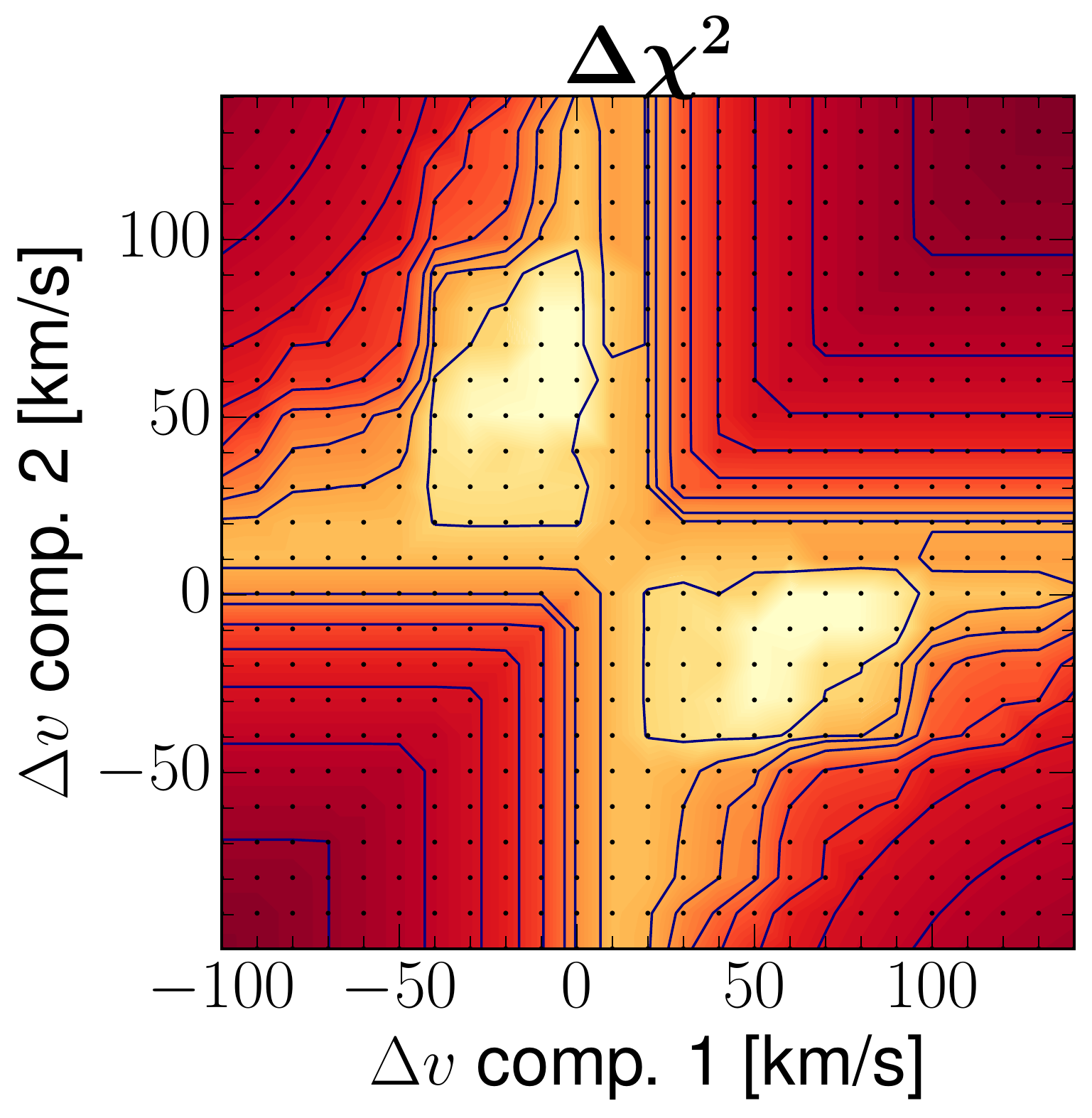}
 \includegraphics[width=0.52\columnwidth]{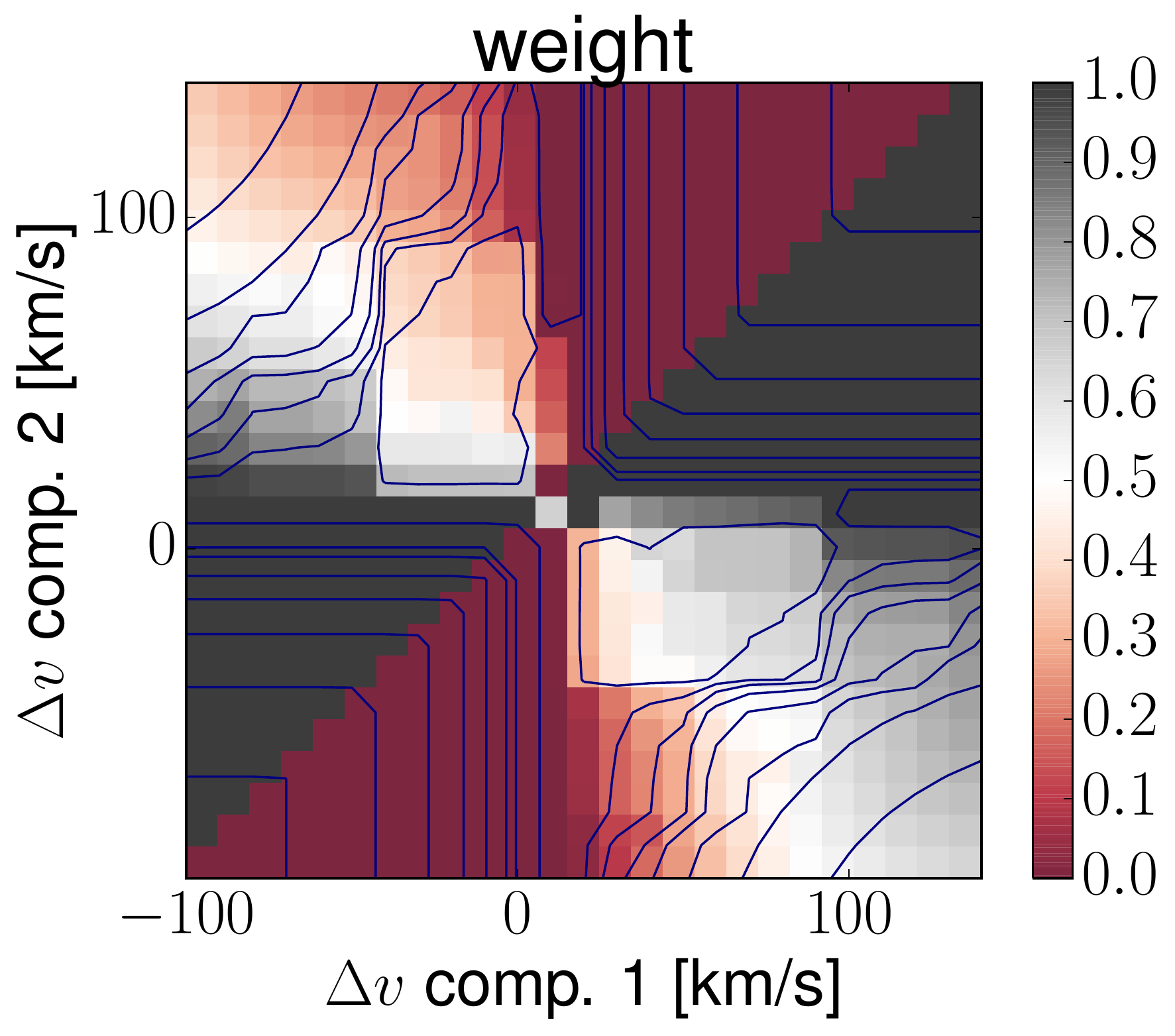}
 \caption{{\em Left:} $\Delta\chi^2$ as function of the input velocities for the two kinematic components. The contour lines are at $\Delta\chi^2 = 1, 4, 9$, after that the value doubles. The cross-like structure is the single component solution.  The best-fitting solutions are the bright areas with the lowest $\Delta\chi^2$ values.  {\em Right:} Weights for one kinematic component as function of the input velocities for the two kinematic components, with $\Delta\chi^2$ contours plotted. A true two component solution is fitted in those regions where both components have a non zero weight, i.e. brighter regions that are perpendicular to the one to one relation. The regions with the lowest $\Delta\chi^2$ fall onto those areas where both components contribute. The afore mentioned single component trails fall on regions where mostly one component contributes. For a better visualisation these plots have a slightly higher sampling in velocity than our analysis, but this does not affect our results.}
 \label{fig:two-comp-chi2}
\end{figure}

\begin{figure}
 \includegraphics[width=0.49\columnwidth]{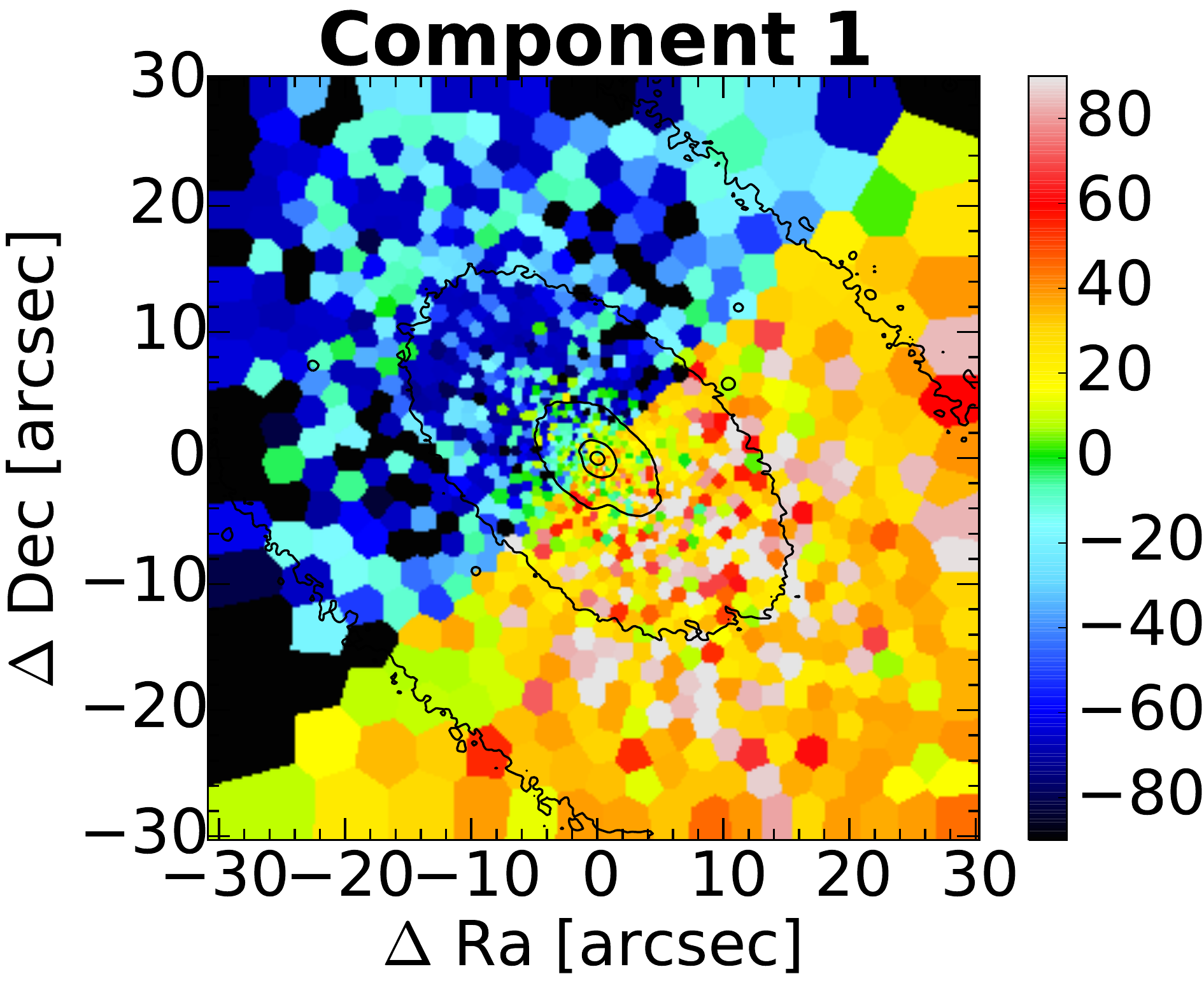}
 \includegraphics[width=0.49\columnwidth]{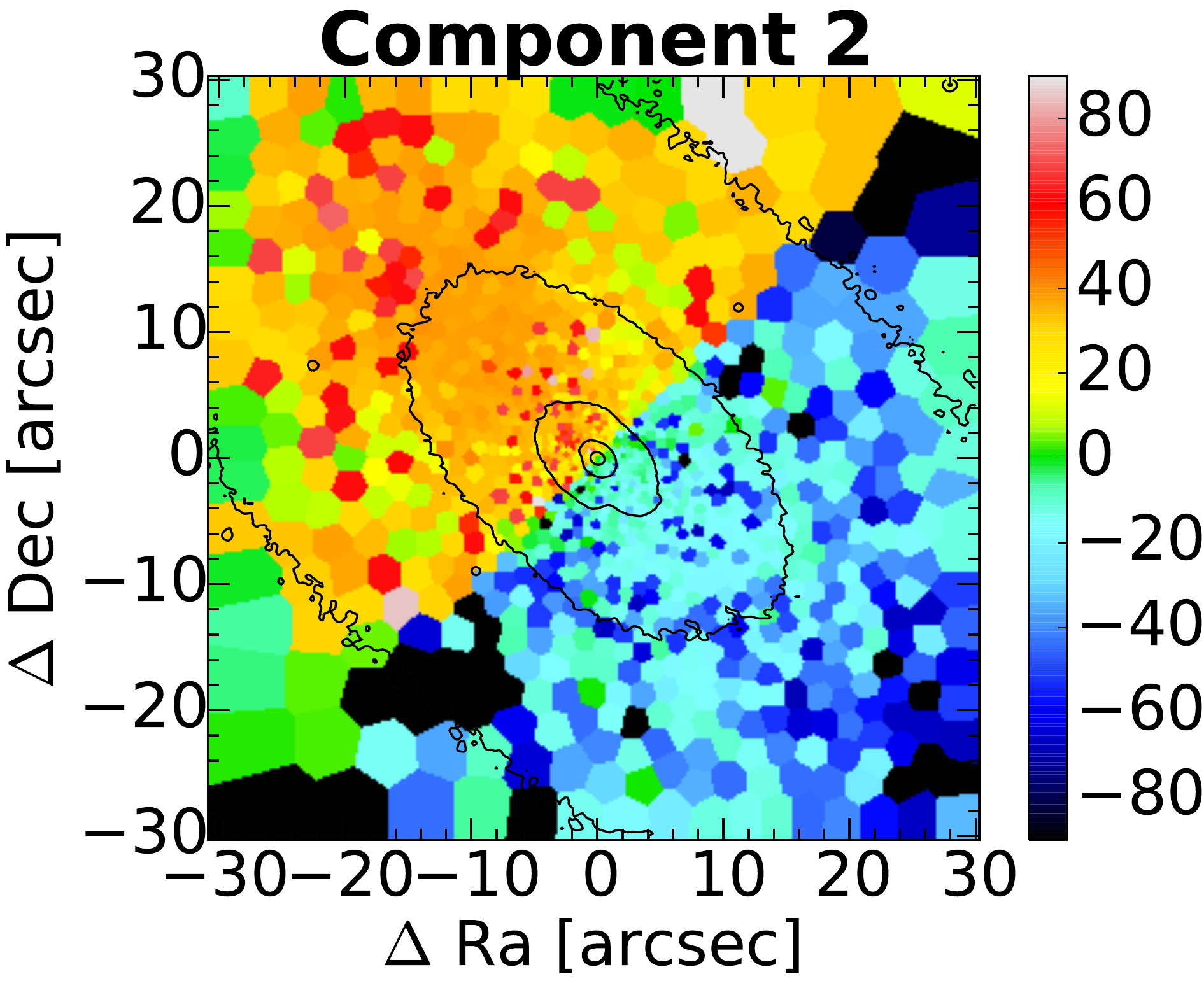}
 \caption{{\em Left:} Velocity filed of component 1 {$[$}km\,s{$^{-1}]$}. {\em Right:} Velocity field of component 2 {$[$}km\,s{$^{-1}]$}.   Black lines are the isophotes of the galaxy's integrated light.}
 \label{fig:two-comp_vel}
\end{figure}

The result of this fitting are two dimensional maps of $\Delta\chi^2$ (see left panel of Fig.~\ref{fig:two-comp-chi2}) and for each component the weight (see right panel of Fig.~\ref{fig:two-comp-chi2}), velocity and dispersion. The $\Delta\chi^2$ map in the left panel of Fig.~\ref{fig:two-comp-chi2} actually shows the cross-like structure centred on the single-component solution. The interpretation of this cross-like structure as the single component solution is supported by the fact that in these regions mostly one component contributes to the fit, as the weight map indicates. The lowest $\Delta\chi^2$ region  is symmetrically above and below the one-to-one relation. In those regions the weight map reveals that both components contribute to the fit. 

We select the 2 fits with the lowest $\Delta\chi^2$ values (because the fit should be symmetric to the one-to-one relation). 
The further analysis is complicated by the fact that \texttt{pPXF} outputs two results for each fit, one for each kinematic component, but the assignment of numerical component 1 and 2 is not related to the two physical components. 
This means for each fit we need to decide to which physical component (i.e. clockwise and counter clockwise rotating) the two numerical components belong. 

We sort in such a way that the difference of the velocities of the two physical components $V_1 - V_2$ has the same sign as the major axis coordinate. After sorting velocity, velocity dispersion and weights we adopt the mean of the two values from the two fits with the lowest $\Delta\chi^2$ as the final value.

In Fig.~\ref{fig:two-comp_vel} we show the velocity fields. These plots show a clear separation of the two components and thus prove the existence of two counter rotating populations. Component~1 rotates faster than component~2. Component~2 rotates in the same direction as the \ion{H}{I} gas \citep{1993A&A...269...15V, 2015MNRAS.452.3139K}. Here the separation of the to components is purely based on a weighted superposition of the two kinematic templates. In Sec. \ref{Mass-decomposition-of-counter-rotating-discs} we determine the mass fractions of the two components based on our dynamical modelling.

\section{Population fitting}
\label{sec:Population_fitting}
We extract the stellar population from the MUSE data using \texttt{pPXF} to perform full-spectrum fitting and a regularization of the weights solution.
To make use of the full wavelength range provided by the MUSE spectrograph (4750--9340\,\AA)\ we used the MIUSCAT \citep{2012MNRAS.424..157V} population models. 
These models combine the observational stellar libraries MILES \citep{2006MNRAS.371..703S, 2011A&A...532A..95F}, Indo-U.S. \citep{2004ApJS..152..251V} and CaT \citep{2001MNRAS.326..959C} to predict stellar populations over the wavelength range from  3465 to 9469\,\AA. 
MILES and CaT libraries are used in their respective wavelength ranges and Indo-U.S. in the gaps and at the blue and red extensions \citep{2012MNRAS.424..157V}

We use stellar populations based on the Padova isochrones \citep{2000A&AS..141..371G} and use as a reference a Salpeter initial mass function (IMF). We restrict to the safe parameter range \citep{2012MNRAS.424..157V} and obtain a set of population models covering the age range from 0.1 to 17.8\,Gyr spaced into 46 logarithmically equidistant steps and six metallicites ([M/H] = $-1.71$, $-1.31$, $-0.71$, $-0.4$, $0.0$, $+0.22$). All template spectra are scaled with one scalar to have a median value of $1$. The same is done for the galaxy spectrum.

The whole MUSE wavelength range is used for the population fitting, no masking is applied.
We simultaneously fit the stellar kinematics, population and the gas emission lines of 
H$\alpha$, H$\beta$, 
[\ion{O}{I}]~$\lambda\lambda$~6300,6364, 
[\ion{O}{III}]~$\lambda\lambda$~4959,5007, 
[\ion{N}{II}]~$\lambda\lambda$~6548,6583, 
[\ion{S}{II}]~$\lambda~6716$ and 
[\ion{S}{II}]~$\lambda~6731$.
The flux ratios of the doublets are fixed to $1/3$ as predicted by atomic physics. All gas emission lines have a common kinematic ($V$ and $\sigma$) solution, while the fluxes of the seven gas components (three doublets and four lines) are freely scaled.

We assume the errors on the spectrum to be constant with wavelength. 
We use the $\chi^2$ per degree of freedom ($\chi^2$/DOF) provided by the best fit to renormalise the error to give a $\chi^2$/DOF\,$=$\,1:
\begin{align}
 \varepsilon_{\rm norm} = \varepsilon\times\sqrt{\chi^2/\rm DOF}
 \label{eq:error_normalisation}
\end{align}
After the unregularised fit we perform a regularised fit, with the \texttt{pPXF} regularisation parameter "REGUL\,$=$\,100".

\begin{figure*}
 \includegraphics[width=0.75\textwidth]{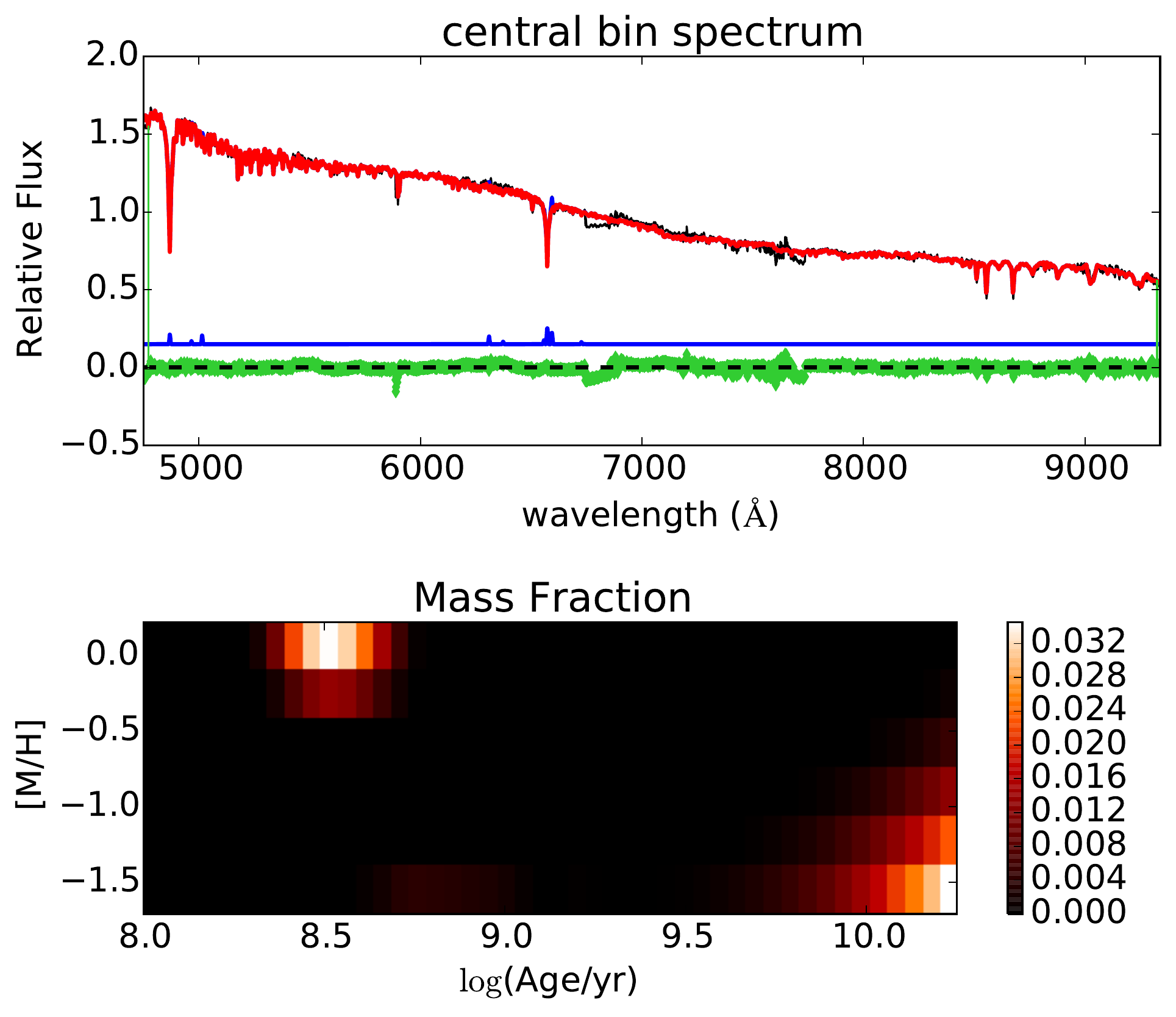}
 \caption{{\em Top}: The spectrum of a centre bin is shown in black. The red line shows the best regularised population fit to each spectrum. The blue line gives the best fit solution for the gas component. The green line shows the residuals (data - stellar model - gas model).  {\em Bottom}: two dimensional grid in age and metallicity spanning the parameter range of the stellar SSP models. The colour coding gives the weight each model contributes to the fit in the upper panel. This gives the distribution of stellar parameters. }
 \label{fig:fit_full_bin_1_regularized}
\end{figure*}

\begin{figure*}
 \includegraphics[width=0.75\textwidth]{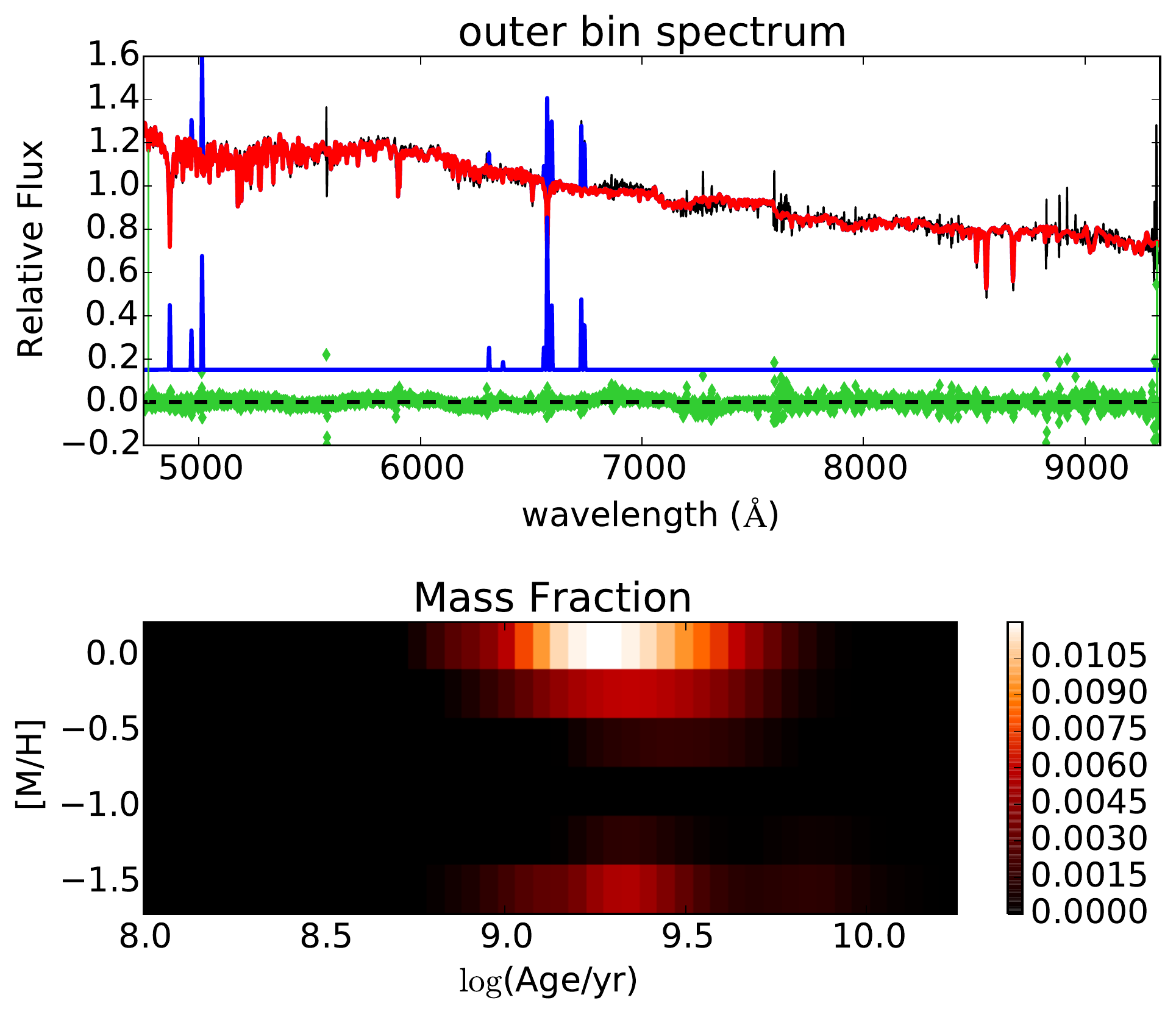}
 \caption{As Fig. \ref{fig:fit_full_bin_1_regularized}, but for an outer Voronoi-bin at 30\farcs8 from the centre.}
 \label{fig:fit_full_bin_2_regularized}
\end{figure*}

Two examples of the regularised fit are shown in Fig.~\ref{fig:fit_full_bin_1_regularized} and \ref{fig:fit_full_bin_2_regularized} for a central bin and one at a major axis radius of $r=30\farcs8$, slightly more than one $R_e$. 
In the centre bin two distinct populations emerge: an  old, metal poor one with $[M/H] < -1.0$ and a  $\sim 0.3$\,Gyr roughly solar metallicity ($[M/H] > -0.5$) one. In the outer bin the separation is less prominent, even though there is an indication for a separation in metallicity. 

For each Voronoi bin the average age and metallicity are calculated as the weighted sums of the individual simple stellar population values 
\begin{align}
 \log(\text{Age}) &=  \frac{\sum_i w_i \log(\text{Age}_i)}{\sum_i w_i} \\
 [M/H] &=  \frac{\sum_i w_i [M/H]_i}{\sum_i w_i}.
\end{align}
The index $i$ runs over the simple stellar populations and $w_i$ is the weight \texttt{pPXF} assigned to the $i$-th population. 
Since MIUSCAT populations are normalized to an initial mass of 1\,M$_{\sun}$, these mean values are mass weighted.
Mass to light ratios are computed from the MIUSCAT mass \citep{2012MNRAS.424..157V} and photometry predictions \citep{2012MNRAS.424..172R} following eq.(2) of \citet{2013MNRAS.432.1862C}
\begin{align}
 (M_{\ast}/L)_{\rm Salp} = \frac{\sum_i w_i M_i }{\sum_i w_i L_i}.
\end{align}

\begin{figure*}
 \includegraphics[width=0.32\textwidth]{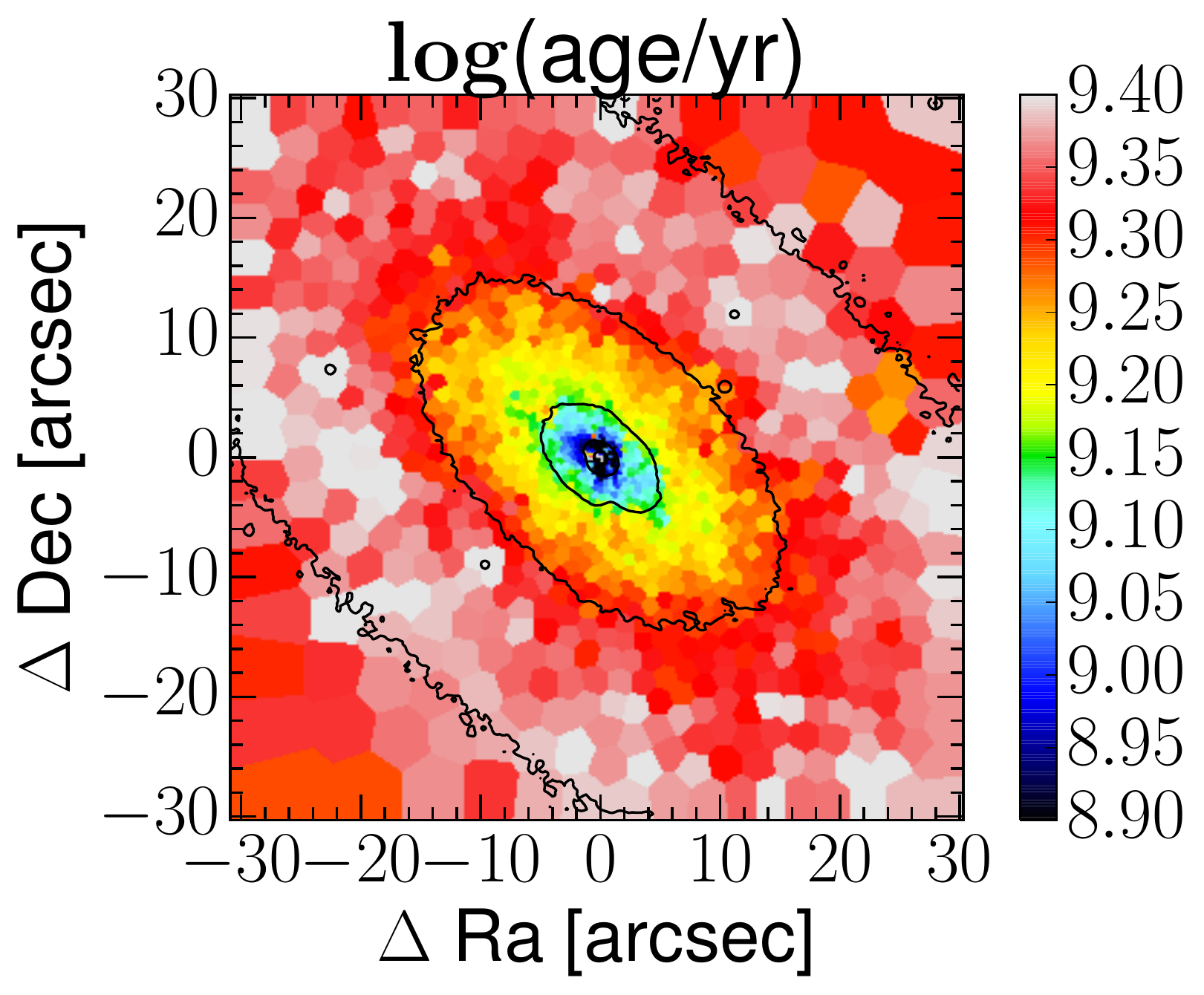}
 \includegraphics[width=0.32\textwidth]{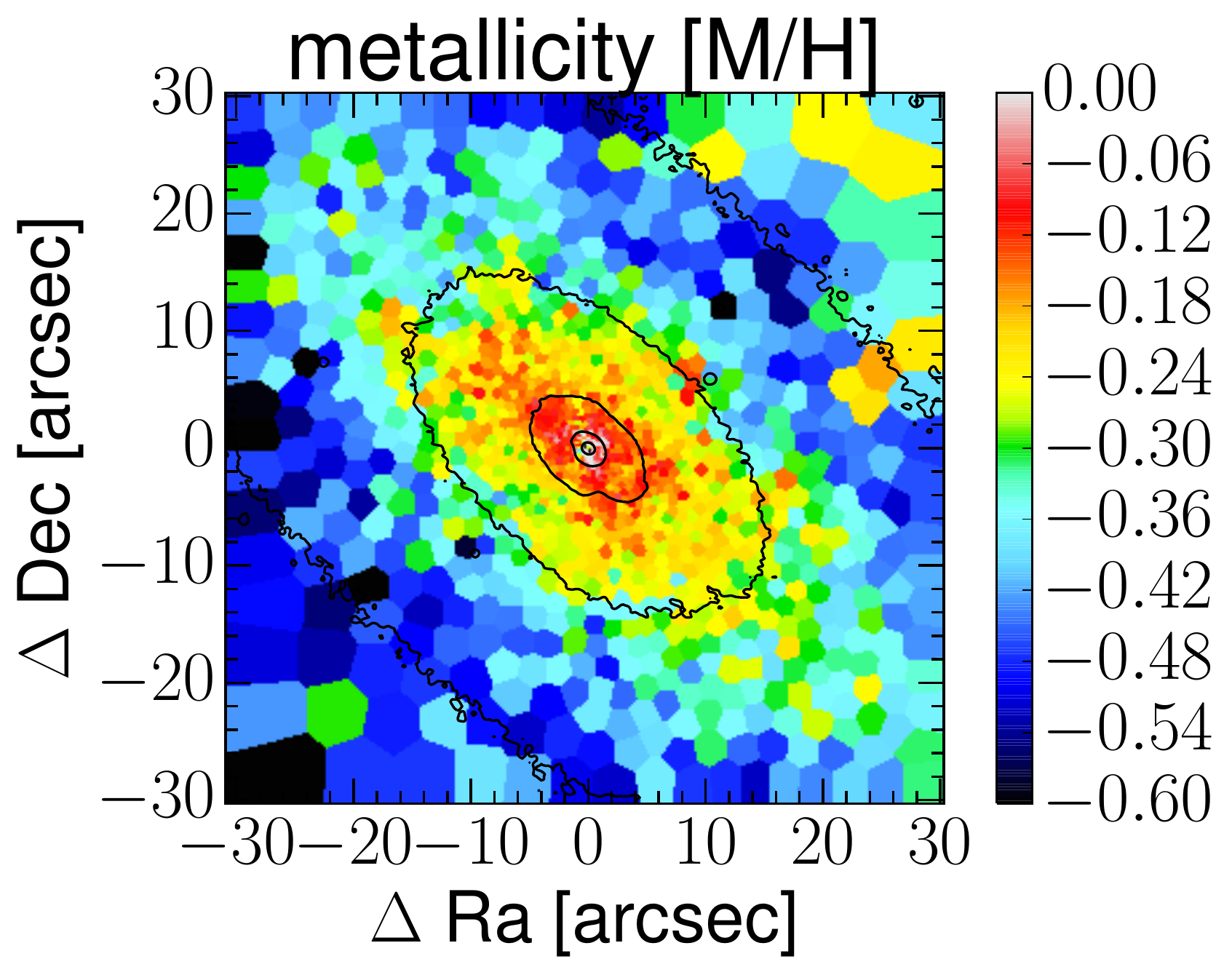}
 \includegraphics[width=0.32\textwidth]{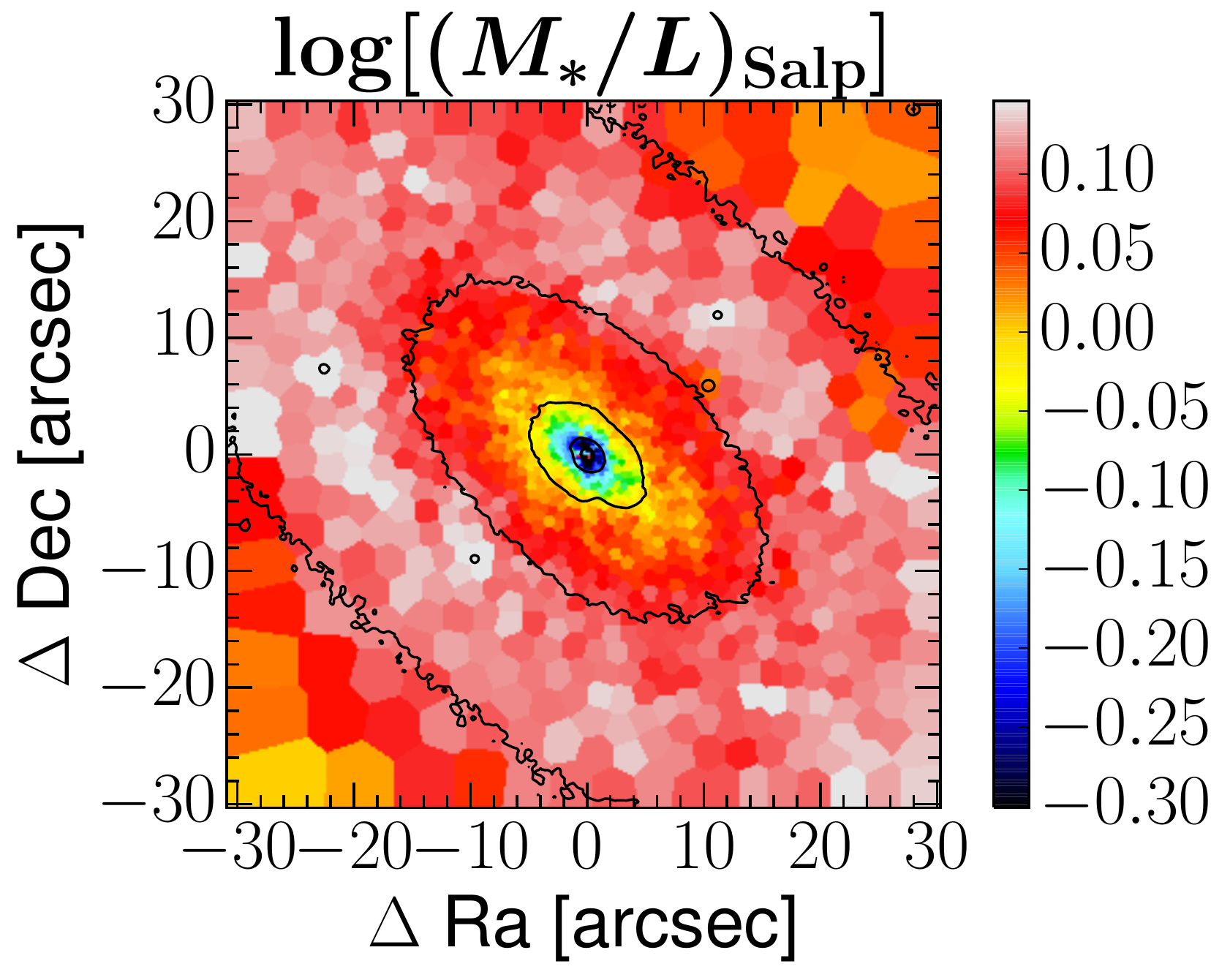}
 \caption{{\em Left:} Regularized mass weighted $\log$(Age) {$[\log(\rm yr)]$} field.  {\em Top Right:} Regularized mass weighted metallicity {$[M/H]$} field. {\em Right:} Regularized $(M_{\ast}/L)_{\rm Salp}$ field.  Black lines are the isophotes of the galaxy integrated light.}
 \label{fig:regularized_population_fit}
\end{figure*}

Fig.~\ref{fig:regularized_population_fit} shows the mean age, metallicity and $(M_{\ast}/L)_{\rm Salp}$ maps of NGC~5102.
The age map shows  a young ($\sim 0.8$\,Gyr) population in the centre which is  in line with the literature \citep[e.g.][]{2015ApJ...799...97D}. 
A relatively flat mean age of $\sim 2$\,Gyr is observed in the outer parts, as reported by \citet[][]{2015ApJ...799...97D} for the bulge of NGC~5102. 
There is a slight indication that the mean age decreases at the largest radii. 
The metallicity map reveals that the highest metallicity population is found in the centre while towards the outer parts the value drops from slightly below solar to $[M/H] = -0.5$. The high metallicity population we see in the centre of NGC~5102 is in stark contrast to previous studies, since \citet{2015ApJ...799...97D} finds a metallicity of $Z=0.004$ ($[M/H]\sim -0.7$) for the nucleus.

\section{Photometry}

There are two selection criteria for the photometric data for the dynamical models: to properly describe the distribution of the kinematic tracer the photometry has to be within the MUSE wavelength range and the photometric system needs to be well calibrated.
In the HST archive there are two image sets that fulfil these criteria, both are taken with the Wide Filed and Planetary Camera 2 (WFPC2) in the F569W and the F547M filter. 
The F569W data are preferred for two reasons: the photometry is deeper because of the larger bandwidth and the longer integration time and is a mosaic of two fields. 
The problem with this data set is that the galaxy centre is saturated. 
We therefore use the F547M Planetary Camera image and add the F569W at radii larger than 6\arcsec\ from the galaxy centre to avoid the saturated pixels.

\begin{figure}
 \includegraphics[width=0.45\textwidth]{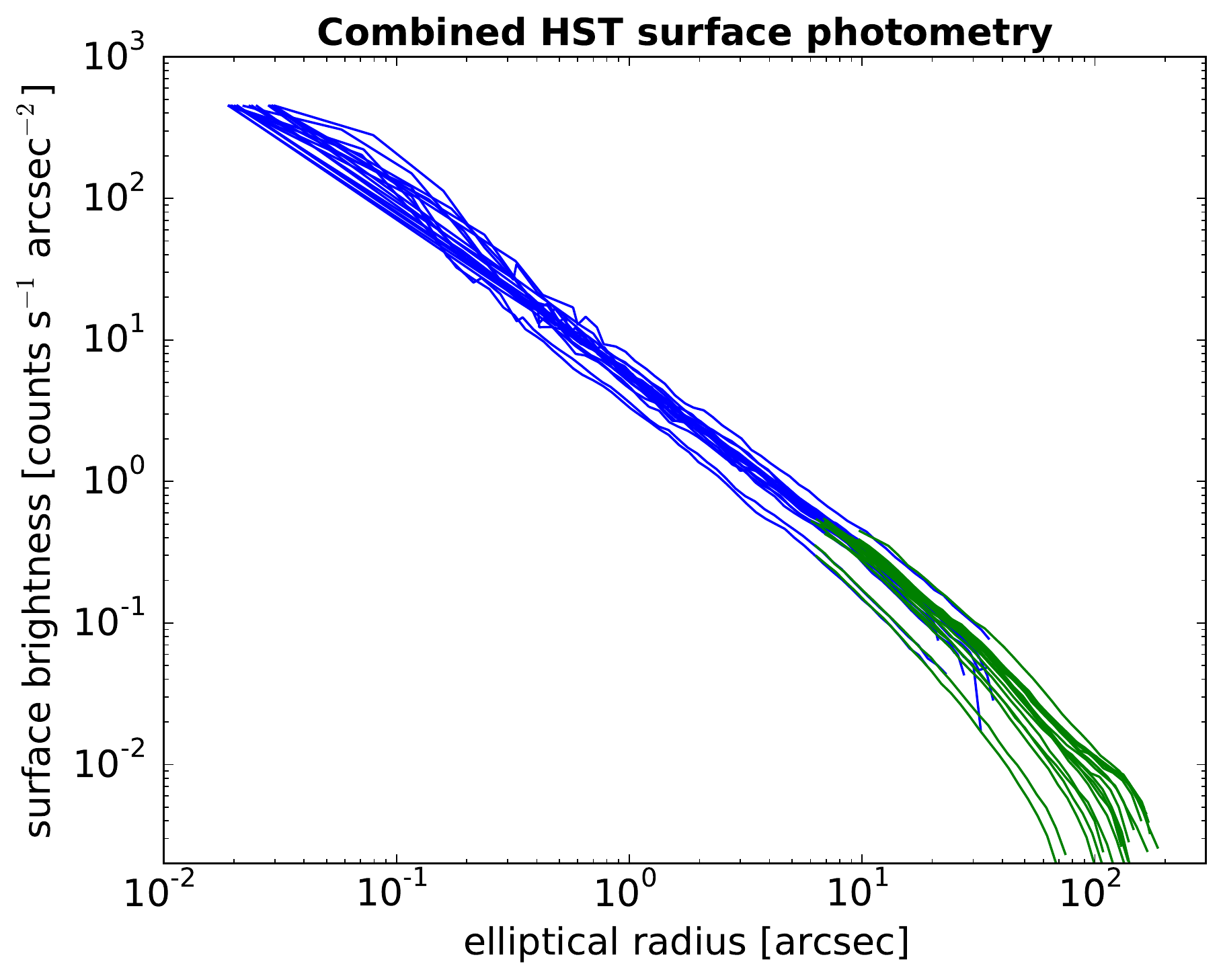}
 \caption{Photometry of two filters/instruments is shown: F547M PC photometry in blue and F569W WF in green. The different lines are the surface photometry from different sectors. The match of the photometry from the two filters/instruments shows that the scaling works.}
 \label{fig:F547M_PC_F569W_WF_scale}
\end{figure}

When extracting the surface photometry of NGC~5102 on the F547M images we discovered that the sky was over subtracted by the standard pipeline. 
We determined the amount of over subtraction by requiring  the profiles to be a power law at large radii. 
For the PC image the sky level is $-0.14$\,counts\,s$^{-1}$ and the resulting surface photometry is shown in blue in  Fig.~\ref{fig:F547M_PC_F569W_WF_scale}. For the F569W WF image we found a sky level of $-0.01$\,counts\,s$^{-1}$. 

The WFPC\,2 PC PSF is modelled using the dedicated software \texttt{TinyTim} version 7.5 \citep{1993ASPC...52..536K, 2011SPIE.8127E..0JK}. 
The input spectrum is chosen to be a black body of 5700\,K, because this approximates the NGC~5102 spectrum in the wavelength range of the F547M filter. The resulting PSF-image is approximated by four 2 dimensional circular Gaussian functions using the Multi-Gaussian Expansion method \citep[MGE;][]{1994A&A...285..723E, 2002MNRAS.333..400C}.

For the dynamical models we parametrize the galaxy surface brightness using the MGE. 
We use the \texttt{FIND\_GALAXY} \texttt{Python} routine$^2$
\citep{2002MNRAS.333..400C} to determine the galaxy centre and position angle.
The surface photometry is measured in sectors equally sampled in eccentric anomaly ($5\degr$) and logarithmically sampled in radius, using the \texttt{MGE\_FIT\_SECTORS}$^2$
software of \citet{2002MNRAS.333..400C}.
The photometry of the four quadrants is averaged by the code. This allows for a simple combination of photometry from different images. To combine the different photometric data sets we have to scale the F569W photometry to match the F547M calibration. This scaling takes the differences in pixels size, sensitivity and bandwidth of the filters into account. 
We found that the F569W fluxes need to be multiplied by 0.13 to match the F547M photometry. The resulting combined photometry is shown in Fig.~\ref{fig:F547M_PC_F569W_WF_scale}.

\begin{figure}
 \centering
 \includegraphics[width=0.45\textwidth]{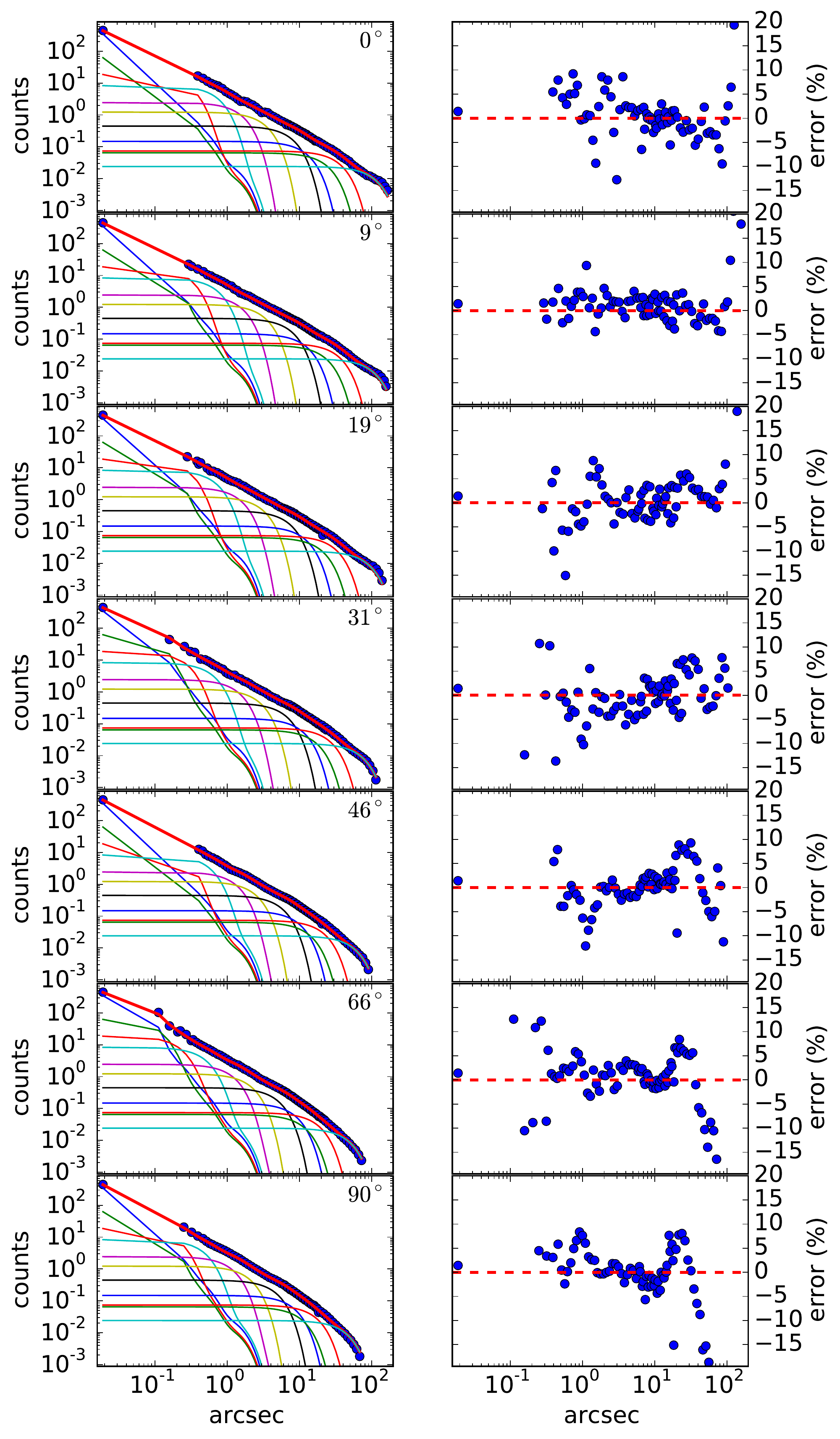}
 \caption{{\em Right column}: For a number of sectors the combined F547M PC and F569W WFPC2 photometry (blue points) and MGE model is shown. {\em Left column}: the residuals for the photometry are shown. There are no consistent structures in the residuals visible - indicating that the MGE model approximates the data well.}
 \label{fig:mge_secors_F547M_F569W_combined}
\end{figure}

\begin{figure}
 \centering
 \includegraphics[width=0.45\textwidth]{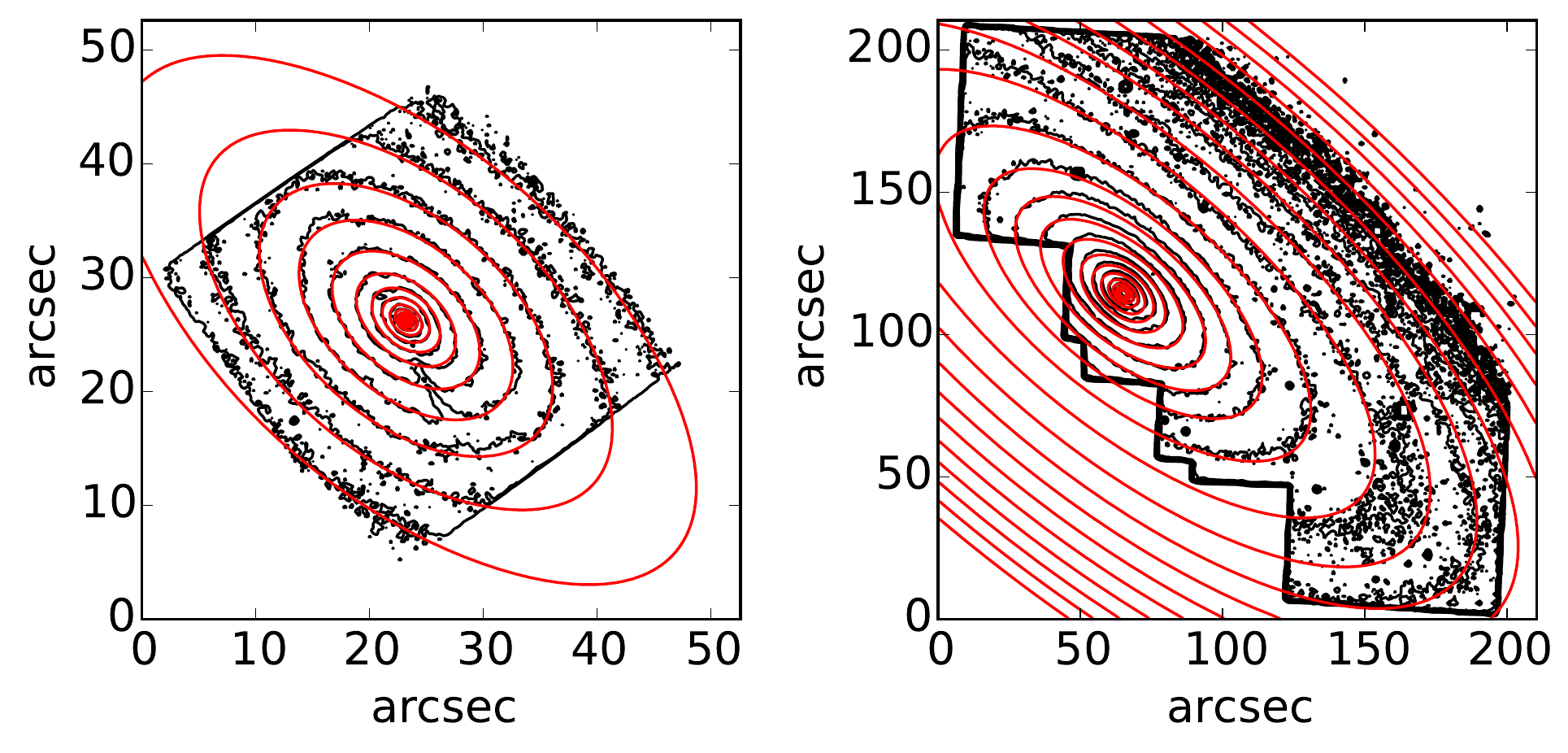}
 \caption{Isophotes from the data (black) and the MGE model (red) are shown. On the left panel for the F547M PC image and on he right panel for the  F569W WFPC2 images. }
 \label{fig:mge_residuals_F547M_F569W_combined}
\end{figure}

To correct the MGE model for PSF effects, we use the intensities and standard deviations of the MGE extraction of the PSF-image  \citep[for details see ][]{2002MNRAS.333..400C}.
The regularised\footnote{The regularisation aims at finding the roundest solution that is in agreement with the errors on the photometry by first iteratively increasing the minimum boundary for the $q$ value, to remove very flat components that are not justified by the data, and then lowering the upper bound on $q$ to remove very round low surface brightness components that do not contribute to the $\chi^2$ as in \citet{2013MNRAS.432.1894S}.} MGE expansion along selected angular sections of the combined photometry is shown in Fig.~\ref{fig:mge_secors_F547M_F569W_combined} and the isophotes of the data and the model are compared in Fig.~\ref{fig:mge_residuals_F547M_F569W_combined}. 
Both plots show that the extraction did work very well, even out to the largest radii no problems are visible.

The output of the MGE (total counts, standard deviation and axial ratio for each Gaussian) is transformed into peak surface densities (L$_{\sun}$\,pc$^{-2}$) using the HST photometric calibration of \citet{2009PASP..121..655D} for gain\,$=$\,14. The $V-I = 0.93$ colour is computed from the $V=11.18$ \citep[mean value, ][]{1978ApJ...223..707S} and $I = 10.25$ \citep{2005MNRAS.361...34D} magnitudes listed on NASA/IPAC Extragalactic Database (NED). The F547M data were read out with gain\,$=$\,15, which is accounted for by using the gain ratio term of \citet{1995PASP..107.1065H}. We also correct for the galactic extinction toward NGC~5102 using $A_V = 0.151$, the value given by NED \citep{2011ApJ...737..103S}. For the transformation from surface brightness to surface density the absolute V-band magnitude of the sun is needed and we use the value given by \citet{2007AJ....133..734B} M$_{\sun,\rm V} = 4.78$.

\begin{table}
  \caption{MGE parametrization of NGC~5102 V-band stellar surface brightness.}
 \label{tab:mge_model_parameters_photometry_extraction}
 \begin{tabular}{c c c }\hline
  $\log$(surface density) & $\log(\sigma)$  & $q$ \\
  $[\log({\rm L}_{\sun}\,{\rm pc}^{-2})]$ & [$\log$(\arcsec)] &  \\\hline
$6.932$  &  $-1.491$  &  $0.783$  \\
$5.807$  &  $-1.096$  &  $0.900$  \\
$5.133$  &  $-0.667$  &  $0.666$  \\
$4.722$  &  $-0.278$  &  $0.636$  \\
$4.164$  &  $+0.055$  &  $0.739$  \\
$3.856$  &  $+0.380$  &  $0.609$  \\
$3.415$  &  $+0.754$  &  $0.601$  \\
$2.926$  &  $+0.958$  &  $0.663$  \\
$2.587$  &  $+1.232$  &  $0.453$  \\
$2.636$  &  $+1.413$  &  $0.473$  \\
$2.145$  &  $+1.902$  &  $0.400$  \\\hline
 \end{tabular}

\end{table}

The resulting MGE parametrization of NGC~5102 is given in table~\ref{tab:mge_model_parameters_photometry_extraction}. Following eq. (11) of \citet{2013MNRAS.432.1709C} we compute from the MGE a total apparent V-band magnitude of $M_V=9.74$\,mag and a circularized effective radius of $R_e = 27\farcs4$. The RC3 value for the total apparent V-band magnitude is $M_V=9.63$\,mag.

\begin{figure}
 \includegraphics[width=0.45\textwidth]{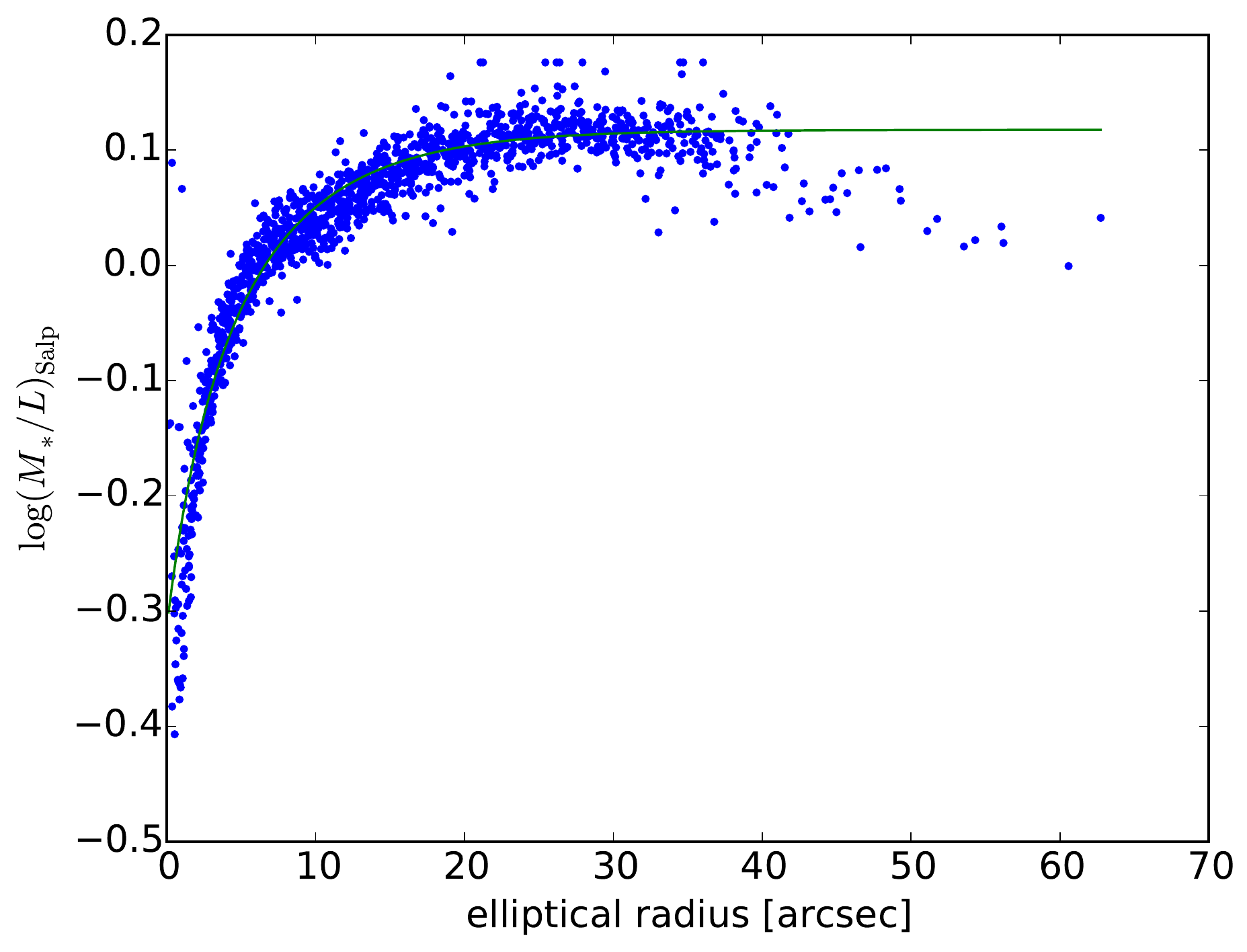}
 \caption{$(M_{\ast}/L)_{\rm Salp}$ from the regularised population fit (see Fig. \ref{fig:regularized_population_fit}) as function of the major axis radius. The best-fitting model to the data is shown in green.}
 \label{fig:fit_MoverL}
\end{figure}

The results from the population fitting showed that NGC~5102 has a prominent $(M_{\ast}/L)_{\rm Salp}$ trend in the sense that the light distribution is more strongly peaked than the mass distribution. To calculate a prediction of the {\em stellar mass} distribution, we multiply the observed surface brightness by $(M_{\ast}/L)_{\rm Salp}$. We use the $(M_{\ast}/L)_{\rm Salp}$ values displayed in Fig.~\ref{fig:regularized_population_fit} and compute the elliptical radius for each value according to
\begin{align}
 r_{\rm ell} = \sqrt{x^2 + y^2/\bar{q}^2}
 \label{eq:elliptical_radius}
\end{align}
where the $x$-axis is aligned with the major axis and the coordinates are centred on the galaxy centre. $\bar{q}$ is the mean axial ratio from the MGE light extraction in table~\ref{tab:mge_model_parameters_photometry_extraction}. We fit these data with the following expression:
\begin{align}
M_{\ast}/L(r_{\rm ell}) =  (M_{\ast}/L)_0 + \Delta(M_{\ast}/L)\left[1 - \exp(-r_{\rm ell}/\tau)\right]
\label{eq:MoverL_correction}
\end{align}
with three free, positive parameters: the difference in the mass to light ratio between the centre and the outer parts is $\Delta(M_{\ast}/L) = 0.825$, the mass to light ratio in the centre is $(M_{\ast}/L)_0 = 0.486$ and the scaling factor is $\tau = 6.758$. The best fit is shown in  Fig. \ref{fig:fit_MoverL}, underlining that this simple formula is actually a good approximation of the measured  $(M_{\ast}/L)_{\rm Salp}$ values. 
The surface photometry obtained by \texttt{sectors\_photometry} is sampled within a set of radial sectors. At any given $r_{\rm ell}$ (eq.~\ref{eq:elliptical_radius}), we multiply the derived surface brightness by the $(M_\ast/L)_{\rm Salp}(r_{\rm ell}	)$ (eq. \ref{eq:MoverL_correction}) and fit a MGE model with the \texttt{mge\_fit\_sectors} procedure.
In this way the MGE represents the stellar mass, rather than luminosity, assuming a Salpeter IMF. The MGE stellar mass parametrization of NGC~5102 is given in table~\ref{tab:mge_model_parameters_mass_extraction}.

\begin{table}
  \caption{MGE parametrization of the NGC~5102 stellar mass surface density. The surface mass is computed by multiplying the photometry with the $(M_{\ast}/L)_{\rm Salp}$ parametrization, see eq.~\ref{eq:MoverL_correction}.}
 \label{tab:mge_model_parameters_mass_extraction}
 \begin{tabular}{c c c }\hline
  $\log$(surface density) & $\log(\sigma)$  & $q$ \\
  $[\log({\rm M}_{\sun}\,{\rm pc}^{-2})]$ & [$\log$(\arcsec)] &  \\\hline
$6.659$  &  $-1.491$  &  $0.650$  \\
$5.720$  &  $-1.138$  &  $0.650$  \\
$4.941$  &  $-0.695$  &  $0.650$  \\
$4.451$  &  $-0.280$  &  $0.650$  \\
$3.869$  &  $+0.103$  &  $0.650$  \\
$3.626$  &  $+0.387$  &  $0.601$  \\
$3.317$  &  $+0.794$  &  $0.564$  \\
$2.950$  &  $+0.985$  &  $0.627$  \\
$2.631$  &  $+1.262$  &  $0.447$  \\
$2.736$  &  $+1.414$  &  $0.472$  \\
$2.263$  &  $+1.902$  &  $0.400$  \\\hline
 \end{tabular}

\end{table}

\section{Jeans anisotropic modelling}

We use the Jeans Anisotropic Modelling$^2$
\citep[JAM, ][]{2008MNRAS.390...71C} approach to compare our kinematic measurements with the predictions from modelling the mass distribution of the galaxy. This modelling approach allows the inclusion of the stellar mass inferred from the MGE, dark matter haloes and multiple kinematic components to constrain the potential of the galaxy. The JAM method makes some observationally motivated, but non general assumptions on the dynamics of the models. This may in principle affect our conclusions on the total mass profile. However, the JAM approach was shown to be able to recover total mass profiles with high accuracy and negligible bias, both using realistic, high-resolution N-body simulations \citep{2012MNRAS.424.1495L}, cosmological hydrodynamical simulations \citep{2016MNRAS.455.3680L} and when compared to more general dynamical models \citep{2015ApJ...804L..21C}. The mass profiles recovered by JAM at large radii in \citet{2015ApJ...804L..21C} were also recently independently confirmed, with remarkable accuracy, using HI gas kinematics \citet{2016MNRAS.tmp..782S}. All these tests motivate our use of JAM models in this paper.

\subsection{Modelling approach}
\label{sec:modelling_approach}

We use JAM models to predict the second velocity moment from the photometry and compare to the 
\begin{align}
V_{\rm rms} = \sqrt{V^2 + \sigma^2} 
\end{align}
measured from the kinematics. 
 The error on the observed $V_{\rm rms}$ are computed via error propagation as
\begin{align}
\varepsilon_{V_{\rm rms}}^{} = \frac{1}{V_{\rm rms}} \sqrt{(V \varepsilon_{V})^2 + (\sigma \varepsilon_{\sigma})^2}. 
\end{align}
To prevent the model from being too strongly influenced by the high-S/N spaxels in the nucleus we do not use the true kinematics errors, instead we proceed as follows: The error on the velocity is initially assumed to be constant with $\varepsilon_{V} = 5$\,km\,s$^{-1}$  and the error on the velocity dispersion to be a constant fraction $\varepsilon_{\sigma} = 0.05\sigma$. To ensure a proper normalization of the model uncertainties the errors are scaled after the best fit was obtained. We scale by a constant to have $\chi^2/{\rm DOF}=1$ for the best-fitting model (model e, see below), following eq.~\ref{eq:error_normalisation}.

The large number of $V_{\rm rms}$ measurements from the MUSE data  challenge the interpretation of the results: As  \citet{2009MNRAS.398.1117V} note, in these cases the standard deviation of $\chi^2$ itself  becomes non negligible. For this reason, we follow their approach and conservatively increase $\Delta\chi^2$ required to reach a given confidence level, taking the $\chi^2$ uncertainty into account. We note that this approach is not statistically rigorous, but appears necessary to avoid unrealistically small errors in modelling fits with thousands of observables and gives sensible results. However, as our errors are based on MCMC (see below) rather than $\chi^2$ contours, we need to scale the errors to reach the same effect as increasing the $\Delta\chi^2$ level. The standard deviation of $\chi^2$ is $\sqrt{(2(N-M)}$ with $N$ data points and $M$ free parameters. To include this uncertainty in the MCMC sampling the error needs to be increased in such a way that a 'miss fit' with a $\Delta\chi^2=\sqrt{2N}$ with the original errors results in a fit with $\Delta\chi^2=1$ with the increased errors. Considering that multiplying all errors by $\varepsilon$, the $\chi^2$ decreases by $\varepsilon^2$, a decrease of the $\chi^2$ by $\sqrt{2N}$ is obtained by multiplying all errors by $(2N)^{1/4}$. This scaling is exactly equivalent to redefining the $\Delta\chi^2$ confidence level.

We use five different sets of models: 

(a) self-consistent JAM model: we assume that the mass follows the light distribution, described by the MGE surface brightness extraction (see table~\ref{tab:mge_model_parameters_photometry_extraction}). This model has three free parameters that are fitted to match the observed $V_{\rm rms}$:  The anisotropy parameter\footnote{In a cylindrical coordinate system with the cylinder axis aligned with the rotation axis of the galaxy, $\sigma_z$ is the velocity dispersion parallel to the cylinder axis and $\sigma_R$ is the velocity dispersion parallel to the radial axis.} $\beta_z = 1 - \sigma_z^2/\sigma_R^2$ and the inclination $i$ which uniquely define the shape of the second velocity moment and the mass to light ratio $(M/L)_{\rm dyn}$, which is a linear scaling parameter used to match the JAM model to the measured $V_{\rm rms}$. 

(b) stars-only JAM model: In this model we make an explicit distinction between (i) the distribution of the kinematic tracer population, which is parametrized using the MGE fitted to the stellar surface brightness in table~\ref{tab:mge_model_parameters_photometry_extraction} (as in model (a)) and (ii) the stellar mass distribution, which is described using the MGE in table~\ref{tab:mge_model_parameters_mass_extraction}, which accounts for the spatial variation in $(M_{\ast}/L)_{\rm Salp}$. This model is motivated by the observed strong $(M_\ast/L)_{\rm Salp}$ gradient, that makes the assumption of a constant $(M_{\ast}/L)$ in model (a) quite inaccurate. Again this model has three free parameters: the anisotropy $\beta_z$ and the inclination $i$ that shape the second velocity moment and a global scaling factor $\alpha=(M/L)_{\rm dyn} / (M_{\ast}/L)_{\rm Salp}$, with $(M/L)_{\rm dyn}$ being the {\em total mass}-to-light ratio in the dynamical JAM model.

(c) JAM model with NFW dark matter halo: Like in model (b), we distinguish (i) the distribution of the kinematic tracer population, which is parametrized using the MGE in table~\ref{tab:mge_model_parameters_photometry_extraction} and (ii) the stellar mass distribution, which is described using the MGE in table~\ref{tab:mge_model_parameters_mass_extraction}. However, we now also include the contribution of an NFW dark matter halo \citep{1996ApJ...462..563N}. Assuming the dark matter (DM) halo is spherical, the double power law NFW profile has two free parameters: the virial mass $M_{200}$ and the concentration $c_{200}$.  We use the virial mass-concentration $M_{200}-c_{200}$ scaling relation \citet{2011ApJ...740..102K} derived from simulations to reduce the number of free parameters for the halo to one. The details of the scaling relation do not influence our results because the break radius of the halo lies at much larger radii than where we have our kinematics, and this implies that we would obtain essentially the same results if our halo was a single power law.
The JAM model with an NFW dark matter halo has then four free parameters: the anisotropy $\beta_z$, the inclination $i$, the IMF normalisation $\alpha_{\ast} = (M_{\ast}/L)_{\rm dyn} / (M_{\ast}/L)_{\rm Salp}$  and the virial mass of the NFW halo $M_{200}$. Given that the stellar mass was computed from the spectral fits assuming a Salpeter IMF, any extra mass scaling can be interpreted as a difference in the IMF mass normalization.

(d) JAM model with generalised gNFW dark matter halo \citep[e.g. ][]{2001ApJ...555..504W}: The only difference between this model and model (c) is that we replace the NFW dark matter halo by a generalised NFW profile. Instead of forcing the inner slope of the halo to be $-1$, we introduce a free inner halo exponent $\gamma$, but force the outer slope to be $-3$. The matter density $\rho$ is then described by
\begin{align}
 \rho(r) = \rho_s\left(\frac{r}{r_s}\right)^{\gamma}\left(\frac{1}{2} + \frac{1}{2}\frac{r}{r_s}\right)^{-\gamma-3}
 \label{eq:power_law_density}
\end{align}
where $\rho_s$ is the normalisation of the halo and $r_s$ is the break radius. The break radius is outside the range covered by our kinematic data and should not affect the results at all. In line with \citet{2013MNRAS.432.1709C} we choose a break radius of $r_s=20$\,kpc. This model has five free parameters: the anisotropy $\beta_z$, the inclination $i$, the inner halo slope $\gamma$, the dark matter fraction $f_{\rm DM}$ and the IMF normalisation $\alpha_{\ast}$.

(e) JAM model with a power law {\em total} mass density: 
In this approach we assume the stars (parametrized by the MGE in table~\ref{tab:mge_model_parameters_photometry_extraction}) are just a tracer population within a total mass distribution which we assume to be spherical and described by a power-law, within the region where we have kinematic information (and with a break at large radii). In particular we do not make the standard assumption that the total mass is the sum of a luminous and dark component with specific parametrizations. In practise we describe the {\em total} mass density by equation \ref{eq:power_law_density}. We are not aware of the use of this approach in stellar dynamical studies, but it was extensively used before in strong gravitational lensing studies of density profiles \citep[e.g.][]{2009ApJ...703L..51K}. The conceptual advantage of this approach is that it does not require any assumption regarding the {\em stellar mass} distribution, which can be quite uncertain due to the variations in $M_{\ast}/L$ of the stellar population, including possible radial variations in the IMF. A motivation for its use comes from the finding that, even when the dark halo is allowed to be quite general, the {\em total} density profiles of ETGs are well approximated by a single power-law out to about 4~$R_{\rm e}$ \citep{2015ApJ...804L..21C}. This model has four free parameters: the anisotropy $\beta_z$, the inclination $i$, the inner {\em total} mass density slope $\gamma$ and the {\em total} mass density at 1\,kpc $\rho(r=1$\,kpc). We choose $\rho(r=1$\,kpc) as free parameter to reduce the degeneracy between the halo normalisation and the halo slope. Numerically it is straightforward to replace $\rho_s$ in equation~\ref{eq:power_law_density} with  $\rho(r=1$\,kpc), while we keep $r_s=20$\,kpc.

In all models the central region with a radius of $r=2\arcsec$ is masked. In this region a sharp $\sigma$ peak is observed, that influences the derived model parameters in a non physical way. The necessary detailed modelling of the central black hole is beyond the scope of this paper. For all five models (a) to (e) the JAM model parameters are obtained using a Markov chain Monte Carlo (MCMC) sampling.
The MCMC sampling is done using the \citet{2013PASP..125..306F} \texttt{emcee} \texttt{Python} code, an implementation of \citet{2010AMaCS...5....65G} affine invariant Markov chain Monte Carlo ensemble sampler. There are basically two inputs to \texttt{emcee}: the prior function $P$(model) and the likelihood function $P$(data|model). We use an uninformative prior function, i.e. within the bounds the likelihood is 1, outside it is zero. Assuming Gaussian errors the likelihood function is:
\begin{align}
 P(\text{data}|\text{model}) \propto \exp\left(-\frac{\chi^2}{2}\right) \\
\chi^2 = \sum \left(\frac{V_{\rm rms} - \langle v_{\rm los}^2\rangle^{1/2}}{\varepsilon_{V_{\rm rms}}^{}}\right)^2.
\label{eq:JAM_chi2}
\end{align}
where $\langle v_{\rm los}^2\rangle$ is the second moment of the JAM model velocity distribution. The posterior distribution is then:
\begin{align}
 P(\text{model}|\text{data}) \propto P(\text{data}|\text{model}) \times P(\text{model})
\end{align}
We use 100 walkers, each performing 500 steps to sample the posterior distribution.

\subsection{JAM models of NGC~5102}

\begin{figure}
 \includegraphics[width=0.45\textwidth]{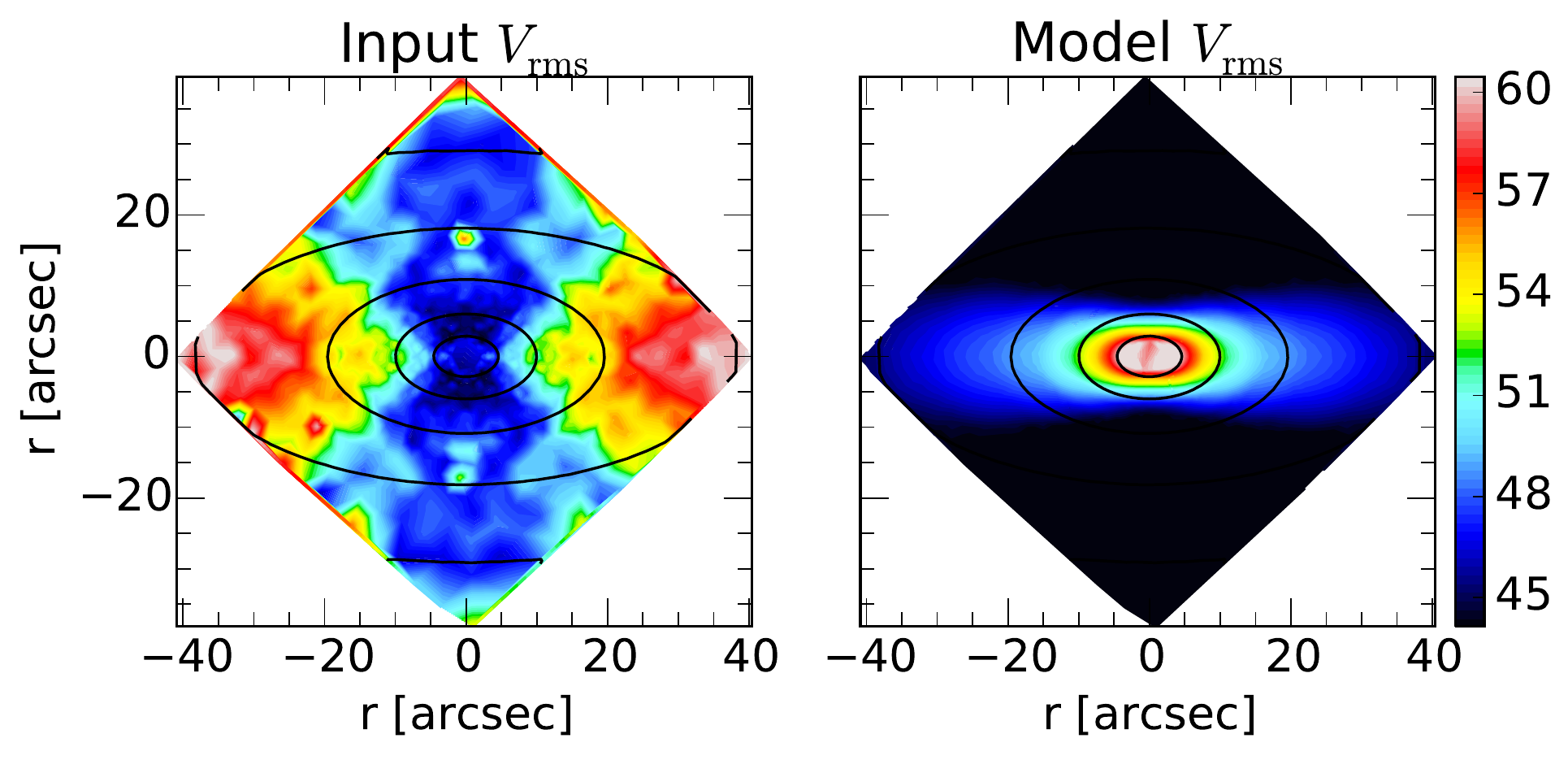}
 \caption{Self-consistent JAM model (a) second velocity moment (right hand panel) is compared to the measured $V_{\rm rms}$ (left hand panel). The model is completely unable to describe the observations. The values on the colour-bar are in km\,s$^{-1}$.}
 \label{fig:bestfit_selfconsistent}
\end{figure}

The best-fitting self-consistent JAM model (a) is compared to the observed $V_{\rm rms}$ in Fig.~\ref{fig:bestfit_selfconsistent}, the median values of the posterior distribution are given in table~\ref{tab:jam_model_parameters}. It is obvious that the model does not at all represent the observed $V_{\rm rms}$. This result is in contrast to the results obtained by \citet{2013MNRAS.432.1709C} who apply the same approach to the 260 ATLAS$^{\rm 3D}$ galaxies (see their Fig. 1) where all galaxies with good kinematic data are well described by the self-consistent model. This qualitatively indicates that NGC~5102, unlike the ATLAS$^{\rm 3D}$ ETGs, is dominated by dark matter. It also shows that the dark matter must be more shallow than the stars, since otherwise the self-consistent model would still be able to produce an acceptable fit. These qualitative results will be quantified in the following.

\begin{table*}
 \caption{JAM model parameters. The table gives the median values from the JAM model MCMC posterior distribution. The errors are the larger of the two intervals: 16th to 50th and 50th to 84th percentile.}
 \label{tab:jam_model_parameters}
 \begin{tabular}{l c c c c c c c c}\hline
  Model & $\beta_z$ & $i$ & $\log(M/L)$ & $\log(\alpha)^a$  & $\gamma$ & $f_{\rm DM}$ &  $\log(\rho(r=1\rm{\,kpc}))$ & $\chi^2$/DOF$^b$ \\
        &   --    & [\degr] & $[\log(\rm M_{\sun}/L_{\sun})]$ & -- & -- & -- & $[\log(\rm M_{\sun}\,pc^{-3})]$ & -- \\
  bounds &  [0.0, 0.5] & [70., 90.] & [$-0.6$, 0.3] & [$-0.6$, 0.3] & [$-2.0$, 0.0] & [0, 1] & [$-4.0$, 1.0] &         \\\hline
  (a)     & $0.09 \pm 0.04$ & $87 \pm 3$ & $0.05 \pm 0.01$ & -- & -- & --    & --    & 14.84 \\
  (b)     & $0.08 \pm 0.03$ & $87 \pm 3$ & -- & $0.14 \pm 0.01$ & -- & --    & -- & 5.42 \\
  (c)     & $0.18 \pm 0.04$ & $86 \pm 3$ & -- & $-0.05 \pm 0.03$ & --    & $0.37\pm0.04$ & --  & 1.13 \\
  (d)     & $0.22 \pm 0.05$ & $86 \pm 4$ & -- & $-0.21 \pm 0.21$ & $-1.4 \pm 0.3$ & $0.58 \pm 0.16$ & --    & 1.02  \\
  (e)     & $ 0.27 \pm 0.04$ & $86 \pm 4$  & -- & -- & $-1.75 \pm 0.04$ & --    & $-0.75 \pm 0.03$ & 1.00  \\\hline
  \multicolumn{9}{p{\textwidth}}{$^a$ For models (c) and (d) $\alpha_{\ast} = (M_{\ast}/L)_{\rm dyn} / (M_{\ast}/L)_{\rm Salp}$, with $(M_{\ast}/L)_{\rm dyn}$ ratio of dynamical {\em stellar} mass to light ratio. For model (b) $\alpha=(M/L)_{\rm dyn} / (M_{\ast}/L)_{\rm Salp}$ with $(M/L)_{\rm dyn}$ the {\em total} mass in the dynamical JAM model.}  \\
  \multicolumn{9}{p{\textwidth}}{$^b$ The $\chi^2$/DOF values stated here give $\chi^2/\text{DOF}=1$ for the best-fitting model (e). This means these $\chi^2$/DOF values are {\em not} scaled to account for the effects of the standard deviation of the $\chi^2$ itself (see Sec.~\ref{sec:modelling_approach}).}
 \end{tabular}
\end{table*}

\begin{figure}
 \includegraphics[width=0.45\textwidth]{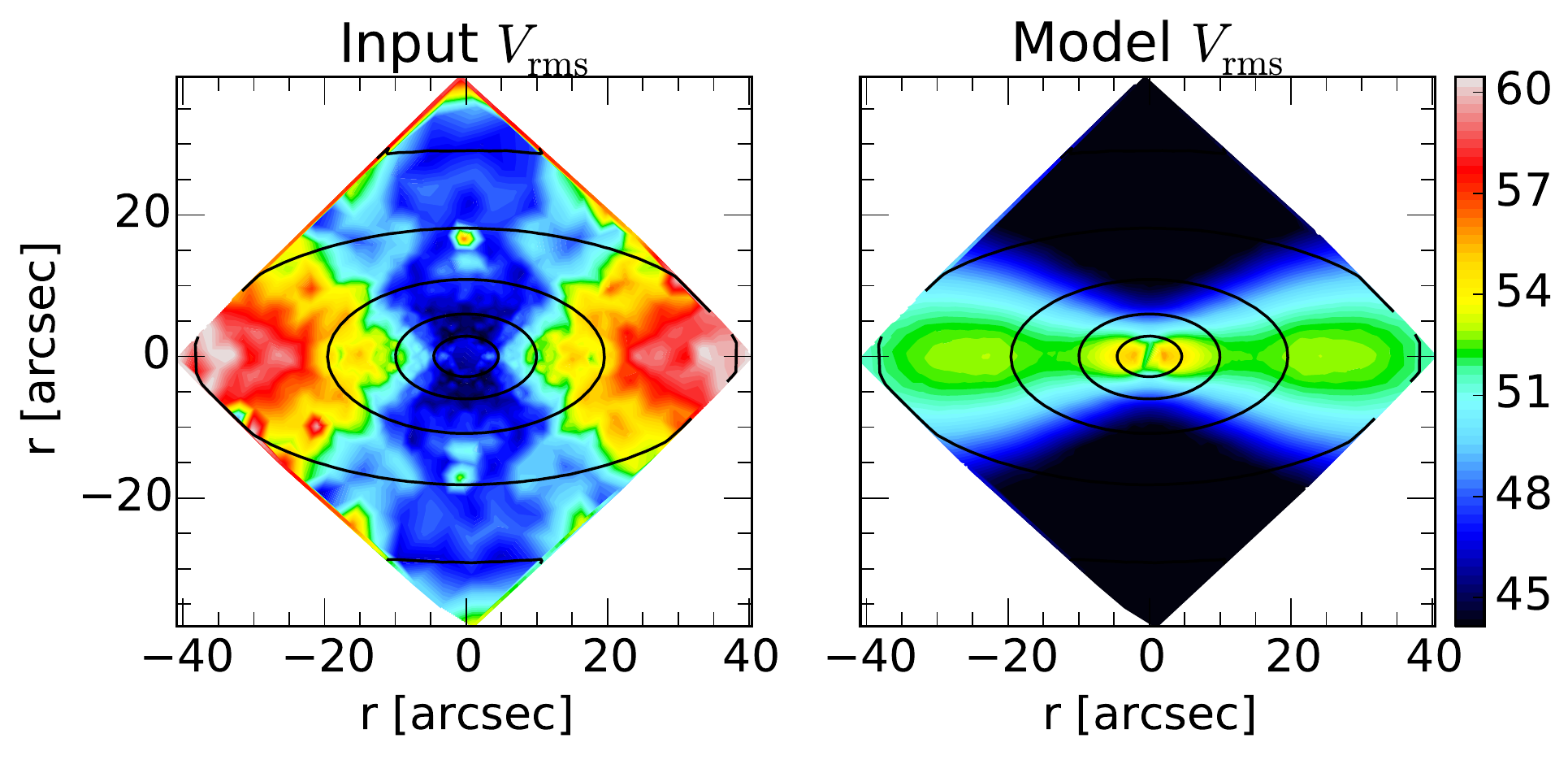}
 \caption{JAM model (b): the stars-only model second velocity moment (right hand panel) is compared to the measured $V_{\rm rms}$ (left hand panel). The fit is slightly better than in Fig.~\ref{fig:bestfit_selfconsistent}, but still the model is unable to describe the observations. The values on the colour-bar are in km\,s$^{-1}$.}
 \label{fig:bestfit_selfconsistent_MoverL_correct}
\end{figure}

The stars-only JAM model (b) shown in Fig.~\ref{fig:bestfit_selfconsistent_MoverL_correct} differs noticeably from model (a) and represents a clear improvement but still at a poor overall level. Using the MGE {\em stellar mass} prediction instead of the simple mass follows light assumption moves the model into the right direction, but does not solve a more fundamental difference between data and model prediction. The median values of the posterior distribution are given in table~\ref{tab:jam_model_parameters}. Even though the $\chi^2$/DOF of this model indicates a fair improvement over model (a), the quality of the JAM model is poor compared to the self-consistent models for the ATLAS$^{\rm 3D}$ galaxies \citep{2013MNRAS.432.1709C}. It is obvious that the shallower mass profile used for this model can only reduce the dark matter needed to explain the kinematics of this galaxy, because the mass profile is shallower than the light profile and the DM halo has an even shallower slope than the mass profile.

\begin{figure}
 \includegraphics[width=0.45\textwidth]{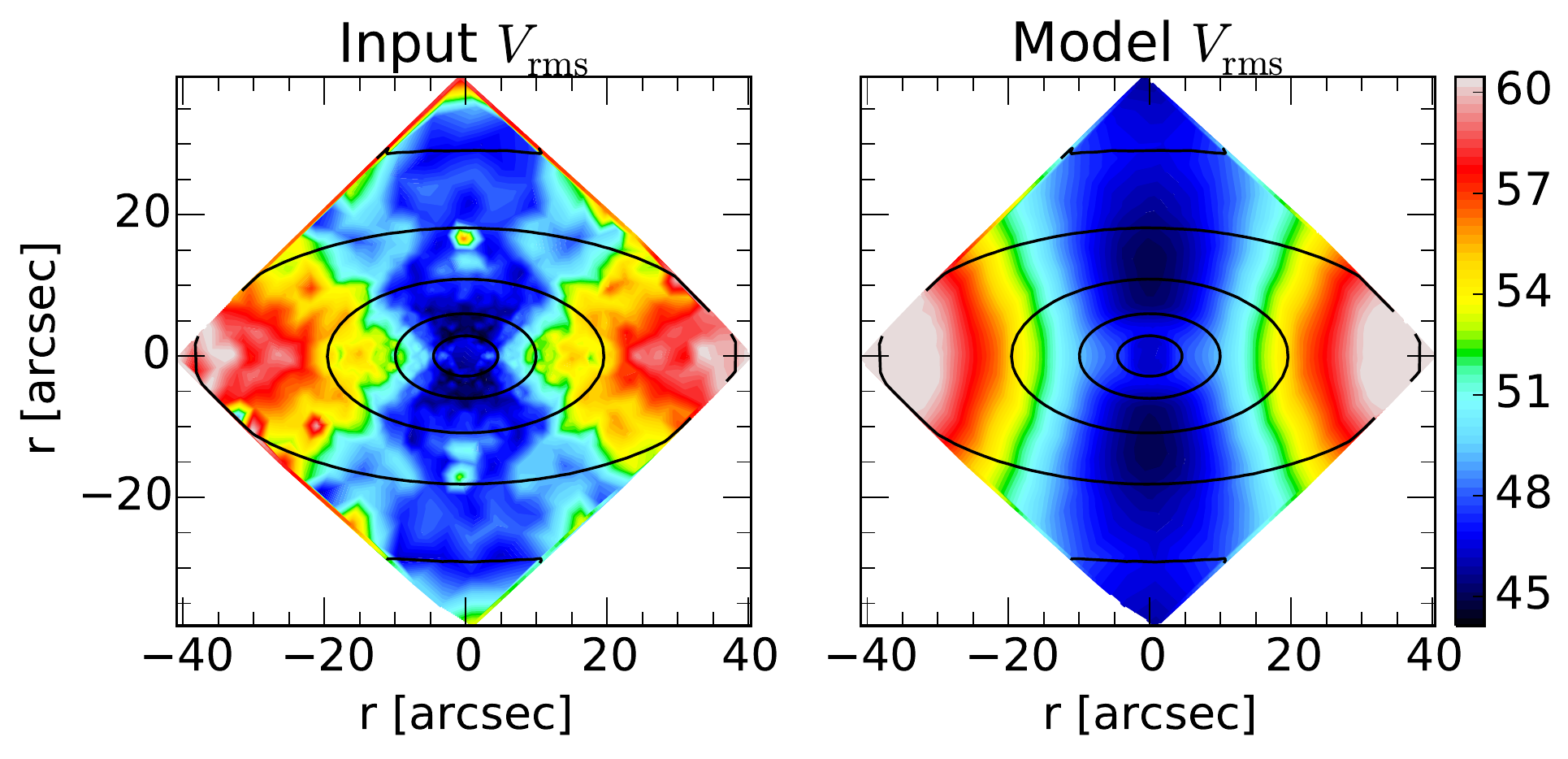}
 \includegraphics[width=0.45\textwidth]{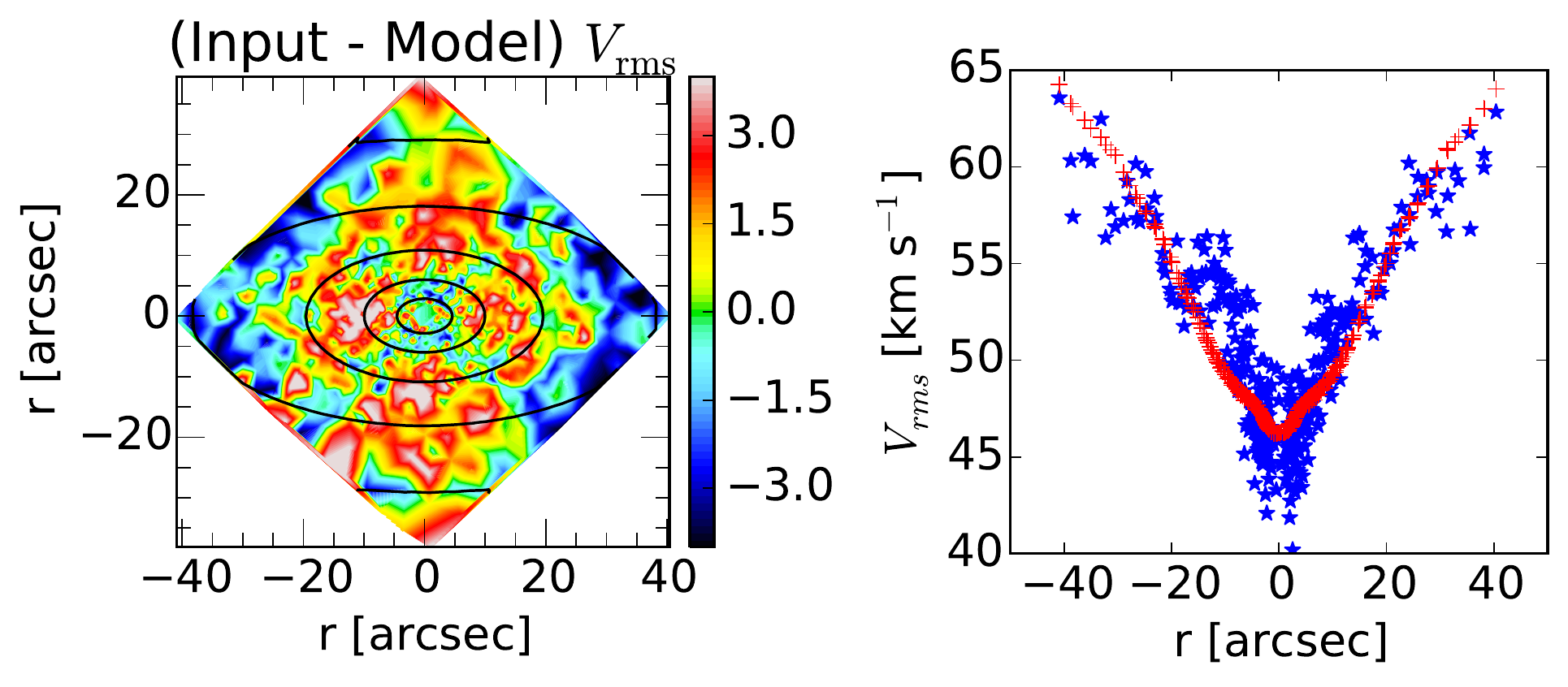}
 \caption{JAM model (c) (NFW halo and MGE stellar mass) second velocity moment. {\em Top row:} The symmetrized measured $V_{\rm rms}$ (left) is compared to the JAM model prediction (right). {\em Bottom left:} The plot shows the difference between measured and model $V_{\rm rms}$. {\em Bottom right:} A cut through the $V_{\rm rms}$ plane is shown, plotting all points with minor axis distances  $-2.5 \leq r_{\rm min} \leq 2.5$. The blue stars are the measurements, the red crosses the JAM model. The values on the colour-bar are in km\,s$^{-1}$.}
 \label{fig:bestfit_nfw_dm_halo_MoverL_correct}
\end{figure}

Fitting model (c) with an NFW dark matter halo to the observed $V_{\rm rms}$ dramatically improves the quality of the JAM model. Fig.~\ref{fig:bestfit_nfw_dm_halo_MoverL_correct} compares the best-fitting JAM model with the NFW dark matter halo. Now the rise in velocity dispersion along the major axis and the flat $V_{\rm rms}$ along the minor axis are reproduced well by the model.
The median model parameters from the posterior distribution are again summarised in table~\ref{tab:jam_model_parameters}. The mass $M(r)$ of an axisymmetric MGE enclosed within a sphere of radius $r$ can be obtained as
\begin{align}
 M(r)=4\pi\int_0^r r^2\rho_{\rm tot}(r) dr,
\end{align}
where the density $\rho_{\rm tot}(r)$ was given in footnote 11 of \citet{2015ApJ...804L..21C}. Using the same notation we obtain:
\begin{align}
M(r) &= \sum_j M_j \left[ {\rm erf}\left(h_j\right) - \frac{1}{e_j} \exp\left(- h_j^2q_j^2\right) {\rm erf} \left(h_je_j\right) \right] 
\end{align}
with
\begin{align}
e_j &\equiv \sqrt{1-q_j^2} \\
h_j &\equiv \frac{r}{\sigma_j q_j\sqrt{2}} 
\end{align}
where the index $j$ runs over the Gaussian components of the MGE model. Inside a sphere of radius $r=R_{\rm e}$ the dark matter fraction is $M_{\rm DM}(r)/M_{\rm tot}(r) = 0.37\pm0.04$.

\begin{figure*}
 \includegraphics[width=0.85\textwidth]{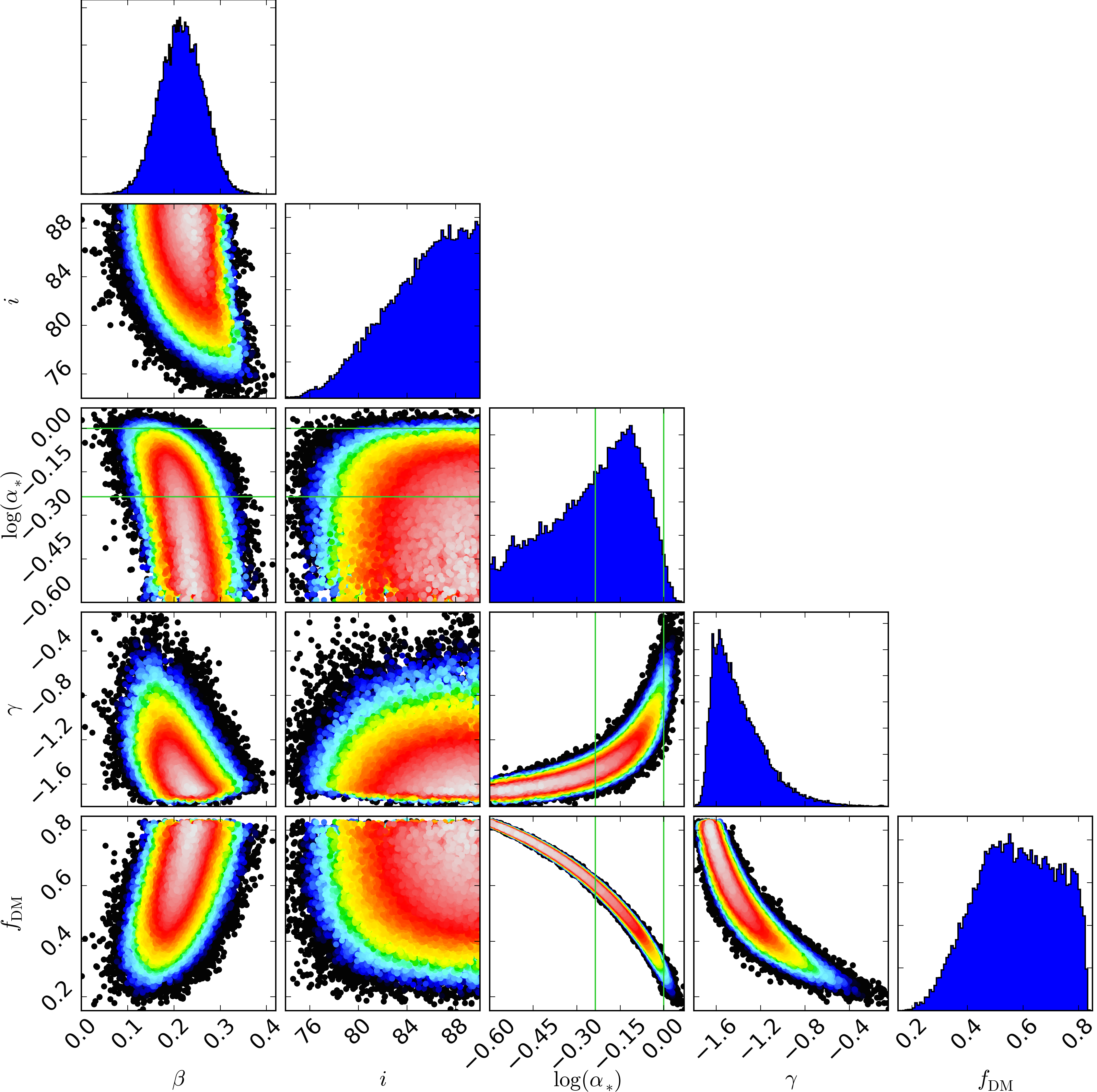}
 \caption{The posterior distribution of model (d) with an MGE mass model and a  gNFW dark matter halo, i.e. a power law halo with free inner slope and outer slope fixed to $-3$. The green lines correspond to a Salpeter IMF ($\log(\alpha_{\ast})=0.$) and a Chabrier IMF ($\log(\alpha_{\ast})=-0.236$).  The colour coding gives the likelihood for the model parameters, with black points being disfavoured at $3\sigma$ or more. The sharp truncation of the dark matter fraction at $f_{\rm DM}\approx 0.83$ is cause by the lower boundary on the IMF normalisation $\alpha_{\ast}$, effectively introducing a lower limit for the stellar mass fraction.}
 \label{fig:lum_and_dm_power_law_corner}
\end{figure*}

For model (d) the posterior distribution from the MCMC sampling is shown in Fig.~\ref{fig:lum_and_dm_power_law_corner} and the median values are summarised in table~\ref{tab:jam_model_parameters}. 
This plot shows the expected degeneracy between the IMF normalisation, the dark matter halo slope $\gamma$ and the dark matter fraction. Due to the degeneracy a range of parameter combinations gives nearly equally acceptable fits: from models with reasonable IMF normalisations to models that are mostly dark matter dominated (see also model e, as this is the limiting case of $\log(\alpha_{\ast})$ going to $-$infinity and $f_{\rm DM}$ to 1). A Salpeter IMF with an inner halo slope $\gamma=-1$ (i.e. model (c)) seems to be less likely. A model with a Chabrier IMF (i.e. a typical IMF for a low mass galaxy) is in the range of high likelihood and requires  a dark matter fraction of $f_{\rm DM}\approx0.6$. In Fig.~\ref{fig:lum_and_dm_power_law_corner} we see a sharp truncation of the dark matter fraction at $f_{\rm DM}\approx 0.83$. We emphasizes that this truncation is caused by our lower boundary on the IMF normalisation $\alpha_{\ast}$. The lower boundary on $\alpha_{\ast}$ is chosen at half the mass of a Chabrier IMF.

\begin{figure}
 \includegraphics[width=0.45\textwidth]{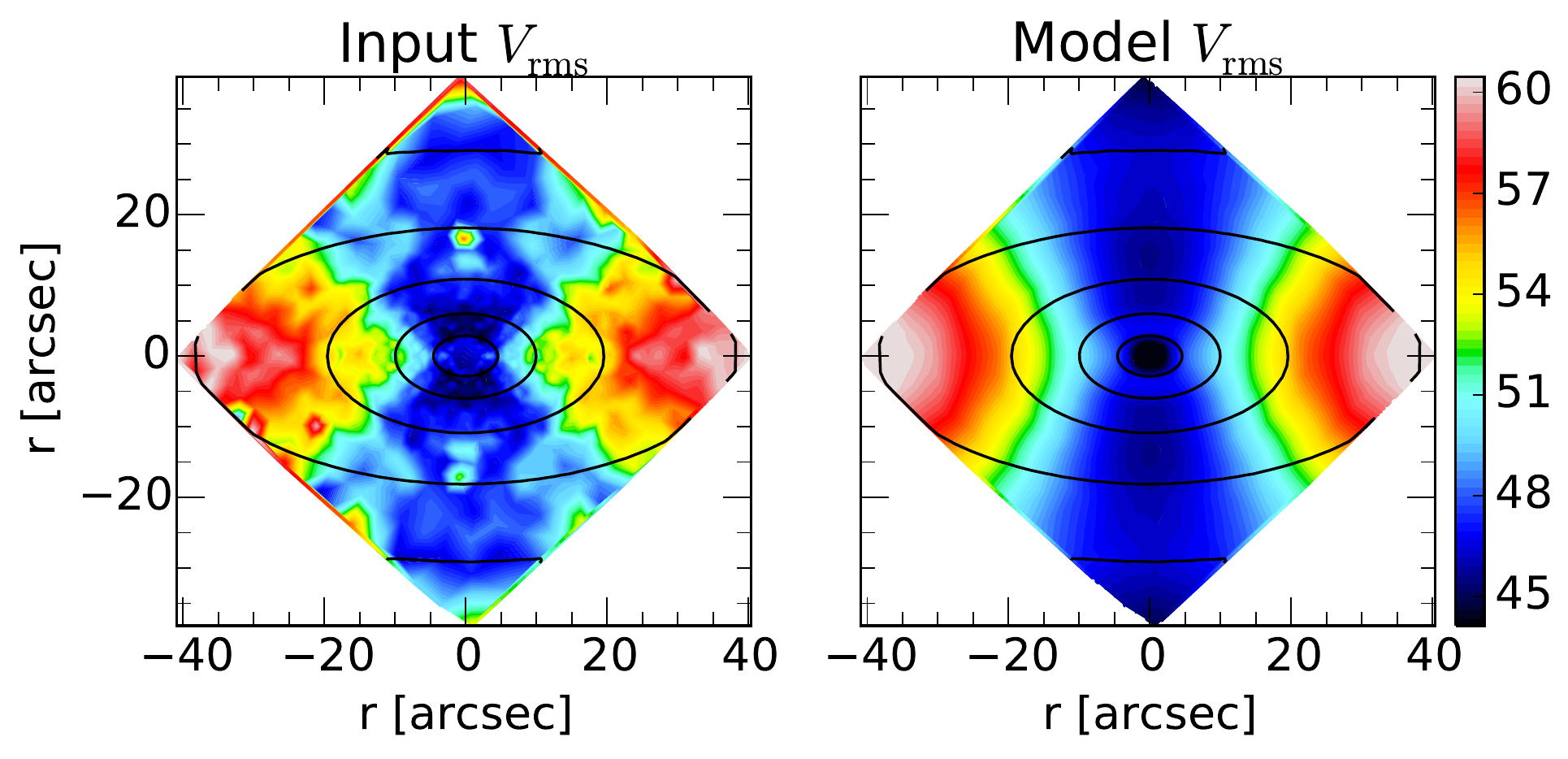}
 \includegraphics[width=0.45\textwidth]{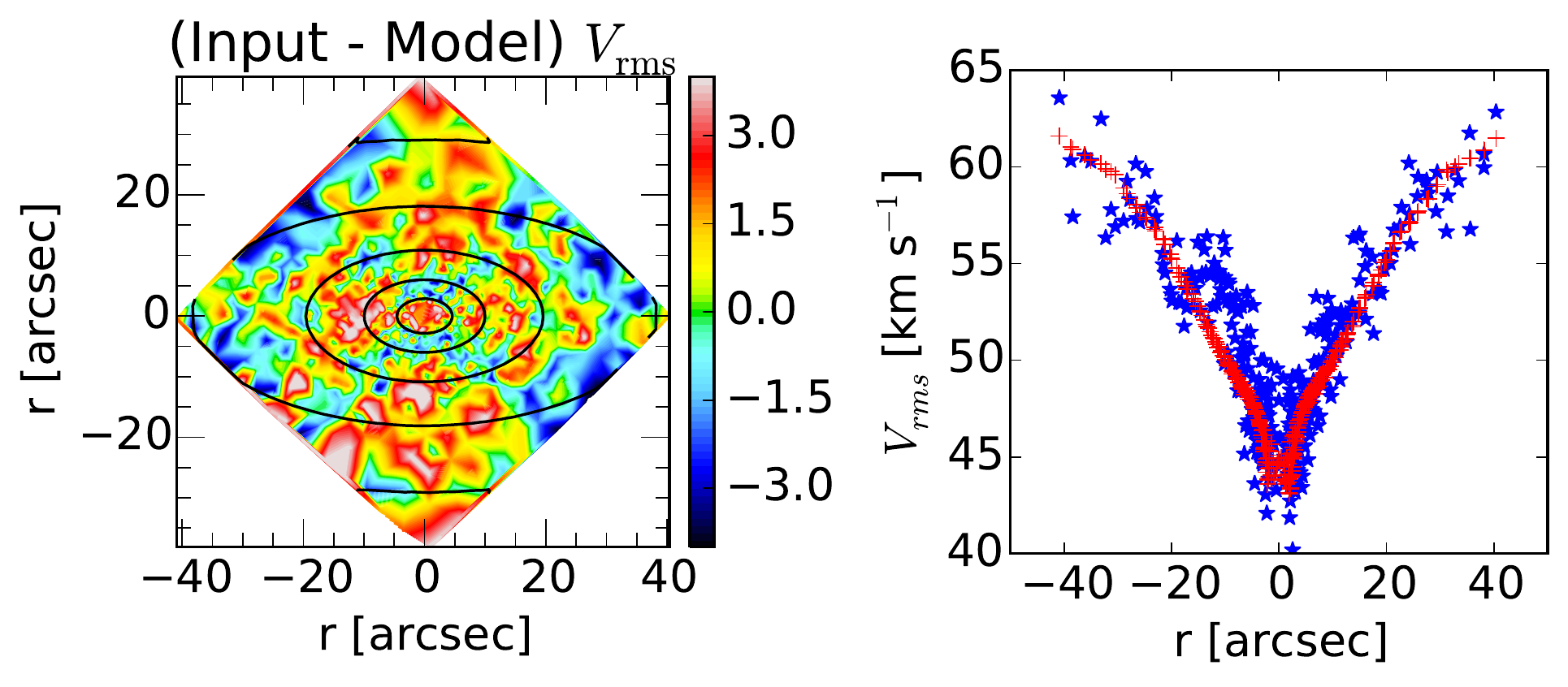}
 \caption{JAM model (e) (total mass is parametrized by a spherical power law) second velocity moment. {\em Top row:} The symmetrized measured $V_{\rm rms}$ (left) is compared to the JAM model prediction (right). {\em Bottom left:} The plot shows the difference between measured and model $V_{\rm rms}$. {\em Bottom right:} A cut through the $V_{\rm rms}$ plane is shown, plotting all points with minor axis distances  $-2.5 \leq r_{\rm min} \leq 2.5$. The blue stars are the measurements, the red crosses the JAM model. The values on the colour-bar are in km\,s$^{-1}$.}
 \label{fig:bestfit_power_law_halo}
\end{figure}

In model (e) we assume the {\em total} matter distribution can be described by a power law. Figure~\ref{fig:bestfit_power_law_halo} shows that this model actually gives the best description of the observed $V_{\rm rms}$, even though the improvement over the NFW model (c) is minor (see table~\ref{tab:jam_model_parameters}).  This underlines that the simple assumption of a power law matter distribution with spherical symmetry describes the observed $V_{\rm rms}$ well. The slope for the {\em total} mass density is a very robust result, in contrast to the slope of the dark halo that is degenerate with the IMF normalisation. The slope $\gamma=-1.75\pm0.04$ we find for NGC~5102 is shallower than the slope of an isothermal halo ($\gamma_{\rm iso}=-2$). \citet{2015ApJ...804L..21C} measure an average slope of $\langle \gamma \rangle = -2.19\pm 0.03$  for the 14 fast rotating galaxies out to large radii of $4R_{\rm e}$. The authors do not find a strong dependence of the slope on the radius and report a marginally smaller slope when limiting the radial range to $r\leq R_{\rm e}$. The smaller slope we find for the low mass galaxy NGC~5102 agrees with the decreasing {\em total} mass slope reported by \citet{2016ARA&A..54..597C}: Fig.~22c shows a trend as a function of $\sigma$ (dashed lines, with values given in Fig.~20). The lowest $\sigma$ in ATLAS$^{\rm 3D}$\ is $\sim50$\,km/s, which is comparable to $\sigma_{\rm e} = 48$ of this galaxy. At that $\sigma$ level the slope in ATLAS$^{\rm 3D}$ is $\sim-1.9$, so this galaxy would be almost consistent with that.

\subsection{Decomposing the power-law profile}

The advantage of model (e) is that it allows us to measure the {\em total} density profile, which is an important observable in itself, without the need to make {\em any} assumption about the dark matter contribution and its parametrization. This is the same situation as in studies of the circular velocity curves of ETGs with gas discs \citep[e.g.][]{2008MNRAS.383.1343W}. However, once the total density or rotation curve has been obtained, in both cases extra assumptions are required to make inferences about the luminous and dark matter.

When measuring dark matter via this two-steps process (modelling of the total mass followed by mass decomposition), one should expect results that are generally similar to what one can obtain via the standard route of assuming a dark halo parametrization during the JAM fitting itself. However the details of the two approaches are sufficiently different that this allow one to test for the robustness of the halo results.

The stellar mass density profile is computed from the axisymmetric MGE using Note 11 in \citet{2015ApJ...804L..21C}. 
The NFW density is computed following the approach detailed in model (c) in Sec.~\ref{sec:modelling_approach} and the power law density is derived from the analytic expression in equation~\ref{eq:power_law_density}.
This fitting has two free parameters: the logarithm of the IMF normalisation $\log(\alpha_{\ast})$ to scale the stellar mass density profile and the logarithm of the  virial mass $\log(M_{200})$ of the NFW halo. For each individual fit we assume constant relative errors on the total density, with arbitrary normalization. We emphasizes that we are using the NFW parametrization for the dark matter halo. From the results of the JAM models one might prefer to use the gNFW parametrization, but that would not work: In the inner part (where we can constrain the model with our kinematic data) the gNFW is simply a power-law with free slope and normalisation. Because we parametrized our {\em total} mass density with a power-law, this could be reproduced by the gNFW, without the need for any luminous matter - a result clearly in contrast to our observations.

\begin{figure}
 \includegraphics[width=0.45\textwidth]{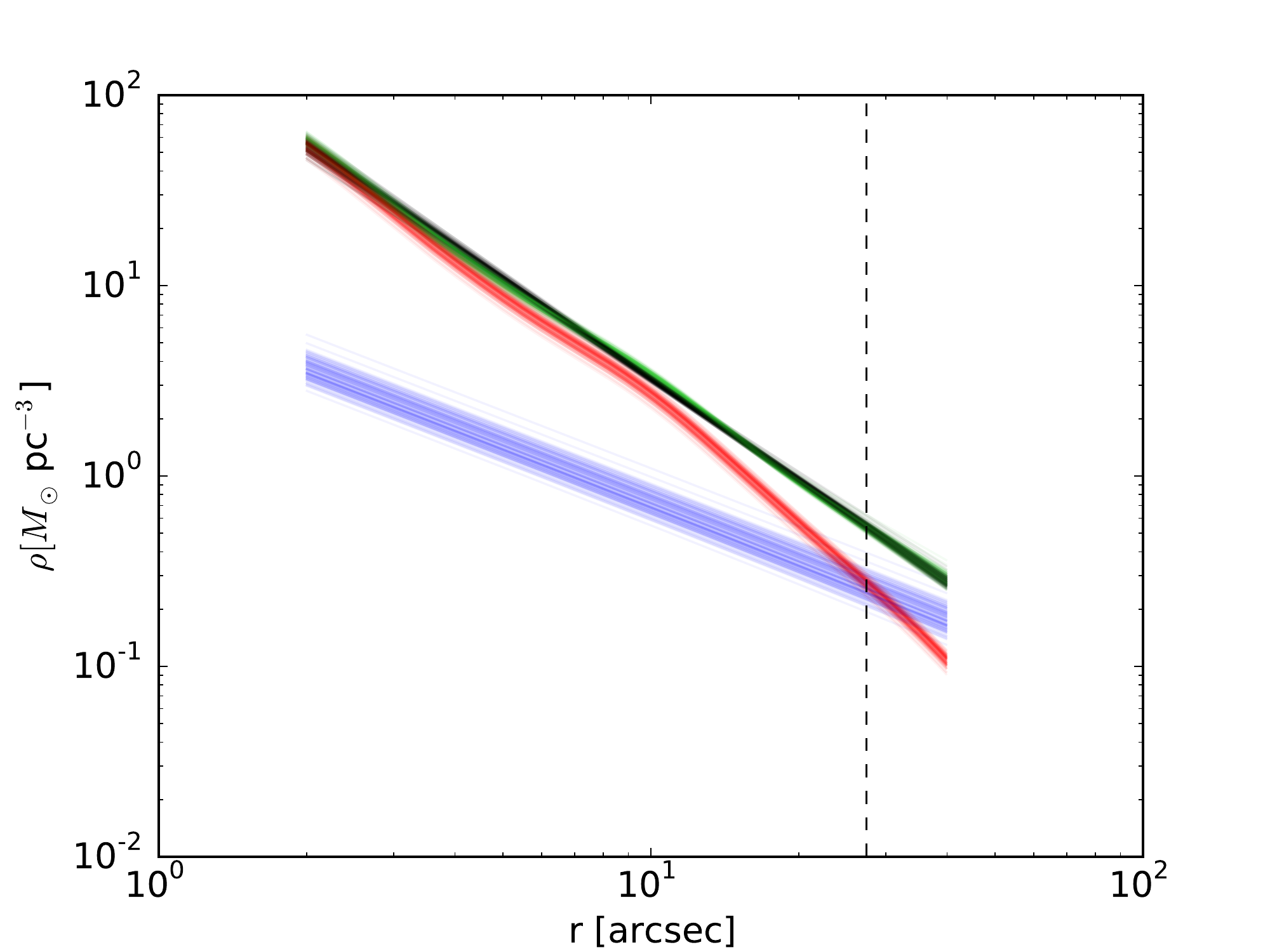}
 \caption{The solid black lines are 100 randomly selected power law models from the posterior distribution of the JAM model (e). These are fitted with a combination of stellar mass density (red) and NFW dark matter halo (blue). The green line is the sum of luminous and dark component. We fit the power law models in the range $2\arcsec \leq r \leq 40\arcsec$, where the lower limit is identical to the central masking radius used for the JAM modelling and the outer limit is the largest radius sampled by our kinematic data. The vertical (black, dashed) line marks $1R_{\rm e}$.}
 \label{fig:fit_power_law_halo_NFW_log_sampling_MoverL_correct}
\end{figure}

\begin{figure}
 \includegraphics[width=0.45\textwidth]{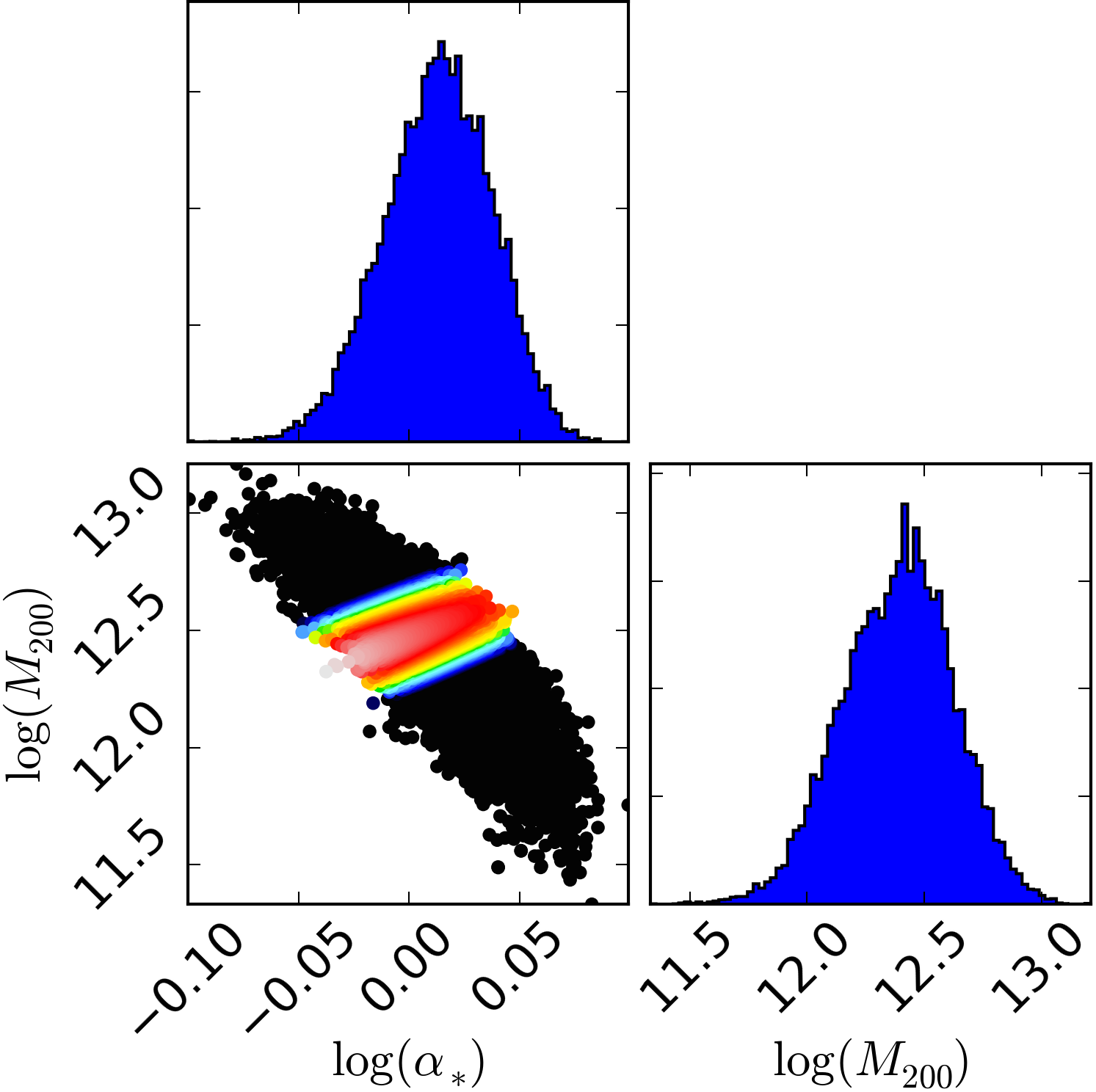}
 \caption{Corner plot showing the posterior distribution of $M_{200}$ and IMF normalisation $\alpha_{\ast}$ values as obtained by decomposing the JAM power law model (e) into stellar and dark matter. The colour coding gives the likelihood for the model parameters, with black points being disfavoured at $3\sigma$ or more.}
 \label{fig:fit_power_law_halo_NFW_log_sampling_MoverL_correct_corner}
\end{figure}

We fit all realisations from the MCMC posterior distribution of the JAM model (e) to obtain a distribution in the two free parameters $\log(\alpha_{\ast})$ and  $\log{M_{200}}$. Fig.~\ref{fig:fit_power_law_halo_NFW_log_sampling_MoverL_correct} shows 100 randomly chosen fits from the posterior distribution. The distribution of the parameters is shown in Fig.~\ref{fig:fit_power_law_halo_NFW_log_sampling_MoverL_correct_corner}. From this posterior distribution we obtain $\log(\alpha_{\ast}) = 0.015\pm 0.026$ and $\log(M_{200}) = 12.4\pm 0.3$, which means the power law decomposition prefers a slightly higher stellar mass then the JAM model (c) (table~\ref{tab:jam_model_parameters}) and accordingly a slightly lower dark matter mass. From the decomposition of the power law model we measure a dark matter fraction of $f_{\rm DM} = 0.30\pm0.04$, compared to $f_{\rm DM} = 0.37\pm0.04$ for model (c). Given the differences in the two methods to infer the dark matter fraction (JAM modelling and the power law decomposition) the two values are in a good agreement and more important they give a sense for the systematic uncertainties that are difficult to estimate otherwise.

The colour coding in Fig.~\ref{fig:fit_power_law_halo_NFW_log_sampling_MoverL_correct_corner} gives the likelihood for the models with black points being disfavoured at $3\sigma$ or more. The fact that the coloured area has a different shape than the compete posterior sample indicates that not all power law realisations can be described by the stellar mass distribution and an NFW dark matter halo. On the other hand the best-fitting stellar plus NFW dark halo model seems to be poorly described by a power law. This nicely emphasizes the systematic differences between the models.

\subsection{Mass decomposition of counter-rotating discs}
\label{Mass-decomposition-of-counter-rotating-discs}

To study the mass distribution we only need the $V_{\rm rms}$. In this way we can ignore how the stellar kinematics separates into ordered and random motions, and we do not need to make any assumptions about the tangential anisotropy. However, in order to study the mean velocity we need to make a choice for the tangential anisotropy. 

In the JAM approach the tangential anisotropy can be specified in two ways: (i) either by giving the $\sigma_{\phi}/\sigma_{R}$ ratio \citep[eq.~34 of][]{2008MNRAS.390...71C} or (ii) by quantifying the ratio $\kappa$ between mean rotation of the model and a model with an oblate velocity ellipsoid \citep[eq.~35 of][]{2008MNRAS.390...71C}. 

The anisotropy of a JAM model can be different for each individual Gaussian of the MGE and this allows one to construct models with a nearly arbitrary mean rotation field, by assigning different tangential anisotropies. Here we allow for $\kappa$ to be different and in particular to have different signs for different Gaussians in order to model the counter-rotating discs in this galaxy.

Similar examples of JAM models of a set of six fast rotator ETGs with counter-rotating discs from the ATLAS$^{\rm 3D}$ survey are presented in Fig.~12 of \citet{2016ARA&A..54..597C}.

\begin{figure*}
 \includegraphics[width=1.0\textwidth]{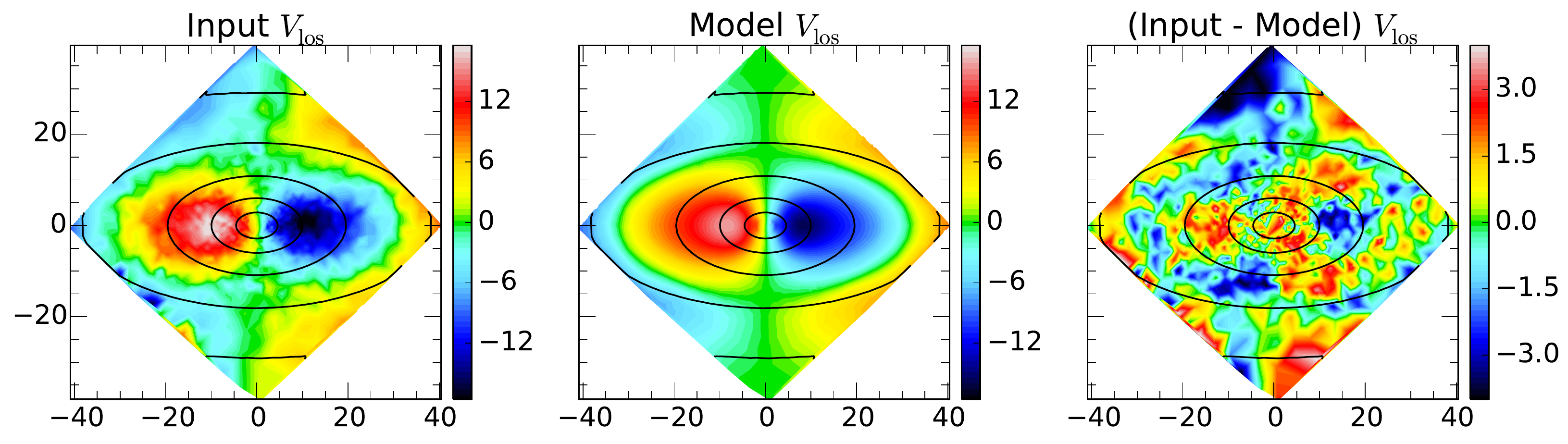}
 \includegraphics[width=1.0\textwidth]{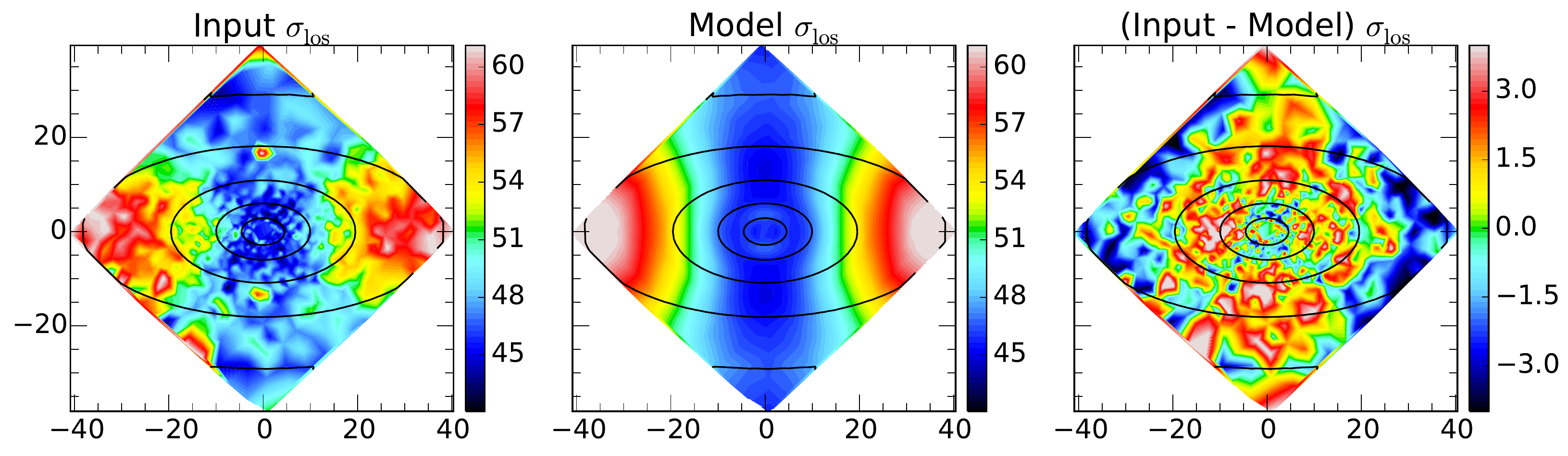}
 \caption{The best-fitting JAM model (c) $V_{\rm rms}$ prediction is decomposed in ordered ($V$) and random ($\sigma$) motion. {\em Top panel:} The measured $V$ (left hand panel, symmetrized version of Fig.~\ref{fig:blue_kinematik_4_mom}) is compared to the predicted velocity field (middle panel). The right panel shows the residuals. {\em Bottom panel:}  The measured $\sigma$ (left hand panel, symmetrized version of Fig.~\ref{fig:blue_kinematik_4_mom}) is compared to the predicted dispersion field (middle) and the residuals are shown in the right panel.  The values on the colour-bar are in km\,s$^{-1}$.}
 \label{fig:nfw_dm_halo__kappa_best_fit_vel} 
\end{figure*}

We use the JAM model (c) (with MGE stellar mass model and NFW dark matter halo) to recover the velocity and dispersion of this model. We use a \texttt{Python} implementation of the MP-fit program \citep{2009ASPC..411..251M} to fit the $\kappa$ value of each Gaussian component, by minimising the squared residuals between the observed and modelled velocity and dispersion fields simultaneously. This fit has 11 free parameters (i.e. the $\kappa$ value of each of the 11 Gaussian components from the photometry given in table~\ref{tab:mge_model_parameters_photometry_extraction}). The resulting velocity and dispersion map are compared to the observed ones in Fig.~\ref{fig:nfw_dm_halo__kappa_best_fit_vel}.

The most prominent disagreement between data and model dispersion is the too steep rise of the model along the major axis. This is the same disagreement one can see in the $V_{\rm rms}$ in Fig.~\ref{fig:bestfit_nfw_dm_halo_MoverL_correct} and  is caused by too much matter in the outer parts of the galaxy and indicative for a too shallow dark matter halo as illustrated by the fact that model (d) prefers a steeper halo slope than NFW.

The decomposition of the JAM $V_{\rm rms}$ in ordered and random motion can be used, with some assumptions, to estimate the mass fraction the two counter rotating discs contribute to the total stellar mass. We start by writing the fitted $\kappa_i$ value of the $i$-th Gaussian component as a linear combination of the intrinsic $\kappa$ values of the two components:
\begin{align}
\kappa_i &= \kappa_i^{(1)} w_i^{(1)} + \kappa_i^{(2)} w_i^{(2)}\\
 1 &=w_i^{(1)} + w_i^{(2)}
\end{align}
To obtain a mass estimate from these two equations we need to know the intrinsic $\kappa$ values of the two discs. Since it is impossible to derive these from the data we assume that the two counter rotating discs have perfect oblate velocity ellipsoids, i.e. their intrinsic $\kappa$-values are $\kappa_i^{(1)} = 1$ and $\kappa_i^{(2)} = -1$. This assumption is motivated by the finding that fast rotator ETGs as a class satisfy this assumption with high accuracy \citet[see Fig.~11 of ][]{2016ARA&A..54..597C}. We then solve for the weights of each Gaussian component:
\begin{align}
  w_i^{(1)} &= \frac{\kappa_i + 1}{2}\\
 w_i^{(2)} &= 1 - w_i^{(1)}.
\end{align}
The masses of the two components are the weighted sums of the total mass $M_i$ of the Gaussians
\begin{align}
 M^{(1)} = \sum_i w_i^{(1)} M_i \\
 M^{(2)} = \sum_i w_i^{(2)} M_i 
\end{align}
We find a mass fraction of $40\%$ for component 1 and $60\%$ for component 2. These fractions refer to the total mass enclosed in our MGE mass model (table~\ref{tab:mge_model_parameters_mass_extraction}) and are quite uncertain. The indication is that the two counter rotating discs have roughly equal masses within the range of our photometry. This nearly equal contribution for the two discs is similar to what was found for the S0 galaxy NGC~4550, which has qualitatively similar kinematics and was also modelled in detail \citep{2007MNRAS.379..418C}.

\section{Discussion}

\subsection{Counter rotating populations}

We reliably showed that the 2$\sigma$ appearance of NGC~5102 is due to two counter rotating discs by disentangling their contribution to the LOSVD. We found that component 2 is more concentrated and rotates slower than component~1, but the resolution of the MUSE data is too low to reliably measure the rotation velocity. The \ion{H}{I} gas has a regular rotation pattern with a flat outer rotation velocity of roughly 95\,km\,s$^{-1}$ \citep{1993A&A...269...15V, 2015MNRAS.452.3139K}. The rotation axis of the gas is essentially aligned with the rotation axis of the stars and the sense of rotation is identical to the more concentrated component 2.

It is very tempting to describe the formation of NGC 5102 by the following scenario: The less concentrated disc (component 1) was in place early on, when a cloud of \ion{H}{I} gas with opposite momentum was swallowed by NGC\,5102. Form this counter rotating gas then the counter rotating stellar disc was formed. This interpretation would need an investigation of the stellar population of the two counter rotating discs, which is not possible with the spectral resolution of the MUSE spectra.

\subsection{Population}

The population fits show a strong gradient in age and metallicity, exhibiting in the centre a mass weighted mean age below $8\times10^8$\,yr. The age gradient flattens at a mass weighted mean age of roughly $2.5\times10^9$\,yr. The metallicity gradient we find is in contrast to previous studies, as e.g. \citet{2015ApJ...799...97D} based on index measurements finds no metallicity gradient.

Our age values are in broad agreement with the ages of 1 and 2\,Gyr \citet{2015ApJ...799...97D} finds for the nuclear and the bulge spectra by fitting SSP models with fixed metallicity of $Z=0.004$ ([Fe/H]\,$\approx -0.7$). In light of the strong metallicity gradient we find the assumption of a constant metallicity questionable, but \citet{2015ApJ...799...97D} already notes that solar metallicity models would result in 0.3\,dex younger ages. 

\citet{2005ApJ...625..785K} find a dominant old (3\,Gyr) super solar ($Z=1.5$) population and a less prominent a young (0.3\,Gyr) metal poor ($Z=0.2$) population in the centre. Neither our work nor the one by \citet{2015ApJ...799...97D} can confirm a dominant metal rich 3\,Gyr population. There are several reasons for the mismatch:  \citet{2005ApJ...625..785K} use archival spectra from four different sources, which they stick together. This might introduce unknown systematic uncertainties in general and especially for NGC~5102: All spectra need to be exposed at exactly the same position, otherwise the strong population gradient makes the concatenated spectrum meaningless. There is not much experience in the literature with fitting spectra from the UV to the NIR, complicating the comparison of our results with the fit of \citet{2005ApJ...625..785K} to the extended wavelength range from 2000 to 9800\,\AA.

The strong population gradients we observe might be influenced by the presence of the two kinematic components. To investigate this a decomposition of the populations of the two counter rotating discs would be necessary. This decomposition works if the two components are well separated in their kinematics or have quiet different populations. Such a decomposition is beyond the scope of this paper, but based on the spatial extend of the age and metallicity gradients we do see some indication that the two kinematic components have different stellar populations. In that case the more concentrated component 2 would have a young and metal rich population, while the more extended component 1 would be older and metal poor. Validating this indication would give interesting insights into the evolution of NGC~5102 and might also explain the strong $(M_{\ast}/L)_{\rm Salp}$ gradient.

\subsection{JAM modelling}

The strong disagreement of the self-consistent JAM models (a) (mass follows light) and (b) (stars only) with the data 
shows that this low mass galaxy has a different dark matter content from most of the higher mass ATLAS$^{\rm 3D}$ ETGs. The JAM models make strong assumptions about the galaxy that might not be fulfilled for real galaxies, especially that the velocity ellipsoid is aligned with cylinder coordinates. The fact that there is a large number of observed fast rotating galaxies that can be approximated quite well, within $1R_{\rm e}$, using the self-consistent JAM models \citep[e.g.][]{2013MNRAS.432.1709C} and further supported from the analysis of simulated galaxies \citep{2012MNRAS.424.1495L, 2016MNRAS.455.3680L} makes it unlikely that anisotropy is driving the observed differences, and rather indicates a difference in the dark matter content.

\begin{figure}
 \includegraphics[width=\columnwidth]{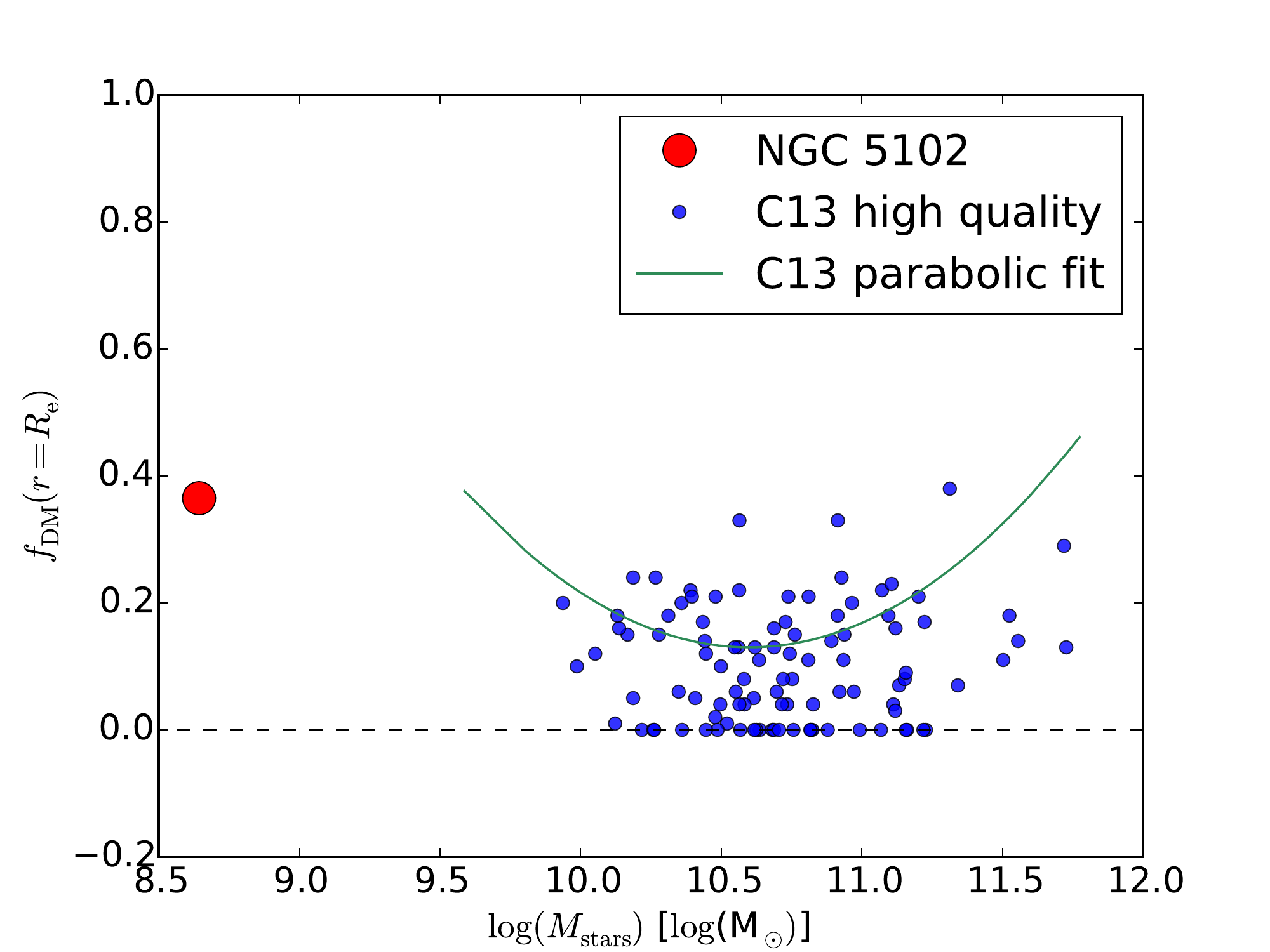}
 \caption{Comparison of the dark matter fraction inside a sphere of radius $r=R_{\rm e}$, assuming the halo has an NFW profile. The NGC~5102 value is from this paper (JAM model (c)), the blue circles (sub sample with high quality kinematic data) are from \citet{2013MNRAS.432.1709C} (mass) and \citet{2013MNRAS.432.1862C} (dark matter fractions). The parabola $f_{\rm DM} = 0.13 + 0.24(\log M_{\rm stars} -10.6)^2$ is the fit by \citet{2013MNRAS.432.1709C} to the sample with NFW haloes, that have masses determined via $M_{200} - M_{\rm stars}$ scaling relations from simulations.}
 \label{fig:dark_matter_frac_comp}
\end{figure}

Including an NFW dark matter halo into the JAM models brings the model and the data into good agreement. The dark matter fraction of $0.37\pm0.04$ we find is relatively high, as shown by the comparison with the dark matter fractions from \citet{2013MNRAS.432.1862C} in Fig.~\ref{fig:dark_matter_frac_comp}. The model with an NFW dark matter halo requires a relatively heavy IMF, an unexpected result for this relatively low mass galaxy. The more general model with a power law dark matter halo with a free inner slope explains this result: The halo slope and the IMF normalisation are degenerate and anti-correlated. These models favour a more Chabrier/Kroupa like IMF and a steeper than NFW dark halo slope, resulting in a larger dark matter fraction inside $1R{\rm e}$. This matches with the results of \citet{2016MNRAS.455..308T} obtained on a sample of 39 Virgo cluster dwarf early type galaxies: dark matter fraction and IMF-normalisation correction are anti-correlated and dwarf early types have on average slightly more massive IMFs than Chabrier. This means the dark matter fraction in Fig.~\ref{fig:dark_matter_frac_comp} represents a lower limit to the dark matter fraction inside a sphere of $1R_{\rm e}$ in NGC~5102. A light IMF would be more consistent with the expectations from the IMF-$\sigma$ relation, as indicated either by similar JAM models \citep{2012Natur.484..485C, 2013MNRAS.432.1862C, 2015MNRAS.446..493P}, or by stellar population models \citep{2012ApJ...760...71C, 2012ApJ...753L..32S, 2013MNRAS.429L..15F}. However the present results do not allow to constrain the IMF normalization with significant accuracy.

Previous literature values for the inclination of NGC~5102 are all based on axial ratios and make an assumption on the intrinsic flattening of the galaxy:  $64.4\degr$ \citep[][ intrinsic flattening unknown]{2008AJ....135.1636D}, $66\degr$ \citep[][ $q_{\rm intr}=0.2$ ]{2010AJ....139..984B} and $70 \pm 6\degr$ \citep[][ $q_{\rm intr}=0.2$ and adding a $3\degr$ offset]{2015MNRAS.452.3139K}. Assuming a flat disc we compute a minimum inclination of $66.3\degr$ from the flattest MGE ($q=0.4$) in table~\ref{tab:mge_model_parameters_photometry_extraction} and \ref{tab:mge_model_parameters_mass_extraction}. Our $q=0.4$ is in good agreement with the value of $q=0.38\pm0.03$ we obtain from the HyperLeda $\text{logr25}=0.42\pm0.04$ \citep{2014A&A...570A..13M}. Even though the flattening of the outermost Gaussian in it self is not very robust, the comparison with the other sources shows that it is reasonable. The problem with inclinations estimated via axial ratios is the assumption of the intrinsic flattening that can not be validated. \citet{2014MNRAS.444.3340W} determine that the mean {\em intrinsic} flattening of the ATLAS$^{\rm 3D}$ fast rotator galaxies is $q_{\rm intr}=0.25$. Under this assumption and taking the HyperLeda observed axial ratio the minimum inclination becomes $73\degr$. This is even more important for dwarf galaxies that tend to be rounder. \citet{2016ApJ...820...69S} determine the mean intrinsic axial ratio of Virgo dwarf galaxies to be $q_{\rm intr}=0.57$ - even rounder than the observed one of NGC~5102. In summary: if the intrinsic axial ratio of NGC~5102 is intermediate between the ATLAS$^{\rm 3D}$ and the Virgo dwarf galaxy value it might well be that we see NGC~5102 edge-on. However our constraints are not very tight and any inclination between $75\degr$ and $90\degr$ would be consistent with the data at $3\sigma$ confidence.

\section{Summary}

We analyse the kinematics and stellar population of the nearby low mass S0 galaxy NGC 5102 with MUSE spectra. The data cover the central part of the galaxy, out to $1R_{\rm e}$ along the major axis. We fit Gaussian (2 moment) and Gauss-Hermite (4 moment) LOSVDs to two different wavelength regions (the overlap between MUSE and the MILES library and the \ion{Ca}{II} triplet). For the first time we reveal that NGC~5102 is a $2\sigma$ galaxy, showing the typical two dispersion peaks along the major axis and at the same position a reversal of the sense of rotation. The results obtained in the two different wavelength regions agree within the expected uncertainties. The four moment extraction finds larger rotational velocity amplitudes, a sign for two counter rotating discs in this galaxy. 

We disentangle the rotation of the two counter rotating discs by fitting the MUSE spectra with two independent Gaussian LOSVDs. The most likely explanation for the evolution of the two counter rotating discs is that the more extended disc (component 1) was in place when gas with opposite momentum was swallowed by NGC~5102. The counter rotating gas has been detected by \citet{1993A&A...269...15V} and most likely created the counter rotating disc. Unfortunately the resolution of the MUSE spectra is too low to fit individual populations for the two counter rotating discs, as this would  help in further constraining the evolution of this galaxy.

We fit the stellar population of NGC~5102. This fitting exhibits strong gradients in the mean stellar age, metallicity and the stellar mass to light ratio. The unusual blue bulge of NGC~5102 has been known since the early 1970th and was attributed to a young nuclear cluster. To our knowledge a strong metallicity gradient has not been discussed in the literature yet. The extent of the young metal rich population does not spatially coincide with the kinematic signature of the counter rotating disc.

We modelled the dynamics of NGC~5102 using the Jeans anisotropic modelling method. 
The JAM models using the simple mass-follows-light assumption, as well as the stars-only model, are not able to reproduce the observed stellar kinematics.

Including dark matter in the JAM models immediately solves the difficulties in reproducing the observed kinematics. The model with an NFW dark halo requires a dark matter fraction of $0.37\pm0.04$ inside a sphere of a radius of one $R_{\rm e}$ and a heavy IMF. Our model with a general NFW dark halo (i.e. a power law halo with a free inner slope) shows that the dark matter fraction is degenerate with the IMF normalisation. Nevertheless these models prefer a light weight IMF, steeper than an NFW dark halo slope and accordingly a higher dark matter fraction. Therefore the indication is that the model with an NFW halo (and a relatively heavy IMF) is a lower limit to the dark matter fraction. 

A more robust result than the dark halo slope is the slope of the {\em total} mass density. We model the mass density as a power law and find excellent agreement with the data for a slope of $-1.75 \pm 0.04$. This slope is shallower than the slope of an isothermal halo ($-2$) and also shallower than the slope \citet{2015ApJ...804L..21C} report for higher mass fast rotating galaxies. This agrees qualitatively with the finding of  \citet{2016ARA&A..54..597C} that the halo slope decreases with decreasing stellar mass.

\section*{Acknowledgements}
This research has made use of the NASA/IPAC Extragalactic Database (NED) which is operated by the Jet Propulsion Laboratory, California Institute of Technology, under contract with the National Aeronautics and Space Administration. We acknowledge the usage of the HyperLeda database (http://leda.univ-lyon1.fr). We thank the anonymous referee for helpful comments. MM is grateful for financial support from the Leibniz Graduate School for Quantitative Spectroscopy in Astrophysics, a joint project of the Leibniz Institute for Astrophysics Potsdam (AIP) and the Institute of Physics and Astronomy of the University of Potsdam (UP). MC acknowledges support from a Royal Society University Research Fellowship. CJW acknowledges support through the Marie Curie Career Integration Grant 303912.




\bibliographystyle{mnras}
\bibliography{Mitzkus_et_al_2016} 

\begin{thebibliography}{}
\makeatletter
\relax
\def\mn@urlcharsother{\let\do\@makeother \do\$\do\&\do\#\do\^\do\_\do\%\do\~}
\def\mn@doi{\begingroup\mn@urlcharsother \@ifnextchar [ {\mn@doi@}
  {\mn@doi@[]}}
\def\mn@doi@[#1]#2{\def\@tempa{#1}\ifx\@tempa\@empty \href
  {http://dx.doi.org/#2} {doi:#2}\else \href {http://dx.doi.org/#2} {#1}\fi
  \endgroup}
\def\mn@eprint#1#2{\mn@eprint@#1:#2::\@nil}
\def\mn@eprint@arXiv#1{\href {http://arxiv.org/abs/#1} {{\tt arXiv:#1}}}
\def\mn@eprint@dblp#1{\href {http://dblp.uni-trier.de/rec/bibtex/#1.xml}
  {dblp:#1}}
\def\mn@eprint@#1:#2:#3:#4\@nil{\def\@tempa {#1}\def\@tempb {#2}\def\@tempc
  {#3}\ifx \@tempc \@empty \let \@tempc \@tempb \let \@tempb \@tempa \fi \ifx
  \@tempb \@empty \def\@tempb {arXiv}\fi \@ifundefined
  {mn@eprint@\@tempb}{\@tempb:\@tempc}{\expandafter \expandafter \csname
  mn@eprint@\@tempb\endcsname \expandafter{\@tempc}}}

\bibitem[\protect\citeauthoryear{{Bacon} et~al.,}{{Bacon}
  et~al.}{2010}]{2010SPIE.7735E..08B}
{Bacon} R.,  et~al., 2010, in Society of Photo-Optical Instrumentation
  Engineers (SPIE) Conference Series. p. 773508, \mn@doi{10.1117/12.856027}

\bibitem[\protect\citeauthoryear{{Beaulieu}, {Freeman}, {Hidalgo}, {Norman}  \&
  {Quinn}}{{Beaulieu} et~al.}{2010}]{2010AJ....139..984B}
{Beaulieu} S.~F.,  {Freeman} K.~C.,  {Hidalgo} S.~L.,  {Norman} C.~A.,
  {Quinn} P.~J.,  2010, \mn@doi [\aj] {10.1088/0004-6256/139/3/984}, \href
  {http://adsabs.harvard.edu/abs/2010AJ....139..984B} {139, 984}

\bibitem[\protect\citeauthoryear{{Beifiori}, {Maraston}, {Thomas}  \&
  {Johansson}}{{Beifiori} et~al.}{2011}]{2011A&A...531A.109B}
{Beifiori} A.,  {Maraston} C.,  {Thomas} D.,   {Johansson} J.,  2011, \mn@doi
  [\aap] {10.1051/0004-6361/201016323}, \href
  {http://adsabs.harvard.edu/abs/2011A%26A...531A.109B} {531, A109}

\bibitem[\protect\citeauthoryear{{Bell} \& {de Jong}}{{Bell} \& {de
  Jong}}{2001}]{2001ApJ...550..212B}
{Bell} E.~F.,  {de Jong} R.~S.,  2001, \mn@doi [\apj] {10.1086/319728}, \href
  {http://adsabs.harvard.edu/abs/2001ApJ...550..212B} {550, 212}

\bibitem[\protect\citeauthoryear{{Bendo} \& {Joseph}}{{Bendo} \&
  {Joseph}}{2004}]{2004AJ....127.3338B}
{Bendo} G.~J.,  {Joseph} R.~D.,  2004, \mn@doi [\aj] {10.1086/420712}, \href
  {http://adsabs.harvard.edu/abs/2004AJ....127.3338B} {127, 3338}

\bibitem[\protect\citeauthoryear{{Blanton} \& {Roweis}}{{Blanton} \&
  {Roweis}}{2007}]{2007AJ....133..734B}
{Blanton} M.~R.,  {Roweis} S.,  2007, \mn@doi [\aj] {10.1086/510127}, \href
  {http://adsabs.harvard.edu/abs/2007AJ....133..734B} {133, 734}

\bibitem[\protect\citeauthoryear{{Cappellari}}{{Cappellari}}{2002}]{2002MNRAS.333..400C}
{Cappellari} M.,  2002, \mn@doi [\mnras] {10.1046/j.1365-8711.2002.05412.x},
  \href {http://adsabs.harvard.edu/abs/2002MNRAS.333..400C} {333, 400}

\bibitem[\protect\citeauthoryear{{Cappellari}}{{Cappellari}}{2008}]{2008MNRAS.390...71C}
{Cappellari} M.,  2008, \mn@doi [\mnras] {10.1111/j.1365-2966.2008.13754.x},
  \href {http://adsabs.harvard.edu/abs/2008MNRAS.390...71C} {390, 71}

\bibitem[\protect\citeauthoryear{{Cappellari}}{{Cappellari}}{2016}]{2016ARA&A..54..597C}
{Cappellari} M.,  2016, \mn@doi [\araa] {10.1146/annurev-astro-082214-122432},
  \href {http://adsabs.harvard.edu/abs/2016ARA%26A..54..597C} {54, 597}

\bibitem[\protect\citeauthoryear{{Cappellari} \& {Copin}}{{Cappellari} \&
  {Copin}}{2003}]{2003MNRAS.342..345C}
{Cappellari} M.,  {Copin} Y.,  2003, \mn@doi [\mnras]
  {10.1046/j.1365-8711.2003.06541.x}, \href
  {http://adsabs.harvard.edu/abs/2003MNRAS.342..345C} {342, 345}

\bibitem[\protect\citeauthoryear{{Cappellari} \& {Emsellem}}{{Cappellari} \&
  {Emsellem}}{2004}]{2004PASP..116..138C}
{Cappellari} M.,  {Emsellem} E.,  2004, \mn@doi [\pasp] {10.1086/381875}, \href
  {http://adsabs.harvard.edu/abs/2004PASP..116..138C} {116, 138}

\bibitem[\protect\citeauthoryear{{Cappellari} et~al.,}{{Cappellari}
  et~al.}{2007}]{2007MNRAS.379..418C}
{Cappellari} M.,  et~al., 2007, \mn@doi [\mnras]
  {10.1111/j.1365-2966.2007.11963.x}, \href
  {http://adsabs.harvard.edu/abs/2007MNRAS.379..418C} {379, 418}

\bibitem[\protect\citeauthoryear{{Cappellari} et~al.,}{{Cappellari}
  et~al.}{2011}]{2011MNRAS.413..813C}
{Cappellari} M.,  et~al., 2011, \mn@doi [\mnras]
  {10.1111/j.1365-2966.2010.18174.x}, \href
  {http://adsabs.harvard.edu/abs/2011MNRAS.413..813C} {413, 813}

\bibitem[\protect\citeauthoryear{{Cappellari} et~al.,}{{Cappellari}
  et~al.}{2012}]{2012Natur.484..485C}
{Cappellari} M.,  et~al., 2012, \mn@doi [\nat] {10.1038/nature10972}, \href
  {http://adsabs.harvard.edu/abs/2012Natur.484..485C} {484, 485}

\bibitem[\protect\citeauthoryear{{Cappellari} et~al.,}{{Cappellari}
  et~al.}{2013a}]{2013MNRAS.432.1709C}
{Cappellari} M.,  et~al., 2013a, \mn@doi [\mnras] {10.1093/mnras/stt562}, \href
  {http://adsabs.harvard.edu/abs/2013MNRAS.432.1709C} {432, 1709}

\bibitem[\protect\citeauthoryear{{Cappellari} et~al.,}{{Cappellari}
  et~al.}{2013b}]{2013MNRAS.432.1862C}
{Cappellari} M.,  et~al., 2013b, \mn@doi [\mnras] {10.1093/mnras/stt644}, \href
  {http://adsabs.harvard.edu/abs/2013MNRAS.432.1862C} {432, 1862}

\bibitem[\protect\citeauthoryear{{Cappellari} et~al.,}{{Cappellari}
  et~al.}{2015}]{2015ApJ...804L..21C}
{Cappellari} M.,  et~al., 2015, \mn@doi [\apjl] {10.1088/2041-8205/804/1/L21},
  \href {http://adsabs.harvard.edu/abs/2015ApJ...804L..21C} {804, L21}

\bibitem[\protect\citeauthoryear{{Cenarro}, {Cardiel}, {Gorgas}, {Peletier},
  {Vazdekis}  \& {Prada}}{{Cenarro} et~al.}{2001}]{2001MNRAS.326..959C}
{Cenarro} A.~J.,  {Cardiel} N.,  {Gorgas} J.,  {Peletier} R.~F.,  {Vazdekis}
  A.,   {Prada} F.,  2001, \mn@doi [\mnras] {10.1046/j.1365-8711.2001.04688.x},
  \href {http://adsabs.harvard.edu/abs/2001MNRAS.326..959C} {326, 959}

\bibitem[\protect\citeauthoryear{{Coccato}, {Morelli}, {Corsini}, {Buson},
  {Pizzella}, {Vergani}  \& {Bertola}}{{Coccato}
  et~al.}{2011}]{2011MNRAS.412L.113C}
{Coccato} L.,  {Morelli} L.,  {Corsini} E.~M.,  {Buson} L.,  {Pizzella} A.,
  {Vergani} D.,   {Bertola} F.,  2011, \mn@doi [\mnras]
  {10.1111/j.1745-3933.2011.01016.x}, \href
  {http://adsabs.harvard.edu/abs/2011MNRAS.412L.113C} {412, L113}

\bibitem[\protect\citeauthoryear{{Conroy} \& {van Dokkum}}{{Conroy} \& {van
  Dokkum}}{2012}]{2012ApJ...760...71C}
{Conroy} C.,  {van Dokkum} P.~G.,  2012, \mn@doi [\apj]
  {10.1088/0004-637X/760/1/71}, \href
  {http://adsabs.harvard.edu/abs/2012ApJ...760...71C} {760, 71}

\bibitem[\protect\citeauthoryear{{Davidge}}{{Davidge}}{2008}]{2008AJ....135.1636D}
{Davidge} T.~J.,  2008, \mn@doi [\aj] {10.1088/0004-6256/135/4/1636}, \href
  {http://adsabs.harvard.edu/abs/2008AJ....135.1636D} {135, 1636}

\bibitem[\protect\citeauthoryear{{Davidge}}{{Davidge}}{2010}]{2010AJ....139..680D}
{Davidge} T.~J.,  2010, \mn@doi [\aj] {10.1088/0004-6256/139/2/680}, \href
  {http://adsabs.harvard.edu/abs/2010AJ....139..680D} {139, 680}

\bibitem[\protect\citeauthoryear{{Davidge}}{{Davidge}}{2015}]{2015ApJ...799...97D}
{Davidge} T.~J.,  2015, \mn@doi [\apj] {10.1088/0004-637X/799/1/97}, \href
  {http://adsabs.harvard.edu/abs/2015ApJ...799...97D} {799, 97}

\bibitem[\protect\citeauthoryear{{Deharveng}, {Jedrzejewski}, {Crane}, {Disney}
   \& {Rocca-Volmerange}}{{Deharveng} et~al.}{1997}]{1997A&A...326..528D}
{Deharveng} J.-M.,  {Jedrzejewski} R.,  {Crane} P.,  {Disney} M.~J.,
  {Rocca-Volmerange} B.,  1997, \aap, \href
  {http://adsabs.harvard.edu/abs/1997A%26A...326..528D} {326, 528}

\bibitem[\protect\citeauthoryear{{Dolphin}}{{Dolphin}}{2009}]{2009PASP..121..655D}
{Dolphin} A.~E.,  2009, \mn@doi [\pasp] {10.1086/600028}, \href
  {http://adsabs.harvard.edu/abs/2009PASP..121..655D} {121, 655}

\bibitem[\protect\citeauthoryear{{Doyle} et~al.,}{{Doyle}
  et~al.}{2005}]{2005MNRAS.361...34D}
{Doyle} M.~T.,  et~al., 2005, \mn@doi [\mnras]
  {10.1111/j.1365-2966.2005.09159.x}, \href
  {http://adsabs.harvard.edu/abs/2005MNRAS.361...34D} {361, 34}

\bibitem[\protect\citeauthoryear{{Eggen}}{{Eggen}}{1971}]{1971QJRAS..12..305E}
{Eggen} O.~J.,  1971, \qjras, \href
  {http://adsabs.harvard.edu/abs/1971QJRAS..12..305E} {12, 305}

\bibitem[\protect\citeauthoryear{{Emsellem}, {Monnet}  \& {Bacon}}{{Emsellem}
  et~al.}{1994}]{1994A&A...285..723E}
{Emsellem} E.,  {Monnet} G.,   {Bacon} R.,  1994, \aap, \href
  {http://adsabs.harvard.edu/abs/1994A%26A...285..723E} {285}

\bibitem[\protect\citeauthoryear{{Falc{\'o}n-Barroso},
  {S{\'a}nchez-Bl{\'a}zquez}, {Vazdekis}, {Ricciardelli}, {Cardiel}, {Cenarro},
  {Gorgas}  \& {Peletier}}{{Falc{\'o}n-Barroso}
  et~al.}{2011}]{2011A&A...532A..95F}
{Falc{\'o}n-Barroso} J.,  {S{\'a}nchez-Bl{\'a}zquez} P.,  {Vazdekis} A.,
  {Ricciardelli} E.,  {Cardiel} N.,  {Cenarro} A.~J.,  {Gorgas} J.,
  {Peletier} R.~F.,  2011, \mn@doi [\aap] {10.1051/0004-6361/201116842}, \href
  {http://adsabs.harvard.edu/abs/2011A%26A...532A..95F} {532, A95}

\bibitem[\protect\citeauthoryear{{Ferreras}, {La Barbera}, {de la Rosa},
  {Vazdekis}, {de Carvalho}, {Falc{\'o}n-Barroso}  \&
  {Ricciardelli}}{{Ferreras} et~al.}{2013}]{2013MNRAS.429L..15F}
{Ferreras} I.,  {La Barbera} F.,  {de la Rosa} I.~G.,  {Vazdekis} A.,  {de
  Carvalho} R.~R.,  {Falc{\'o}n-Barroso} J.,   {Ricciardelli} E.,  2013,
  \mn@doi [\mnras] {10.1093/mnrasl/sls014}, \href
  {http://adsabs.harvard.edu/abs/2013MNRAS.429L..15F} {429, L15}

\bibitem[\protect\citeauthoryear{{Foreman-Mackey}, {Hogg}, {Lang}  \&
  {Goodman}}{{Foreman-Mackey} et~al.}{2013}]{2013PASP..125..306F}
{Foreman-Mackey} D.,  {Hogg} D.~W.,  {Lang} D.,   {Goodman} J.,  2013, \mn@doi
  [\pasp] {10.1086/670067}, \href
  {http://adsabs.harvard.edu/abs/2013PASP..125..306F} {125, 306}

\bibitem[\protect\citeauthoryear{{Foster}, {Arnold}, {Forbes}, {Pastorello},
  {Romanowsky}, {Spitler}, {Strader}  \& {Brodie}}{{Foster}
  et~al.}{2013}]{2013MNRAS.435.3587F}
{Foster} C.,  {Arnold} J.~A.,  {Forbes} D.~A.,  {Pastorello} N.,  {Romanowsky}
  A.~J.,  {Spitler} L.~R.,  {Strader} J.,   {Brodie} J.~P.,  2013, \mn@doi
  [\mnras] {10.1093/mnras/stt1550}, \href
  {http://adsabs.harvard.edu/abs/2013MNRAS.435.3587F} {435, 3587}

\bibitem[\protect\citeauthoryear{{Gallagher}, {Faber}  \& {Balick}}{{Gallagher}
  et~al.}{1975}]{1975ApJ...202....7G}
{Gallagher} J.~S.,  {Faber} S.~M.,   {Balick} B.,  1975, \mn@doi [\apj]
  {10.1086/153948}, \href {http://adsabs.harvard.edu/abs/1975ApJ...202....7G}
  {202, 7}

\bibitem[\protect\citeauthoryear{{Girardi}, {Bressan}, {Bertelli}  \&
  {Chiosi}}{{Girardi} et~al.}{2000}]{2000A&AS..141..371G}
{Girardi} L.,  {Bressan} A.,  {Bertelli} G.,   {Chiosi} C.,  2000, \mn@doi
  [\aaps] {10.1051/aas:2000126}, \href
  {http://adsabs.harvard.edu/abs/2000A%26AS..141..371G} {141, 371}

\bibitem[\protect\citeauthoryear{{Goodman} \& {Weare}}{{Goodman} \&
  {Weare}}{2010}]{2010AMaCS...5....65G}
{Goodman} J.,  {Weare} J.,  2010, \mn@doi [Comm. App. Math and Comp. Sci.]
  {10.2140/camcos.2010.5.65}, 5, 65

\bibitem[\protect\citeauthoryear{{Holtzman}, {Burrows}, {Casertano}, {Hester},
  {Trauger}, {Watson}  \& {Worthey}}{{Holtzman}
  et~al.}{1995}]{1995PASP..107.1065H}
{Holtzman} J.~A.,  {Burrows} C.~J.,  {Casertano} S.,  {Hester} J.~J.,
  {Trauger} J.~T.,  {Watson} A.~M.,   {Worthey} G.,  1995, \mn@doi [\pasp]
  {10.1086/133664}, \href {http://adsabs.harvard.edu/abs/1995PASP..107.1065H}
  {107, 1065}

\bibitem[\protect\citeauthoryear{{Johnston}, {Merrifield},
  {Arag{\'o}n-Salamanca}  \& {Cappellari}}{{Johnston}
  et~al.}{2013}]{2013MNRAS.428.1296J}
{Johnston} E.~J.,  {Merrifield} M.~R.,  {Arag{\'o}n-Salamanca} A.,
  {Cappellari} M.,  2013, \mn@doi [\mnras] {10.1093/mnras/sts121}, \href
  {http://adsabs.harvard.edu/abs/2013MNRAS.428.1296J} {428, 1296}

\bibitem[\protect\citeauthoryear{{Kamphuis}, {J{\'o}zsa}, {Oh}, {Spekkens},
  {Urbancic}, {Serra}, {Koribalski}  \& {Dettmar}}{{Kamphuis}
  et~al.}{2015}]{2015MNRAS.452.3139K}
{Kamphuis} P.,  {J{\'o}zsa} G.~I.~G.,  {Oh} S.-.~H.,  {Spekkens} K.,
  {Urbancic} N.,  {Serra} P.,  {Koribalski} B.~S.,   {Dettmar} R.-J.,  2015,
  \mn@doi [\mnras] {10.1093/mnras/stv1480}, \href
  {http://adsabs.harvard.edu/abs/2015MNRAS.452.3139K} {452, 3139}

\bibitem[\protect\citeauthoryear{{Karachentsev} et~al.,}{{Karachentsev}
  et~al.}{2002}]{2002A&A...385...21K}
{Karachentsev} I.~D.,  et~al., 2002, \mn@doi [\aap]
  {10.1051/0004-6361:20020042}, \href
  {http://adsabs.harvard.edu/abs/2002A%26A...385...21K} {385, 21}

\bibitem[\protect\citeauthoryear{{Karachentsev} et~al.,}{{Karachentsev}
  et~al.}{2007}]{2007AJ....133..504K}
{Karachentsev} I.~D.,  et~al., 2007, \mn@doi [\aj] {10.1086/510125}, \href
  {http://adsabs.harvard.edu/abs/2007AJ....133..504K} {133, 504}

\bibitem[\protect\citeauthoryear{{Klypin}, {Trujillo-Gomez}  \&
  {Primack}}{{Klypin} et~al.}{2011}]{2011ApJ...740..102K}
{Klypin} A.~A.,  {Trujillo-Gomez} S.,   {Primack} J.,  2011, \mn@doi [\apj]
  {10.1088/0004-637X/740/2/102}, \href
  {http://adsabs.harvard.edu/abs/2011ApJ...740..102K} {740, 102}

\bibitem[\protect\citeauthoryear{{Koopmans} et~al.,}{{Koopmans}
  et~al.}{2009}]{2009ApJ...703L..51K}
{Koopmans} L.~V.~E.,  et~al., 2009, \mn@doi [\apjl]
  {10.1088/0004-637X/703/1/L51}, \href
  {http://adsabs.harvard.edu/abs/2009ApJ...703L..51K} {703, L51}

\bibitem[\protect\citeauthoryear{{Kraft}, {Nolan}, {Ponman}, {Jones}  \&
  {Raychaudhury}}{{Kraft} et~al.}{2005}]{2005ApJ...625..785K}
{Kraft} R.~P.,  {Nolan} L.~A.,  {Ponman} T.~J.,  {Jones} C.,   {Raychaudhury}
  S.,  2005, \mn@doi [\apj] {10.1086/429982}, \href
  {http://adsabs.harvard.edu/abs/2005ApJ...625..785K} {625, 785}

\bibitem[\protect\citeauthoryear{{Krajnovi{\'c}}, {Cappellari}, {de Zeeuw}  \&
  {Copin}}{{Krajnovi{\'c}} et~al.}{2006}]{2006MNRAS.366..787K}
{Krajnovi{\'c}} D.,  {Cappellari} M.,  {de Zeeuw} P.~T.,   {Copin} Y.,  2006,
  \mn@doi [\mnras] {10.1111/j.1365-2966.2005.09902.x}, \href
  {http://adsabs.harvard.edu/abs/2006MNRAS.366..787K} {366, 787}

\bibitem[\protect\citeauthoryear{{Krajnovi{\'c}} et~al.,}{{Krajnovi{\'c}}
  et~al.}{2011}]{2011MNRAS.414.2923K}
{Krajnovi{\'c}} D.,  et~al., 2011, \mn@doi [\mnras]
  {10.1111/j.1365-2966.2011.18560.x}, \href
  {http://adsabs.harvard.edu/abs/2011MNRAS.414.2923K} {414, 2923}

\bibitem[\protect\citeauthoryear{{Krist}}{{Krist}}{1993}]{1993ASPC...52..536K}
{Krist} J.,  1993, in {Hanisch} R.~J.,  {Brissenden} R.~J.~V.,   {Barnes} J.,
  eds,  Astronomical Society of the Pacific Conference Series Vol. 52,
  Astronomical Data Analysis Software and Systems II. p.~536

\bibitem[\protect\citeauthoryear{{Krist}, {Hook}  \& {Stoehr}}{{Krist}
  et~al.}{2011}]{2011SPIE.8127E..0JK}
{Krist} J.~E.,  {Hook} R.~N.,   {Stoehr} F.,  2011, in Society of Photo-Optical
  Instrumentation Engineers (SPIE) Conference Series. p. 81270J,
  \mn@doi{10.1117/12.892762}

\bibitem[\protect\citeauthoryear{{Lablanche} et~al.,}{{Lablanche}
  et~al.}{2012}]{2012MNRAS.424.1495L}
{Lablanche} P.-Y.,  et~al., 2012, \mn@doi [\mnras]
  {10.1111/j.1365-2966.2012.21343.x}, \href
  {http://adsabs.harvard.edu/abs/2012MNRAS.424.1495L} {424, 1495}

\bibitem[\protect\citeauthoryear{{Li}, {Li}, {Mao}, {Xu}, {Long}  \&
  {Emsellem}}{{Li} et~al.}{2016}]{2016MNRAS.455.3680L}
{Li} H.,  {Li} R.,  {Mao} S.,  {Xu} D.,  {Long} R.~J.,   {Emsellem} E.,  2016,
  \mn@doi [\mnras] {10.1093/mnras/stv2565}, \href
  {http://adsabs.harvard.edu/abs/2016MNRAS.455.3680L} {455, 3680}

\bibitem[\protect\citeauthoryear{{Makarov}, {Prugniel}, {Terekhova}, {Courtois}
   \& {Vauglin}}{{Makarov} et~al.}{2014}]{2014A&A...570A..13M}
{Makarov} D.,  {Prugniel} P.,  {Terekhova} N.,  {Courtois} H.,   {Vauglin} I.,
  2014, \mn@doi [\aap] {10.1051/0004-6361/201423496}, \href
  {http://adsabs.harvard.edu/abs/2014A%26A...570A..13M} {570, A13}

\bibitem[\protect\citeauthoryear{{Markwardt}}{{Markwardt}}{2009}]{2009ASPC..411..251M}
{Markwardt} C.~B.,  2009, in {Bohlender} D.~A.,  {Durand} D.,   {Dowler} P.,
  eds,  Astronomical Society of the Pacific Conference Series Vol. 411,
  Astronomical Data Analysis Software and Systems XVIII. p.~251 (\mn@eprint
  {arXiv} {0902.2850})

\bibitem[\protect\citeauthoryear{{McMillan}, {Ciardullo}  \&
  {Jacoby}}{{McMillan} et~al.}{1994}]{1994AJ....108.1610M}
{McMillan} R.,  {Ciardullo} R.,   {Jacoby} G.~H.,  1994, \mn@doi [\aj]
  {10.1086/117181}, \href {http://adsabs.harvard.edu/abs/1994AJ....108.1610M}
  {108, 1610}

\bibitem[\protect\citeauthoryear{{Navarro}, {Frenk}  \& {White}}{{Navarro}
  et~al.}{1996}]{1996ApJ...462..563N}
{Navarro} J.~F.,  {Frenk} C.~S.,   {White} S.~D.~M.,  1996, \mn@doi [\apj]
  {10.1086/177173}, \href {http://adsabs.harvard.edu/abs/1996ApJ...462..563N}
  {462, 563}

\bibitem[\protect\citeauthoryear{{Posacki}, {Cappellari}, {Treu}, {Pellegrini}
  \& {Ciotti}}{{Posacki} et~al.}{2015}]{2015MNRAS.446..493P}
{Posacki} S.,  {Cappellari} M.,  {Treu} T.,  {Pellegrini} S.,   {Ciotti} L.,
  2015, \mn@doi [\mnras] {10.1093/mnras/stu2098}, \href
  {http://adsabs.harvard.edu/abs/2015MNRAS.446..493P} {446, 493}

\bibitem[\protect\citeauthoryear{{Pritchet}}{{Pritchet}}{1979}]{1979ApJ...231..354P}
{Pritchet} C.,  1979, \mn@doi [\apj] {10.1086/157198}, \href
  {http://adsabs.harvard.edu/abs/1979ApJ...231..354P} {231, 354}

\bibitem[\protect\citeauthoryear{{Ricciardelli}, {Vazdekis}, {Cenarro}  \&
  {Falc{\'o}n-Barroso}}{{Ricciardelli} et~al.}{2012}]{2012MNRAS.424..172R}
{Ricciardelli} E.,  {Vazdekis} A.,  {Cenarro} A.~J.,   {Falc{\'o}n-Barroso} J.,
   2012, \mn@doi [\mnras] {10.1111/j.1365-2966.2012.21178.x}, \href
  {http://adsabs.harvard.edu/abs/2012MNRAS.424..172R} {424, 172}

\bibitem[\protect\citeauthoryear{{Rix}, {Franx}, {Fisher}  \&
  {Illingworth}}{{Rix} et~al.}{1992}]{1992ApJ...400L...5R}
{Rix} H.-W.,  {Franx} M.,  {Fisher} D.,   {Illingworth} G.,  1992, \mn@doi
  [\apjl] {10.1086/186635}, \href
  {http://adsabs.harvard.edu/abs/1992ApJ...400L...5R} {400, L5}

\bibitem[\protect\citeauthoryear{{Rubin}, {Graham}  \& {Kenney}}{{Rubin}
  et~al.}{1992}]{1992ApJ...394L...9R}
{Rubin} V.~C.,  {Graham} J.~A.,   {Kenney} J.~D.~P.,  1992, \mn@doi [\apjl]
  {10.1086/186460}, \href {http://adsabs.harvard.edu/abs/1992ApJ...394L...9R}
  {394, L9}

\bibitem[\protect\citeauthoryear{{S{\'a}nchez-Bl{\'a}zquez}
  et~al.,}{{S{\'a}nchez-Bl{\'a}zquez} et~al.}{2006}]{2006MNRAS.371..703S}
{S{\'a}nchez-Bl{\'a}zquez} P.,  et~al., 2006, \mn@doi [\mnras]
  {10.1111/j.1365-2966.2006.10699.x}, \href
  {http://adsabs.harvard.edu/abs/2006MNRAS.371..703S} {371, 703}

\bibitem[\protect\citeauthoryear{{S{\'a}nchez-Janssen}
  et~al.,}{{S{\'a}nchez-Janssen} et~al.}{2016}]{2016ApJ...820...69S}
{S{\'a}nchez-Janssen} R.,  et~al., 2016, \mn@doi [\apj]
  {10.3847/0004-637X/820/1/69}, \href
  {http://adsabs.harvard.edu/abs/2016ApJ...820...69S} {820, 69}

\bibitem[\protect\citeauthoryear{{Sandage} \& {Visvanathan}}{{Sandage} \&
  {Visvanathan}}{1978}]{1978ApJ...223..707S}
{Sandage} A.,  {Visvanathan} N.,  1978, \mn@doi [\apj] {10.1086/156305}, \href
  {http://adsabs.harvard.edu/abs/1978ApJ...223..707S} {223, 707}

\bibitem[\protect\citeauthoryear{{Schlafly} \& {Finkbeiner}}{{Schlafly} \&
  {Finkbeiner}}{2011}]{2011ApJ...737..103S}
{Schlafly} E.~F.,  {Finkbeiner} D.~P.,  2011, \mn@doi [\apj]
  {10.1088/0004-637X/737/2/103}, \href
  {http://adsabs.harvard.edu/abs/2011ApJ...737..103S} {737, 103}

\bibitem[\protect\citeauthoryear{{Scott} et~al.,}{{Scott}
  et~al.}{2013}]{2013MNRAS.432.1894S}
{Scott} N.,  et~al., 2013, \mn@doi [\mnras] {10.1093/mnras/sts422}, \href
  {http://adsabs.harvard.edu/abs/2013MNRAS.432.1894S} {432, 1894}

\bibitem[\protect\citeauthoryear{{Serra}, {Oosterloo}, {Cappellari}, {den
  Heijer}  \& {J{\'o}zsa}}{{Serra} et~al.}{2016}]{2016MNRAS.tmp..782S}
{Serra} P.,  {Oosterloo} T.,  {Cappellari} M.,  {den Heijer} M.,   {J{\'o}zsa}
  G.~I.~G.,  2016, \mn@doi [\mnras] {10.1093/mnras/stw1010}, \href
  {http://adsabs.harvard.edu/abs/2016MNRAS.tmp..782S} {}

\bibitem[\protect\citeauthoryear{{Spiniello}, {Trager}, {Koopmans}  \&
  {Chen}}{{Spiniello} et~al.}{2012}]{2012ApJ...753L..32S}
{Spiniello} C.,  {Trager} S.~C.,  {Koopmans} L.~V.~E.,   {Chen} Y.~P.,  2012,
  \mn@doi [\apjl] {10.1088/2041-8205/753/2/L32}, \href
  {http://adsabs.harvard.edu/abs/2012ApJ...753L..32S} {753, L32}

\bibitem[\protect\citeauthoryear{{Tonry}, {Dressler}, {Blakeslee}, {Ajhar},
  {Fletcher}, {Luppino}, {Metzger}  \& {Moore}}{{Tonry}
  et~al.}{2001}]{2001ApJ...546..681T}
{Tonry} J.~L.,  {Dressler} A.,  {Blakeslee} J.~P.,  {Ajhar} E.~A.,  {Fletcher}
  A.~B.,  {Luppino} G.~A.,  {Metzger} M.~R.,   {Moore} C.~B.,  2001, \mn@doi
  [\apj] {10.1086/318301}, \href
  {http://adsabs.harvard.edu/abs/2001ApJ...546..681T} {546, 681}

\bibitem[\protect\citeauthoryear{{Tortora}, {La Barbera}  \&
  {Napolitano}}{{Tortora} et~al.}{2016}]{2016MNRAS.455..308T}
{Tortora} C.,  {La Barbera} F.,   {Napolitano} N.~R.,  2016, \mn@doi [\mnras]
  {10.1093/mnras/stv2250}, \href
  {http://adsabs.harvard.edu/abs/2016MNRAS.455..308T} {455, 308}

\bibitem[\protect\citeauthoryear{{Tully}, {Libeskind}, {Karachentsev},
  {Karachentseva}, {Rizzi}  \& {Shaya}}{{Tully}
  et~al.}{2015}]{2015ApJ...802L..25T}
{Tully} R.~B.,  {Libeskind} N.~I.,  {Karachentsev} I.~D.,  {Karachentseva}
  V.~E.,  {Rizzi} L.,   {Shaya} E.~J.,  2015, \mn@doi [\apjl]
  {10.1088/2041-8205/802/2/L25}, \href
  {http://adsabs.harvard.edu/abs/2015ApJ...802L..25T} {802, L25}

\bibitem[\protect\citeauthoryear{{Valdes}, {Gupta}, {Rose}, {Singh}  \&
  {Bell}}{{Valdes} et~al.}{2004}]{2004ApJS..152..251V}
{Valdes} F.,  {Gupta} R.,  {Rose} J.~A.,  {Singh} H.~P.,   {Bell} D.~J.,  2004,
  \mn@doi [\apjs] {10.1086/386343}, \href
  {http://adsabs.harvard.edu/abs/2004ApJS..152..251V} {152, 251}

\bibitem[\protect\citeauthoryear{{Vazdekis}, {Ricciardelli}, {Cenarro},
  {Rivero-Gonz{\'a}lez}, {D{\'{\i}}az-Garc{\'{\i}}a}  \&
  {Falc{\'o}n-Barroso}}{{Vazdekis} et~al.}{2012}]{2012MNRAS.424..157V}
{Vazdekis} A.,  {Ricciardelli} E.,  {Cenarro} A.~J.,  {Rivero-Gonz{\'a}lez}
  J.~G.,  {D{\'{\i}}az-Garc{\'{\i}}a} L.~A.,   {Falc{\'o}n-Barroso} J.,  2012,
  \mn@doi [\mnras] {10.1111/j.1365-2966.2012.21179.x}, \href
  {http://adsabs.harvard.edu/abs/2012MNRAS.424..157V} {424, 157}

\bibitem[\protect\citeauthoryear{{Weijmans}, {Krajnovi{\'c}}, {van de Ven},
  {Oosterloo}, {Morganti}  \& {de Zeeuw}}{{Weijmans}
  et~al.}{2008}]{2008MNRAS.383.1343W}
{Weijmans} A.-M.,  {Krajnovi{\'c}} D.,  {van de Ven} G.,  {Oosterloo} T.~A.,
  {Morganti} R.,   {de Zeeuw} P.~T.,  2008, \mn@doi [\mnras]
  {10.1111/j.1365-2966.2007.12680.x}, \href
  {http://adsabs.harvard.edu/abs/2008MNRAS.383.1343W} {383, 1343}

\bibitem[\protect\citeauthoryear{{Weijmans} et~al.,}{{Weijmans}
  et~al.}{2014}]{2014MNRAS.444.3340W}
{Weijmans} A.-M.,  et~al., 2014, \mn@doi [\mnras] {10.1093/mnras/stu1603},
  \href {http://adsabs.harvard.edu/abs/2014MNRAS.444.3340W} {444, 3340}

\bibitem[\protect\citeauthoryear{{Weilbacher}, {Streicher}, {Urrutia}, {Jarno},
  {P{\'e}contal-Rousset}, {Bacon}  \& {B{\"o}hm}}{{Weilbacher}
  et~al.}{2012}]{2012SPIE.8451E..0BW}
{Weilbacher} P.~M.,  {Streicher} O.,  {Urrutia} T.,  {Jarno} A.,
  {P{\'e}contal-Rousset} A.,  {Bacon} R.,   {B{\"o}hm} P.,  2012, in Society of
  Photo-Optical Instrumentation Engineers (SPIE) Conference Series. p. 84510B,
  \mn@doi{10.1117/12.925114}

\bibitem[\protect\citeauthoryear{{Wyithe}, {Turner}  \& {Spergel}}{{Wyithe}
  et~al.}{2001}]{2001ApJ...555..504W}
{Wyithe} J.~S.~B.,  {Turner} E.~L.,   {Spergel} D.~N.,  2001, \mn@doi [\apj]
  {10.1086/321437}, \href {http://adsabs.harvard.edu/abs/2001ApJ...555..504W}
  {555, 504}

\bibitem[\protect\citeauthoryear{{van Woerden}, {van Driel}, {Braun}  \&
  {Rots}}{{van Woerden} et~al.}{1993}]{1993A&A...269...15V}
{van Woerden} H.,  {van Driel} W.,  {Braun} R.,   {Rots} A.~H.,  1993, \aap,
  \href {http://adsabs.harvard.edu/abs/1993A%26A...269...15V} {269, 15}

\bibitem[\protect\citeauthoryear{{van den Bergh}}{{van den
  Bergh}}{1976}]{1976AJ.....81..795V}
{van den Bergh} S.,  1976, \mn@doi [\aj] {10.1086/111956}, \href
  {http://adsabs.harvard.edu/abs/1976AJ.....81..795V} {81, 795}

\bibitem[\protect\citeauthoryear{{van den Bosch} \& {van de Ven}}{{van den
  Bosch} \& {van de Ven}}{2009}]{2009MNRAS.398.1117V}
{van den Bosch} R.~C.~E.,  {van de Ven} G.,  2009, \mn@doi [\mnras]
  {10.1111/j.1365-2966.2009.15177.x}, \href
  {http://adsabs.harvard.edu/abs/2009MNRAS.398.1117V} {398, 1117}

\makeatother
\end{thebibliography}




\appendix


\bsp	
\label{lastpage}
\end{document}